\documentclass[12pt]{article}
\pdfoutput=1
\usepackage{amsmath,amssymb,amsthm,bm,graphicx,float,array,multirow,multicol,rotfloat,caption,subcaption,hyperref,cleveref,enumerate,geometry,mathdots,adjustbox,booktabs,parskip,mathtools,tikz,tikz-cd,pdflscape,csquotes,lscape,rotating,empheq,mathrsfs,longtable}
\usepackage{braket,mleftright}
\usepackage[para]{threeparttable}
\usepackage[all]{xy}
\usepackage[normalem]{ulem}
\usepackage[aligntableaux=center]{ytableau}
\usepackage[numbers,sort&compress]{natbib}
\numberwithin{equation}{section}
\allowdisplaybreaks
\makeatletter

\definecolor{darkgreen}{rgb}{0,0.5,0}
\definecolor{darkred}{rgb}{0.5,0,0}
\definecolor{darkblue}{rgb}{0,0,0.6}
\definecolor{purple}{rgb}{0.4,.2,0.7}

\newcommand{\la}{\ell_A}
\newcommand{\lb}{\ell_B}
\newcommand{\lc}{\ell_C}
\newcommand{\ld}{\ell_D}
\newcommand{\lk}{\ell_\kappa}
\newcommand{\lgm}{\ell_{\gamma}}
\newcommand{\lgp}{\ell_{\gamma'}}
\newcommand{\ltg}{\ell_{\tilde \gamma}}
\newcommand{\lgc}{\ell_{\gamma_C}}
\newcommand{\lgd}{\ell_{\gamma_D}}
\newcommand{\lgk}{\ell_{\gamma_\kappa}}
\newcommand{\lgg}{\ell_{\gamma_{\gamma}}}
\newcommand{\lgcd}{\ell_{\gamma_{CD}}}
\newcommand{\lgkd}{\ell_{\gamma_{\kappa D}}}

\newcommand{\zmax}{z_{\max}}

\DeclareMathOperator{\Tr}{Tr}

\numberwithin{lemma}{section}

\usetikzlibrary{arrows,shapes.misc,positioning,decorations.pathmorphing,decorations.markings,decorations.pathreplacing,matrix,patterns,backgrounds}
\tikzset{rectangle/.style={draw,circle,inner sep=1pt}, 
    big arrow/.style={decoration={markings,mark=at position 1 with {\arrow[scale=1.5,#1]{>}}}, postaction={decorate}, shorten >=0.4pt}, 
    scale cd/.style={every label/.append style={scale=#1}, cells={nodes={scale=#1}}},
    brace/.style={decoration={brace, mirror},decorate}}

\begin{document}
\begin{titlepage}
\begin{center}
\vspace{2.2cm}
{\LARGE\bfseries Signals of multiparty entanglement\\ and holography\\}
\vspace{1.6cm}

{\large
Vijay Balasubramanian$^{1,\,2,\,3}$, Monica Jinwoo Kang$^1$, Chitraang Murdia$^1$, and Simon F. Ross$^{4}$\\}
\vspace{.7cm}
{$^1$ Department of Physics and Astronomy, University of Pennsylvania,}\\
{Philadelphia, PA 19104, U.S.A.}\par
\vspace{.2cm}
{$^2$  Theoretische Natuurkunde, Vrije Universiteit Brussel and International Solvay Institutes,}\\ 
{Pleinlaan 2, B-1050 Brussels, Belgium} \par
\vspace{.2cm}
{$^3$  Rudolf Peierls Centre for Theoretical Physics, University of Oxford,}\\
{Oxford OX1 3PU, U.K.}\par
\vspace{.2cm}
{$^4$ Centre for Particle Theory, Department of Mathematical Sciences, Durham University,}\\
{Durham DH1 3LE, U.K.}\par
\vspace{.6cm}

\scalebox{.84}{\tt vijay@physics.upenn.edu,  monica6@sas.upenn.edu, murdia@sas.upenn.edu, s.f.ross@durham.ac.uk}\par
\vspace{1.6cm}

\textbf{Abstract}
\end{center}

\noindent We study multiparty entanglement signals, which are functions of a quantum state that are non-zero only when the state has multiparty entanglement. 
We consider known signals of three- and four-party entanglement, and propose new signals for four- and higher-party entanglement. We make some remarks on their general properties, but mainly focus on using holographic states in AdS$_3/$CFT$_2$ as a test case to explore their properties. For both the AdS vacuum and multiboundary wormhole states, we find that the multiparty entanglement signals are generically non-zero and of order one in units of the central charge of the dual CFT, revealing substantial multiparty entanglement. In the large-horizon limit of multiboundary wormhole states, however, the signals for three or more parties become small indicating the short-range nature of entanglement. 

\vfill 
\end{titlepage} 
\tableofcontents

\newpage
\section{Introduction}

Entanglement is a fundamental aspect of quantum systems, and plays a central role in our understanding of the emergence of spacetime in holography. Much of our detailed understanding of entanglement focuses on the bipartite case, where we divide a system into two parts, and study the entanglement between them. However, a system with many degrees of freedom can be entangled in ways that are not captured by considering simply the bipartite entanglement between different subsystems. In recent years, there have been efforts to extend our understanding of such multiparty entanglement, see \cite{Walter:2016lgl} for a review. There are a rich variety of possible structures of multiparty entanglement, and no comprehensive classification is yet available. In the present work, our aim is more modest: to identify quantities that signal when a state possesses multiparty entanglement. That is, considering a state of a system made up of $n$ subsystems, we seek to identify functions of the state that vanish if the state only involves entanglement between collections of $n-1$ subsystems or fewer.\footnote{Our focus is thus somewhat different from some previous holographic studies of multipartite quantities such as \cite{Bao:2018gck,Bao:2019zqc,DeWolfe:2020vjp,Gadde:2022cqi}.} We focus our discussion on pure states; if the state of interest is mixed, we can always purify it by adjoining an additional purifying subsystem. 

Our work is inspired by recent developments in holography, and a substantial fraction of our work will explore the behavior of such signals in a holographic context, specifically in the context of AdS$_3/$CFT$_2$ holography. This is an interesting context because there are rich families of explicit solutions where we can explore these questions, describing connected spatial wormholes with multiple boundaries. These are dual to pure states in multiple copies of the CFT. 
The exploration of multiparty entanglement in this holographic context was initiated in \cite{Balasubramanian:2014hda}, which sparked a rich set of subsequent developments. 

In \cite{Cui:2018dyq}, it was noted that the triple information $I_3$ provides a signal of four-party entanglement: for a state on four systems the triple information between any three vanishes if the entanglement is only between sets of three parties. This is our first example of a multiparty entanglement signal. In \cite{Akers:2019gcv}, a signal of three-party entanglement was proposed, based on the reflected entropy studied in \cite{Dutta:2019gen}. These signals both have the desirable property that they can be computed from geometric features of the holographic duals.\footnote{A recent analysis \cite{Akers:2024pgq} also studied the latter in random tensor networks.}

It is important to note that while these quantities provide us with some information about the multiparty entanglement, they are not entanglement measures. To be an entanglement measure, a quantity should satisfy certain axiomatic properties, such as positivity and monotonicity. The triple information is in general not sign-definite, although it is negative on holographic states \cite{Hayden:2011ag}, and in \cite{Hayden:2023yij}, it was shown that reflected entropy is not monotonic under partial trace. Also, while these quantities vanish on states that do not have three- or four-party entanglement, the converse is not true: there are states with genuine three- or four-party entanglement where these signals vanish. 

The first aim of our work is to extend this story to obtain signals of multiparty entanglement for higher numbers of parties. In Section \ref{signals}, we review the three and four-party quantities and propose extensions to general numbers of parties. We can construct entanglement signals for $n$ parties by a fairly direct extension of the previous constructions, generalizing the triple information for even numbers of parties and generalizing the construction involving the reflected entropy for odd numbers of parties. We also propose alternative signals, constructed by variations of the latter technique. It is useful to have a range of different quantities, as multiparty entanglement has a complex structure, which we would not expect to be captured by a single number.\footnote{Already for the simplest possible case, three qubits, it is known that there are different forms of three-party entanglement. (See e.g. \cite{Walter:2016lgl} for a discussion.)} Also a given entanglement signal can fail in some cases, vanishing on states with genuine $n$-party entanglement, as we will see later. By considering multiple signals we can hope to detect more of the multiparty entanglement structure in the space of states. In particular, signals which are not sign-definite necessarily vanish on a subspace of codimension one in the space of states, so we need to consider multiple signals to have any hope of isolating the true subspace of states with no multiparty entanglement. 

In Section \ref{gen}, we consider some general features of these signals. We first consider their behavior on simple states in systems of three and four qubits. A notable feature is that the reflected entropy signal vanishes on the GHZ state. We then derive some bounds for these quantities. If the subsystems have finite-dimensional Hilbert spaces we can bound the magnitude of the multiparty signals in terms of the dimension of these Hilbert spaces. These bounds are useful for interpreting our signals: when a multiparty signal is non-zero, this signals the existence of 
a component in the state that has multiparty entanglement. The size of the signal determines a lower bound on the dimension of the components of the Hilbert spaces this multiparty entanglement involves. 

We then turn to studying these quantities holographically. In Section \ref{holo}, we review the relation between entanglement and bulk geometry in holography, and derive some general results on the multiparty entanglement signals. We show that the new four-party entanglement signal introduced in Section \ref{higher} is positive holographically. In holography, there are often open regions in the parameter space where for example the mutual information vanishes at leading order in $1/G$. We show that the multiparty entanglement signals exhibit similar behavior, and that as for the mutual information, this vanishing can be related to the connectedness of entanglement wedges. 

We then turn to a detailed consideration of the behavior of the multiparty entanglement signals in particular cases in AdS$_3/$CFT$_2$. In Section \ref{vacads}, we consider the vacuum state in a single CFT, and calculate the multiparty signals when we divide the boundary into $n$ regions. We find that the vacuum has $n$-party entanglement for all values of $n$; the size of the multiparty entanglement signals is generically of order the central charge in the CFT, so this is a significant feature of the state, although the entropy on any single party is dominated by a divergent bipartite contribution. 

In Section \ref{multbound}, we review the construction of multiboundary wormholes. In Section \ref{punc}, we study the multiparty quantities in the puncture limit considered in \cite{Balasubramanian:2014hda}, and in Section \ref{largel}, we consider them in the limit of large horizon sizes considered in \cite{Marolf:2015vma}. There is a rich structure to the behavior of our signals in these wormhole cases, from which a few general lessons emerge: there are regions of the parameter space where the multiparty signals vanish, but in each case there is a region, where the different subsystems are of similar size, where the multiparty entanglement signals are non-zero. For the puncture limit, they are again typically of order the central charge, signaling a significant $n$-party component in the holographic state on $n$ subsystems. In the puncture limit, the dual state effectively lives in finite-dimensional subspaces of the dual CFTs, and in some cases, the multiparty signals saturate the bounds in terms of the dimensions of these subspaces. This indicates that the dual states are purely multipartite in these cases, with essentially no bipartite component. 
For the large horizon size limit, we argue that the four-party signals are small in units of $1/G$, and conjecture that this would continue for higher numbers of parties, consistent with the special structure of the dual state discussed in \cite{Marolf:2015vma}. 

We conclude with a discussion of our results and prospects for future research in Section \ref{disc}. A promising direction for future work is to use our results to constrain and further develop toy models of the holographic states, such as tensor network pictures. This will require further understanding of the behavior of the multipartite entanglement signals in such model systems. 

\section{Signals of multipartite entanglement}
\label{signals}

We are interested in studying the entanglement properties of a pure state $|\psi \rangle$ on some $n$-party system.\footnote{We focus on pure states. However, our results also work for a mixed state via purification, with the purifier being an additional party.}  We want to identify quantities that act as signals of higher-party entanglement; that is, we want quantities that vanish on states with no $n$-party entanglement. We will define the space of states with no $n$-party entanglement for general $n$ in the next subsection. We begin by reviewing known quantities for two, three, and four parties.

\paragraph{Two parties and von Neumann entropy:}
In the simplest case of a two-party system, we have a pure state $|\psi\rangle_{AB}$, where $A$ and $B$ denote the two parties. 
Then, the entanglement or {\it von Neumann entropy} is 
\begin{equation}
    S_A = - \Tr_A \left( \rho_A \ln \rho_A \right) \, ,
    \qquad
    \rho_A = \Tr_B \left( |\psi\rangle \langle \psi | \right) \, ,
\end{equation}
where the subscript on the trace denotes the partial trace over such a subsystem/party. This provides a measure of entanglement between the two parties, as $S_A=0$ for unentangled states, when $|\psi \rangle_{AB} = |\psi_1 \rangle_A \otimes |\psi_2 \rangle_B$. We can also consider the mutual information,
\begin{align}
    I(A:B) = S_A + S_B - S_{AB} \, .
\end{align}
For a pure state, $S_{AB}=0$; it follows that the mutual information is simply 
\begin{align}
    I(A:B) = 2 S_A , 
    \label{eq:mut_inf_pure}
\end{align}
so for pure states, the mutual information reduces to the entanglement entropy. This is a pattern that will repeat in higher numbers of parties. 

\paragraph{Three parties and residual information:}  A natural generalization of the mutual information for three parties is the {\it triple information}, 
\begin{align} \label{eq:I3}
    I_3(A:B:C) = S_A + S_B + S_C - S_{AB} - S_{AC} - S_{BC} + S_{ABC} . 
\end{align}
It is important to note that the triple information vanishes on all three-party pure states. This is similar to the von Neumann entropy, which vanishes on all pure states but is a useful quantity for two-party pure states. Analogously, the triple information is a useful quantity for four-party pure states.

Then what would be a diagnostic quantity for three-party entanglement? We utilize \emph{canonical purification} to construct such a quantity. Consider a density matrix
\begin{equation}
    \rho = \sum_i p_i |\chi_i \rangle \langle \chi_i | \,
\end{equation}
in a Hilbert space $\mathcal{H}$, where $|\chi_i\rangle$ are orthonormal eigenstates. The canonical purification provides the state 
\begin{equation}
| \sqrt{\rho} \rangle  = \sum_i \sqrt{p_i}  |\chi_i \rangle  |\chi_i \rangle \in \mathcal H \otimes \mathcal H^* \, ,
\end{equation}
which is a purification of $\rho$ in the sense that we recover the original density matrix by a partial trace on $\mathcal{H}^*$,
\begin{align}
    \rho=\Tr_{\mathcal{H}^*} \left( |\sqrt{\rho}\rangle \langle \sqrt{\rho}| \right) .
\end{align}
The name \emph{canonical} comes from the fact that this purification does not depend on the choice of basis on $\mathcal{H}$.

Let the pure state of a three-party system be $|\psi\rangle_{ABC}$ and the reduced density matrix on $AB$ be $\rho_{AB}$. 
The canonical purification then provides $| \sqrt{\rho} \rangle_{AA*BB*}$. 
We can define the {\it reflected entropy} as \cite{Dutta:2019gen}
\begin{align}
    S_R(A:B) = S_{AA*} \left( \sqrt{\rho} \right) =  \frac{1}{2} I(AA^*:BB^*) \, ,
\end{align}
where $S_{AA*}\left( \sqrt{\rho} \right)$ is the von Neumann entropy of the $A$ parts of the canonical purification. 
We define the \emph{residual information} as the difference between the reflected entropy and the mutual information such that
\begin{align}
    R_3(A:B) := S_R (A:B) - I(A:B) = \frac{1}{2} I(AA^*:BB^*) - I(A:B) \, .
    \label{eq:residual-info-def}
\end{align}
It was observed that the residual information vanishes on states with purely bipartite entanglement \cite{Akers:2019gcv}. Thus, non-zero $R_3$ is a signal of tripartite entanglement in the state $|\psi\rangle_{ABC}$.   

It is important to note that $R_3$ is not a measure of tripartite entanglement, as it is not monotonic under partial traces. This is similar to the reflected entropy $S_R$, which is not a measure of correlations \cite{Hayden:2023yij}. While $R_3$ is not a measure, it has nice properties as a diagnostic quantity such as
\begin{itemize}
    \item non-negativity \cite{Dutta:2019gen},
    \item additivity under tensor product, 
    i.e.,  if $|\psi \rangle_{A_1 A_2 B_1 B_2 C_1 C_2} =  |\psi_1 \rangle_{A_1 B_1 C_1} \otimes |\psi_2 \rangle_{A_2 B_2 C_2}$, $R_3(A_1 A_2 : B_1 B_2) =R_3 (A_1 : B_1)+ R_3(A_2 : B_2)$.
\end{itemize}
However, it can vanish on states with non-trivial tripartite entanglement and is not permutation invariant: in general, $R_3$ takes different values depending on which party is traced out.

\paragraph{Four parties and triple information:}
For four-party entanglement, the triple information, defined in equation \eqref{eq:I3}, is a useful signal. Considering a pure state $|\psi\rangle_{ABCD}$, without loss of generality, taking the partial trace over $D$ defines a reduced density matrix $\rho_{ABC}$. The triple information $I_3(A:B:C)$ for this reduced density matrix vanishes if $|\psi\rangle_{ABCD}$ has no four-party entanglement; i.e., if $I_3(A:B:C)$ is non-zero, the state must have four-party entanglement \cite{Balasubramanian:2014hda}. 

The triple information is not sign-definite for general quantum systems (although $-I_3$ is non-negative for holographic systems \cite{Hayden:2011ag}). It is additive under tensor product. It is also symmetric under permutations of all four parties $A,B,C,D$ in a pure state $|\psi \rangle_{ABCD}$. Indeed, it follows that
\begin{equation}
    I_3(A:B:C) = S_A + S_B + S_C + S_D - \frac{1}{2} \left( S_{AB} + S_{BC} + S_{CD} + S_{AC} + S_{BD} + S_{AD} \right)  \, ,
\end{equation}
where we used the fact that, in a pure state, the von Neumann entropy of a subsystem and that of its complement are equal.

\subsection{Signals of higher-party entanglement}
\label{higher}

In this subsection, we generalize these quantities to devise diagnostic quantities for $n$-party entanglement. As with bipartite entanglement, $n$-party entanglement is defined by exclusion. First, we define the set of states with only $(n-1)$-party entanglement. A state on $n$ parties involving only $(n-1)$-party entanglement can be written, up to local unitaries on each of the parties $A_i$, as
\begin{align}
    |\psi\rangle_{A_1 \ldots A_n} = \prod_{a=1}^n |\psi_a \rangle_{\sigma_a},
    \label{kprod}
\end{align}
where we write the full state as a product of $n$ states $|\psi_a \rangle$ which each involve $(n-1)$-party entanglement. To construct such states, we introduce a decomposition of each party in terms of components,  $A_i = \prod_{a \neq i} A^a_i$.  Each $|\psi_a\rangle$ is a wavefunction on the subspace $\sigma_a = \prod_{i \neq a} A^a_i$, so each of the $A_i^a$ is involved in only one of the $|\psi_a\rangle$. That is,  $|\psi_a\rangle$ entangles $n-1$ parties, excluding $A_{i=a}$. The entanglement of $A_i$ with different subsets of the other parties is thus distilled into distinct components $A_i^a$ in the various $|\psi_a\rangle$.
For example, when $n=3$, $A_1 = A_1^2 A_1^3$, $A_2 = A_2^1 A_2^3$, and $A_3 = A_3^1 A_3^2$; a generic two-party entangled state is $|\psi\rangle_{ABC} = |\psi_1 \rangle_{A_2^1 A_3^1} \otimes |\psi_2 \rangle_{A_1^2 A_3^2} \otimes |\psi_3 \rangle_{A_1^3 A_2^3}$.  

There are multiple possible forms for $(n-1)$-party entangled states, depending on how we split up the overall Hilbert space into factors $\sigma_a$, so the subspace of $(n-1)$-party entangled states has multiple components. Each such component is a fairly special subspace of the overall space of states; that is, the codimensions of these subspaces in the overall space of states are large. If the $A^a_i$ have dimension $d^a_i$, the $A_i$ have dimension $d_i = \prod_{a \neq i} d^a_i$, the $\sigma_a$ have dimension $d_a = \prod_{i \neq a} d^a_i$, and the overall Hilbert space has dimension $d = \prod_a d_a$. A normalized state in a $d$-dimensional Hilbert space is specified up to an overall phase by $(d-1)$ complex parameters. Hence, if a state is of the form in equation \eqref{kprod}, it is specified by $\sum_a (d_a-1)$ parameters, and in turn, the subspace of such states has real codimension 
\begin{equation}
   2\left( (d-1) - \sum_a (d_a-1)\right) = 2 \left( \prod_a d_a - \sum_a d_a +a-1 \right).  
\end{equation}
The largest $(n-1)$-party entangled subspace consists of states where the smallest factor $A_i$ is entirely disentangled and we take a generic $(n-1)$-party state on the remaining factors. Without loss of generality, we take $A_n$ to be the smallest factor. Then,
\begin{equation}
    |\psi\rangle_{A_1 \ldots A_n}  = |\phi \rangle_{A_1 \ldots A_{n-1}} \otimes |\chi \rangle_{A_n}.
\end{equation}
The first factor has dimension $d' = \prod_{i=1}^{n-1} d_i$, so states of this form are specified by $(d'-1) + (d_n-1)$ complex parameters, so this subspace has codimension 
\begin{equation}
   2 \left( (d' d_n-1) - (d'-1) - (d_n-1)\right) = 2(d'-1)(d_n-1) 
\end{equation}
in the overall space of $n$-party states. For example, if we take all the factors to be single qubits, the space of $2$-party entangled states in the $3$-qubit Hilbert space has codimension $6$, and the space of $3$-party entangled states in the $4$-qubit Hilbert space has codimension $14$. 

\paragraph{$n$-information:}

Consider a state on $n$ parties, $A_1, A_2, \cdots A_n$.  Similar to the triple information $I_3$, we consider \cite{Balasubramanian:2014hda}\footnote{Note that the definition here differs from \cite{Balasubramanian:2014hda} by an overall sign, agreeing with the sign chosen in \cite{Agon:2022efa}.}
\begin{align}
    I_n(A_1: \cdots : A_n) = - \sum_{i \leq n} (-1)^i S_i ,
    \label{eq:k-information}
\end{align}
where $S_i$ is the permutation-invariant combination of all the entropies on sets of $i$ parties. For example, $S_1 = \sum_{i=1}^n S_{A_i}$ and $S_2 = \sum_{i <j} S_{A_i A_j}$. These quantities have been considered before in different contexts. We aim to show that they provide multiparty entanglement signals. 

We first review some basic properties. The $I_n$'s are permutation invariant by construction. They are also additive under the tensor product:  when $|\psi \rangle_{A_1 B_1 \cdots A_n B_n} =  |\psi_A \rangle_{A_1 \cdots A_n} \otimes |\psi_B \rangle_{B_1 \cdots B_n}$,  $I_n(A_1 B_1 : \cdots : A_n B_n) = I_n(A_1 : \cdots : A_n) + I_n(B_1 : \cdots : B_n) $. For a pure state $|\psi\rangle_{A_1A_2\cdots A_n}$, 
\begin{align}
    S_n=0, \qquad S_i=S_{n-i},
\end{align}
and it follows that 
\begin{align}
    I_n(A_1: \cdots : A_n) =\begin{cases}
        0 & n\text{ odd},\\
        2I_{n-1}(A_1: \cdots : A_{n-1}) & n\text{ even} 
    \end{cases}
    \label{eq:Ikoddeven}
\end{align}
where the expression for $n$ even is twice the $n-1$ information computed on the reduced density matrix produced by tracing out the $n$th party from the full $n$-party state.
For example, $I_2 = I(A:B) = 2 S_A$ as in equation \eqref{eq:mut_inf_pure} and $I_4(A:B:C:D) = 2 I_3(A:B:C)$. This relation to $I_4$ also makes the permutation invariance of $I_3$ obvious. The higher quantities are not sign-definite; we will give holographic examples of the failure of sign-definiteness in Section \ref{npos}, generalizing a discussion in \cite{Balasubramanian:2014hda}.

When $n$ is even, $I_n$ is a signal of $n$-party entanglement. That is, $I_n$ vanishes on all states of the form equation \eqref{kprod}. Since $I_n$ is additive under tensor product and permutation invariant, we only need to show that it vanishes on one of the factors in equation \eqref{kprod}. Consider for definiteness $|\psi_n\rangle$. Since $A_n$ is disentangled in this state, 
\begin{align}
   S_{T A_n} = S_{T}
   \qquad
   \forall \, T \subset \{ A_1, \dots A_{n-1} \} .
\end{align}
We can use this relation to show that the permutation symmetric combinations of entropies in equation \eqref{eq:k-information} simplify to
\begin{align}
\begin{aligned} \label{symdis}
    S_i(A_1: \cdots :A_n) &= S_i(A_1 \dots A_{n-1}) + S_{i-1} (A_1: \cdots :A_{n-1}) \, , & \text{for } i<n \, ,\\
    S_n(A_1: \cdots :A_n) &= S_{n-1}(A_1: \cdots :A_{n-1}) \, . &
\end{aligned}
\end{align}
Thus, we see that $I_n$ vanishes
\begin{equation}
\begin{split}
    - I_n(A_1: \cdots : A_n) 
    &= \sum_{i = 1}^{n-1} (-1)^i \Big( S_i(A_1: \cdots :A_{n-1}) + S_{i-1} (A_1: \cdots :A_{n-1}) \Big) \\ 
    & \qquad\qquad \qquad\qquad + (-1)^{n} S_{n-1}(A_1: \cdots :A_{n-1}) \\
    &= \sum_{i = 1}^{n-1} (-1)^i S_{i} (A_1: \cdots :A_{n-1}) + \sum_{i = 1}^{n-2} (-1)^{i+1} S_{i} (A_1: \cdots :A_{n-1})  \\ 
    & \qquad\qquad \qquad\qquad + (-1)^{n} S_{n-1}(A_1: \cdots :A_{n-1}) \\
    &= 0 \, .
\end{split}
\end{equation}
Thus, there must be $n$-party entanglement in a $n$-party pure state for $n$ even if $I_n$ is nonzero. 
Due to the lack of sign-definite property, either positive or negative, $I_n$ is possible; either case indicates $n$-party entanglement. 

\paragraph{$n$-residual information:}
For $n$ odd, $I_n$ vanishes, so this is not a useful signal for $n$-party entanglement. We can instead proceed as for three parties, by considering a canonical purification. Performing a partial trace on one party, we obtain a reduced density matrix $\rho_{A_1 \ldots A_{n-1}}$. We can calculate $ I_{n-1}(A_1 : \cdots : A_{n-1})$ for this density matrix, and defining the canonical purification $|\sqrt \rho \rangle_{A_1 A_1^* \ldots A_{n-1} A_{n-1}^*}$, we can also calculate $I_{n-1}(A_1 A_1^* : \cdots : A_{n-1} A_{n-1}^*)$ in the canonically purified state. We then define an \emph{$n$-residual information} as
\begin{align}
    R_{n}(A_1 : \cdots : A_{n-1} ; A_n) = \frac{1}{2} I_{n-1}(A_1 A_1^* : \cdots : A_{n-1} A_{n-1}^*) - I_{n-1}(A_1 : \cdots : A_{n-1}),
\end{align}
which is similar to the residual information $R_3$ in equation \eqref{eq:residual-info-def}. It is easy to see that if $\rho_{A_1\cdots A_{n-1}}$ is disentangled, both terms in $R_n$ vanish. If $\rho_{A_1\cdots A_{n-1}}$ is a pure state, both terms are identical, and hence $R_n$ vanishes. Thus, for states with only $(n-1)$-party entanglement, $R_n=0$; hence, if $R_n$ is nonzero, the state has $n$-party entanglement. We note that similar to $R_3$, but unlike $I_n$, these quantities are not permutation invariant; thus, $R_n$ can take different values depending on the choice of the party that is traced out. For $n>3$, these quantities are not sign-definite because $I_{n-1}$ is not sign-definite.

We could also compute the $n$-residual information for even $n$, but it does not give anything new. The mutual information $I_{n-1}$ of the canonical purification vanishes due to equation \eqref{eq:Ikoddeven}, because $(n-1)$ is odd and the canonical purification is a pure state. Thus, for even $n$
\begin{equation}
    R_n(A_1 : \cdots : A_{n-1} ; A_n) = - I_{n-1}(A_1 : \cdots : A_{n-1}) = - \frac{1}{2} I_n(A_1 : \cdots : A_n) ,
\end{equation}
recovering the $n$-information.

\paragraph{Residual entropy:}

While $I_4(A:B:C:D) = 2 I_3(A:B:C)$ is a useful signal of four-party entanglement, it is possible to have four-party entanglement when $I_4$ vanishes. 
Consider moving in the state space between states with positive and negative $I_3$.  In between, we have to pass through a state with $I_3=0$. Hence, $I_3$ must vanish on a subspace of states of codimension one, which is higher in dimension than the set of states with no four-party entanglement. In any case, there are many different forms of multi-party entanglement, so we would not expect to be able to describe all of them by a single number. These considerations motivate interest in identifying further signals of four-party entanglement. We will describe an alternative here; another construction is given in Appendix~\ref{further}.

Consider a four-party state $|\psi \rangle_{ABCD}$. Starting from the reduced density matrix $\rho_{AB}$, we can construct the canonical purification $|\sqrt{\rho}_{AB} \rangle_{ABA^*B^*}$, and obtain the reflected entropy $S_R(A:B) = S_{AA^*} = \frac{1}{2} I(AA^*:BB^*)$. Now we take the partial trace over $D$ and get a reduced density matrix $\rho_{ABC}$. Via canonical purification, we obtain $|\sqrt{\rho_{ABC}} \rangle_{ABCA^*B^*C^*}$. Next, we trace out $CC^*$ and obtain a reduced density matrix $\rho_{AA^*BB^*}$. We perform a second canonical purification and obtain $|\sqrt{\rho}_{AA^*BB^*} \rangle_{AA^*BB^*A_*A^*_*B^*B^*_*}$, and calculate $S_{AA^*A_*A^*_*} = \frac{1}{2} I(AA^*A_*A^*_*:BB^*B_*B^*_*)$. We define the \emph{4-party residual entropy} as 
\begin{equation}  \label{eq:q3} 
    Q_4 = \frac{1}{2} I(AA^*A_*A^*_*:BB^*B_*B^*_*) - I(AA^*:BB^*) = S_{AA^*A_*A^*_*} - 2 S_{AA^*}. 
\end{equation}
As with the residual information, there is a dependence on what we chose to trace out.
Hence, there is a collection of such quantities, depending on the parties and the order in which the partial traces are performed.
For example, tracing out $C$ first and then $DD^*$ will, in general, give a different answer to tracing out $D$ first and then $CC^*$. 

To see that $Q_4$ signals four-party entanglement, consider states with three-party entanglement. $Q_4$ is additive under tensor product, so we can consider states with purely one form of three-party entanglement. If we consider a state with either $ACD$ or $BCD$ entangled, both quantities in equation \eqref{eq:q3} vanish, as they are only sensitive to correlations involving both $A$ and $B$. If the state has $ABC$ entangled, the reduced density matrix $\rho_{ABC}$ at the first stage of the construction will be pure, and the canonical purification will be two copies of this state. Thus $\rho_{AA^*BB^*}$ will be the tensor product of two copies of $\rho_{AB}$, and hence $S_{AA^*A_*A^*_*} = 2 S_{AA^*}$, and $Q_4=0$. If the state has $ABD$ entangled, the reduced density matrix is $\rho_{ABC} = \rho_{AB} \otimes |\xi \rangle_C \langle \xi |_C$, and its canonical purification $|\sqrt{\rho}_{ABC} \rangle$ involves the canonical purification of $\rho_{AB}$. Tracing out $CC^*$ now leaves us in a pure state, so the second canonical purification just gives us two copies of the canonical purification of $\rho_{AB}$, and again $S_{AA^*A_*A^*_*} = 2 S_{AA^*}$, so $Q_4=0$. Thus, $Q_4$ is non-zero only for states with genuine four-party entanglement. 

It is interesting to note that $I(A:B)=0$ implies $R_3(A:B)=0$, but it does not imply the vanishing of either $I_3$ or $Q_4$. There are states with four-party entanglement where the two-party reduced density matrix factorizes. An example is a perfect tensor state, see e.g. \cite{Cui:2018dyq}.

\section{General properties}
\label{gen}

\subsection{Behavior on simple states}
\label{simstates}

The residual information $R_3$ is nonnegative \cite{Dutta:2019gen}, and vanishes if the state has purely bipartite entanglement. Naively, one might conjecture that $R_3>0$ for states with multiparty entanglement, but this is false. We demonstrate this by providing a counterexample.

Consider the GHZ state,
\begin{equation}
| \psi \rangle_{ABC} = \frac{1}{\sqrt{2}} (|000 \rangle + |111 \rangle). 
\end{equation}
The reduced density matrix on two parties is simply $\rho_{AB}  = \frac{1}{2} ( |00 \rangle \langle 00| + |11\rangle \langle 11|)$. Then, we have $I(A:B) = \ln 2$. The canonical purification is $| \sqrt{\rho_{AB}} \rangle_{ABA^*B^*} = \frac{1}{\sqrt{2}} (|0000 \rangle + |1111 \rangle)$; hence, $\rho_{AA^*} = \frac{1}{2} ( |00 \rangle \langle 00| + |11\rangle \langle 11|)$, and the reflected entropy is $S_R(A:B) = \ln 2$. Thus, despite the multiparty entanglement in this state, we have $R_3(A:B)=0$, which was also observed in \cite{Akers:2019gcv}. Since tripartite stabilizer states are composed of bipartite and GHZ states up to local unitaries \cite{Bravyi_2006}, this implies $R_3=0$ on all stabilizer states. It would be interesting to understand in general what set of states have $R_3=0$. In Appendix \ref{zeros}, we show that for three qubits, the states with $R_3=0$ are precisely bipartite entangled states and generalized GHZ states. For larger systems, the question is open. 

As discussed above, since $I_3$ is not sign-definite, the set of states with $I_3=0$ will have codimension one; this is higher than the codimension of the set of states with no four-party entanglement, so there will certainly be classes of four-party entangled states with $I_3=0$. We can construct a simple example of such a state using previous results from \cite{Cui:2018dyq}, who noted that the four-party GHZ state
\begin{equation} \label{4ghz}
| \psi \rangle_{ABCD} = \frac{1}{\sqrt{2}} (|0000 \rangle + |1111 \rangle), 
\end{equation}
has $I_3 = \ln 2$, while 
\begin{equation} \label{4state2}
| \psi \rangle_{ABCD} = \frac{1}{\sqrt{2}} (|0000 \rangle +  |0111 \rangle +  |1012 \rangle +  |1103 \rangle), 
\end{equation}
where $D$ is a four-state system (a pair of qubits), has $I_3 = - \ln 2$. Now consider a system where $A,B,C$ are pairs of qubits and $D$ consists of three qubits, and take the state as the tensor product of these two states. Despite having four-party entanglement, this state has $I_3=0$ because $I_n$ is additive under tensor products of states.

Similar remarks apply for $Q_4$, as it is also not sign-definite, and we can similarly construct examples of four-party entangled states with $Q_4=0$.  If we consider the four-party GHZ state as in equation \eqref{4ghz}, the state on $A B A^* B^* A_* B_* A^*_* B^*_*$ in the construction of $Q_4$ is an eight-party pure GHZ state, so it has $S_{AA^*A_* A^*_*} = \ln 2$, and $S_R = \ln 2$, so $Q_4 = - \ln 2$.  For the state in equation \eqref{4state2}, $Q_4= \ln 2$, a state which is the tensor product of equations \eqref{4ghz} and \eqref{4state2} has $Q_4=0$, because $Q_4$ is additive under tensor product of states.  By contrast, consider a variant of $Q_4$ obtained by tracing out $C$ and $B$ rather than $D$ and $C$. This variant $Q'_{4} = 3 \ln 2$ for equation \eqref{4state2}. Hence,  $Q'_{4} = 2 \ln 2$ for the tensor product state, signaling its four-party entangled nature. This illustrates a useful role for different multiparty entanglement signals: their zero sets will in general be different, so by taking their intersection we can hope to have a more refined probe of four-party entanglement. 

\subsection{Bounds}
\label{bounds}

\paragraph{Reflected Entropy and Residual Information: } For the reflected entropy, \cite{Dutta:2019gen} bounded $S_R(A:B) < 2 \min (\ln d_A, \, \ln d_B)$, where $d_{A,B}$ are the dimensions of the respective Hilbert spaces.  Now $I(A:B)=0$ implies $S_R(A:B)=0$,\footnote{ This is because
a state with $I(A:B)=0$ factorizes; $\rho_{AB} = \rho_A \otimes \rho_B$, so the canonical purification factorizes, and hence $S_R(A:B)=0$.} so one might expect that it would be possible to find a tighter bound on $R_3$, but we will see later that there are holographic systems that approach $R_3 = 2 \min (\ln d_A,\, \ln d_B)$, so there cannot be a tighter bound in general. It would be interesting to identify states in low-dimensional systems which saturate this upper bound. We also have $S_R(A:B) = S_{AA^*} = S_{BB^*} \leq 2 \min (S_A, \, S_B)$, so 
\begin{equation}
    R_3(A:B) = S_R(A:B) - I(A:B) \leq S_{AB} - |S_A - S_B| \leq S_{AB}, 
\end{equation}
and as $S_{AB} = S_C \leq \ln d_C$, we can also bound $R_3 \leq \ln d_C$. Thus, we have bounds 
\begin{equation} \label{r3bounds}
    0 \leq R_3(A:B) \leq  2 \min (\ln d_A, \, \ln d_B)
    \quad
    \text{and}
    \quad R_3(A:B) \leq \ln d_C.
\end{equation}
Note the asymmetry in the bounds involving $d_A, d_B, d_C$, which is a clear indication of the lack of permutation invariance in $R_3(A:B)$. 

\paragraph{Triple information: } By writing 
\begin{equation}
    I_3 = I(A:B) + I(A:C) - I(A:BC), 
\end{equation}
we can bound $I_3$ both from above and below. Since the mutual informations are positive, $I_3 \geq - I(A:BC) \geq - 2 \ln d_A$, where $d_A$ is the dimension of $\mathcal H_A$. Since $I_3$ is permutation invariant, this implies $I_3 \geq - 2 \ln d_{min}$, where $d_{min}$ is the minimum of the dimensions of the subsystems $A,B,C,D$. $I(A:BC) \geq I(A:C)$, so $I_3 \leq I(A:B) \leq 2 \ln d_A$, and hence $I_3 \leq 2 \ln d_{min}$. The lower bound is saturated by perfect tensor states \cite{Nezami:2016zni}, and was seen in \cite{Balasubramanian:2014hda} to be saturated by some holographic examples, so this bound is tight, but the upper bound is not tight. We can improve it by considering $I_4$. 

\paragraph{4-information: } We can bound $I_4$ by expressing it as 
\begin{equation}
    I_4 = I(A:BCD) - (I(A:BC) - I(A:B)) - (I(B:AD) - I(B:D)) - (I(A:CD) - I(A:C)). 
\end{equation}
The terms in the brackets are all non-negative, so $I_4 \leq I(A:BCD) \leq 2 \ln d_A$. Since $I_4 = 2 I_3$ on pure states, and these quantities are permutation invariant, this gives a tighter bound on $I_3$, $I_3 \leq \ln d_{min}$. This bound is saturated by a GHZ-like state, a uniform superposition of products of basis states in the smallest subsystem with orthonormal states in the other systems. Thus, 
\begin{equation} \label{i3bounds}
    - \ln d_{min} \leq -I_3 \leq 2 \ln d_{min},
\end{equation}
where both bounds are tight. The upper bound is saturated by perfect tensor states when they exist and in holographic examples, while the lower bound is saturated by the GHZ state, as we saw explicitly for $d_{min}=2$ in Section \ref{simstates}.

\paragraph{Residual entropy: } For $Q_4$, we have the simple bounds 
\begin{equation} \label{q4bounds}
    -4 \min \left(\ln d_A, \, \ln d_{B} \right) \leq Q_4 \leq 4 \min \left( \ln d_A, \, \ln d_{B} \right).
\end{equation}
Unlike $R_3$, we have not been able to obtain a bound relating $Q_4$ to $d_C, d_D$. 
We will see holographic examples that saturate the upper bound, so it is tight. It would be interesting to characterize the structure of states saturating the upper bound. For the perfect tensor state on four qutrits discussed in \cite{Pastawski:2015qua}, we find $Q_4 = 2 \ln 3$. Hence, this does not saturate the upper bound. We do not know if the lower bound is tight and it likely is not. The GHZ state on the four qubits has $Q_4 = - \ln 2$. This is not a lower bound for $Q_4$, which we can verify by examining 10,000 randomly chosen states on four qubits, where we found examples with slightly lower values of $Q_4$ (plot not shown).  Thus, the nature of states saturating the lower bound here is also an open problem.

\section{Multiparty quantities in holography}
\label{holo}

In holographic systems, we can calculate the entanglement entropy from the geometry of the bulk spacetime. Our focus is on the calculation at leading order in $1/G$, and the explicit examples we consider are time-reflection symmetric, so it will suffice for us to consider the original Ryu--Takayanagi formula \cite{Ryu:2006bv}, where the entropy of the reduced density matrix $\rho_A$ associated to a spatial region $A$ in a CFT at a moment in time is calculated by the area of a minimal surface $\gamma$ homologous to $A$ in the bulk spatial slice, 
\begin{equation}
    S_A = \frac{A_\gamma}{4G}. 
\end{equation}
This basic formula allows us to calculate the $n$-informations $I_n$ from knowledge of the bulk geometry.  

For the quantities involving the canonical purification, we need the conjectured bulk dual of the canonical purification proposed in \cite{Engelhardt:2018kcs,Dutta:2019gen} (a replica path integral argument for this prescription was given in \cite{Dutta:2019gen}). In general, the density matrix $\rho_X$ for some spatial region $X$ is dual to the bulk entanglement wedge $EW(X)$. In our time-symmetric case, this entanglement wedge will intersect the bulk spatial slice in a region bounded by $X$ and the minimal surface $\gamma_X$; with a slight abuse of notation we will also refer to this spatial region as $EW(X)$.  The dual of the canonical purification is constructed by gluing together two copies of the spatial surface $EW(X)$  across $\gamma_X$. This spatial surface provides Cauchy data which could be evolved to obtain a full spacetime, but our interest will be in its geometry. We refer to this as the doubled geometry. A multiboundary wormhole example of this construction is illustrated in Figure \ref{fig:4bdpur} later. 

For the reflected entropy $S_R(A:B)$, we consider a case where the spatial subregion in the boundary theory has two disjoint components $A, B$. The doubled geometry then has four boundary components, $A, B, A^*, B^*$, and the reflected entropy is conjectured to be given by the Ryu-Takayanagi surface for $AA^*$ in the doubled geometry \cite{Dutta:2019gen},
\begin{equation}
   S_R(A:B) =  S_{AA^*} = \frac{A_{\gamma'}}{4G}, 
\end{equation}
where $\gamma'$ is the area of the minimal surface homologous to $AA^*$ in the doubled geometry. By the symmetry of this doubled surface, $\gamma'$ consists of two copies of a surface in $EW(AB)$ called the entanglement wedge cross-section \cite{Takayanagi:2017knl}, which is the minimal-area surface that separates $EW(AB)$ into two components, one containing $A$ and the other containing $B$. Thus, $S_R(A:B)$ is given by twice the area of this entanglement wedge cross-section. 

\subsection{Holographic positivity}
\label{sec:holography}

In holography, entanglement entropies obey further restrictions that are not true in general quantum systems. This is referred to as the holographic entropy cone \cite{Bao:2015bfa}. The first example of such a restriction was the observation of \cite{Hayden:2011ag} that while $I_3$ is not sign-definite in general quantum systems, $-I_3$ is non-negative in holographic systems. 

The new four-party quantity $Q_4$ is also holographically non-negative. In the holographic dual description, the dual of the canonical purification $|\sqrt{\rho} \rangle_{ABCA^*B^*C^*}$ is a geometry formed from two copies of $EW(ABC)$. For a multiboundary wormhole case, this is pictured in Figure \ref{fig:4bdpur}.  Within this geometry we consider the entanglement wedge $EW(ABA^*B^*)$ and double this again to obtain the dual description of $|\sqrt{\rho}_{AA^*BB^*} \rangle_{AA^*BB^*A_*A^*_*B^*B^*_*}$. 

Let $\xi$ be the minimal surface in the doubly-reflected geometry that computes $S_{A A^* A_* A^*_*}$. 
Due to entanglement wedge nesting, the doubly-reflected geometry has four copies of $EW(AB)$, and these are disjoint. Thus we have
\begin{equation}
\begin{split}
    4G\, S_{A A^* A_* A^*_*}
    &= A_\xi \\
    & \geq A_{\xi \cap EW(AB)} + A_{\xi \cap EW(A^*B^*)}  + A_{\xi \cap EW(A_*B_*)} + A_{\xi \cap EW(A^*_*B^*_*)} \, .
\end{split}
\end{equation}
The holographic definition of $S_R(A:B)$ tells us that
\begin{equation}
    2G \, S_R(A:B) = \min_{\zeta \in S} A_\zeta \, .
\end{equation}
where $S$ is the set of all surfaces in $EW(AB)$ that separate $A$ and $B$. 
Since the surface $\left(\xi \cap EW(AB) \right) \in S$,
\begin{equation}
    A_{\xi \cap EW(AB)} \geq 2G\,  S_R(A:B) \, .
\end{equation}
A similar argument holds for the other three surfaces, so
\begin{equation}
    S_{A A^* A_* A^*_*} \geq 2 S_R(A:B) \quad \Rightarrow Q_4 =  S_{A A^* A_* A^*_*} -  2 S_R(A:B)  \geq 0 \, . 
\end{equation}
The $n$-informations for $n>3$ are not sign-definite even holographically. A counterexample for $I_4$ was given in the context of multiboundary wormholes in \cite{Balasubramanian:2014hda}. We will extend this to arbitrary $n$ in Section \ref{npos}.  

For the alternative quantities explored in Appendix \ref{further}, we have not been able to give an argument for sign-definiteness or an explicit counterexample.

\subsection{Vanishing conditions}

In holographic theories, one general feature is that there are regions of the parameter space where the entropy and other multiparty quantities vanish at leading order in $1/G$. It is interesting to understand both when the quantities vanish, and how they change from zero to non-zero values. A classic example is the mutual information of two boundary regions in vacuum AdS; this is non-zero if the regions are sufficiently close together and zero if they are far apart. This phase transition is continuous, although it involves a discontinuous change in the Ryu-Takanagi surface. Another example is the entropy of a thermal state; this is zero below a critical temperature and non-zero above the critical temperature (this is the the Hawking-Page phase transition in the bulk). This transition is discontinuous at leading order in $1/G$, due to a change in the topology of the dominant bulk saddle. 

For the higher $I_n$, transitions from zero to non-zero values will similarly be continuous when the transition involves a change in the Ryu--Takayanagi surface but not in the bulk geometry, but discontinuous if it involves a change of dominant solution in the bulk. For the quantities involving canonical purifications, the relevant geometry is the doubled one constructed from the entanglement wedges, so the quantities will change continuously when the transition involves a change in the minimal surface in this doubled geometry, that is, a change in the entanglement wedge cross-section but not the entanglement wedge, and will change discontinuously if the transition involves a change in the entanglement wedge. We will give examples of both behaviors for $R_3$ in the next subsection. 

\subsubsection{Residual information}
\label{sript}

\begin{figure}
    \centering
    \includegraphics[scale=0.6]{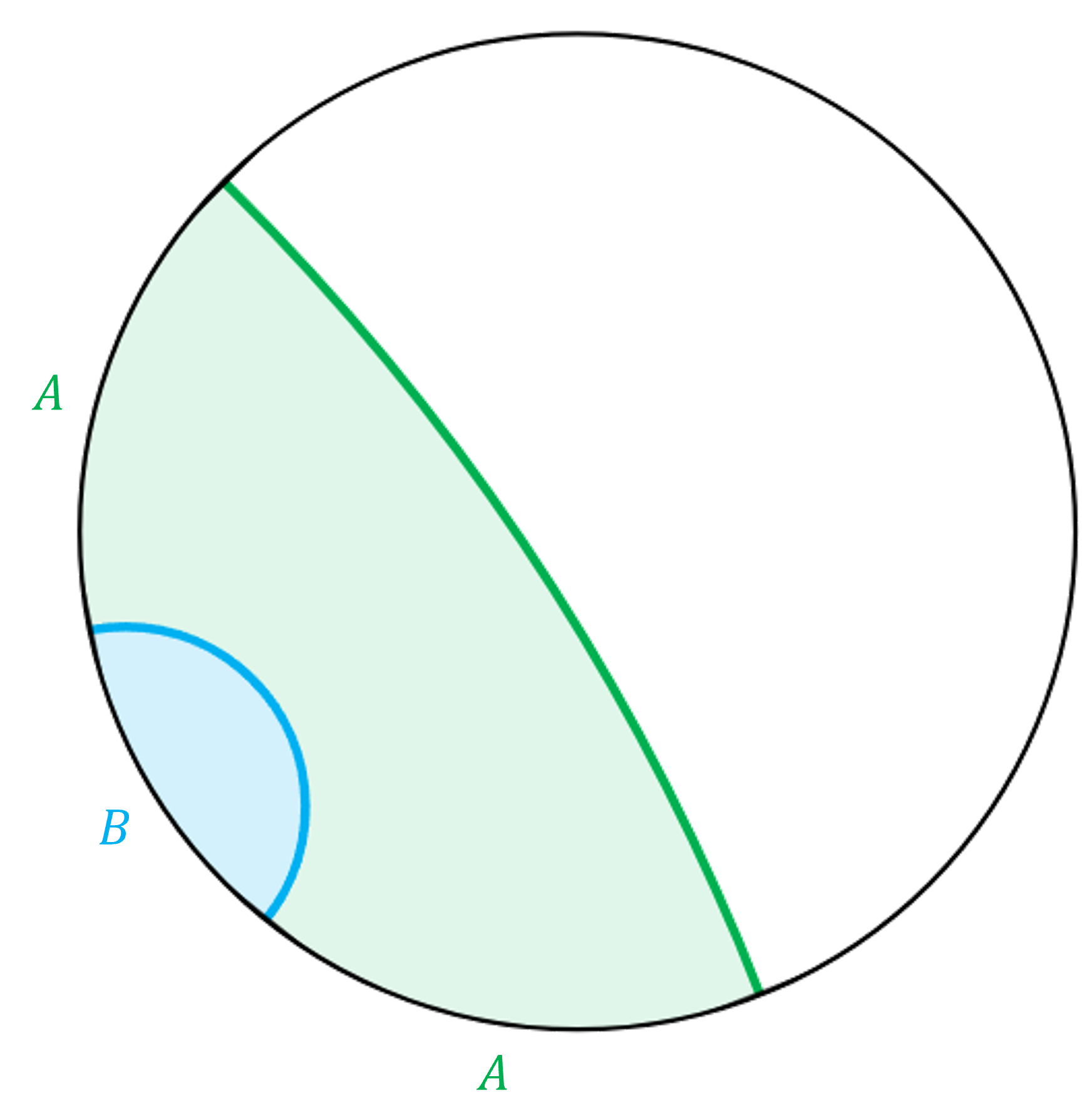}
    \caption{An example where a connected entanglement wedge has $EW(AB) = EW(A) \cup EW(B)$. We consider vacuum AdS$_3$, and take $B$ to be a single interval while $A$ consists of intervals on either side of $B$, chosen large enough that $EW(A)$ is connected. $EW(A)$ is shaded in green and $EW(B)$ is shaded in blue. $S_B$ is given by the length of the blue surface, $S_{AB}$ is given by the green surface, and $S_A$ is given by blue and green surfaces together, so we see that $S_A = S_{AB} + S_B$.}
    \label{fig:conneweg}
\end{figure}

Recall that the residual information defined in equation \eqref{eq:residual-info-def} is a signal of tripartite entanglement in a pure state on three parties, and is computed from the reduced density matrix obtained by tracing over one party.  Suppose we have a pure state on three disjoint boundary regions $A,B,C$ and seek the residual information for $A,B$ after tracing over $C$. Then the residual information $R_3(A:B)$ vanishes at leading order in $1/G$ if the entanglement wedge of $AB$ is equal to the union of the entanglement wedges of $A$ and $B$, $EW(AB) = EW(A) \cup EW(B)$. In proving this, there are two cases to consider: 
\begin{enumerate}
    \item If the $AB$ entanglement wedge is disconnected, then necessarily $EW(AB) = EW(A) \cup EW(B)$, and $S_{AB} = S_A + S_B$, so $I(A:B) = 0$. Since the entanglement wedge is disconnected the minimal cross-section is the empty set so also $S_R(A:B) = 0$. Thus $R_3(A:B) =0$.\footnote{Note that in general (not just in holography) $I(A:B)=0$ implies $S_R(A:B)=0$, but we do not want to rely on the general statement here as the holographic calculation tells us only that $I(A:B)=0$ to leading order in $1/G$, and it is not clear what such approximate vanishing implies for $S_R(A:B)$ in general.}
    \item If the $AB$ entanglement wedge is connected but we still have $EW(AB) = EW(A) \cup EW(B)$, as seen in the example in Figure \ref{fig:conneweg}, then taking  $S_A > S_B$ without loss of generality, we have $S_A = S_{AB} + S_B$,  so $I(A:B) = 2 S_B$, saturating the upper bound on the mutual information. The minimal surface for $B$ is also a cross-section of the entanglement wedge, so $S_R(A:B) \leq 2 S_B$ in general. Since $S_R \geq I$, we must have $S_R(A:B) = 2S_B$ in this case and hence $R_3(A:B) =0$. Thus, $EW(AB) = EW(A) \cup EW(B)$ implies $R_3(A:B) =0$.  
\end{enumerate} 

Furthermore, although $R_3(A:B)$ is not permutation invariant, this condition is. This is because the portion of the spatial slice in $EW(C)$ is the complement of that in $EW(AB)$, so we can restate the condition in the permutation-invariant form that $R_3(A:B)=0$ if $EW(A) \cup EW(B) \cup EW(C)$ covers the whole spatial slice. 

We also have in a sense a converse: $R_3(A:B)=0$ in an open region of the parameter space only if $EW(AB)$ is the union of $EW(A)$ and $EW(B)$. The argument for this is that if $EW(AB)$ is not the union of $EW(A)$ and $EW(B)$, we have $I(A:B) < 2S_B$. If the minimal entanglement wedge cross-section is still the minimal surface for $B$, this implies $R_3(A:B) >0$. If not, the minimal entanglement wedge cross-section is some other surface not involved in the calculation of $I(A:B)$, and generically its area will be different from that of the surfaces involved in $I(A:B)$, so for generic moduli $R_3(A:B)$ is different from zero. \\

\begin{figure}[H]
    \centering
    \includegraphics[scale=0.6]{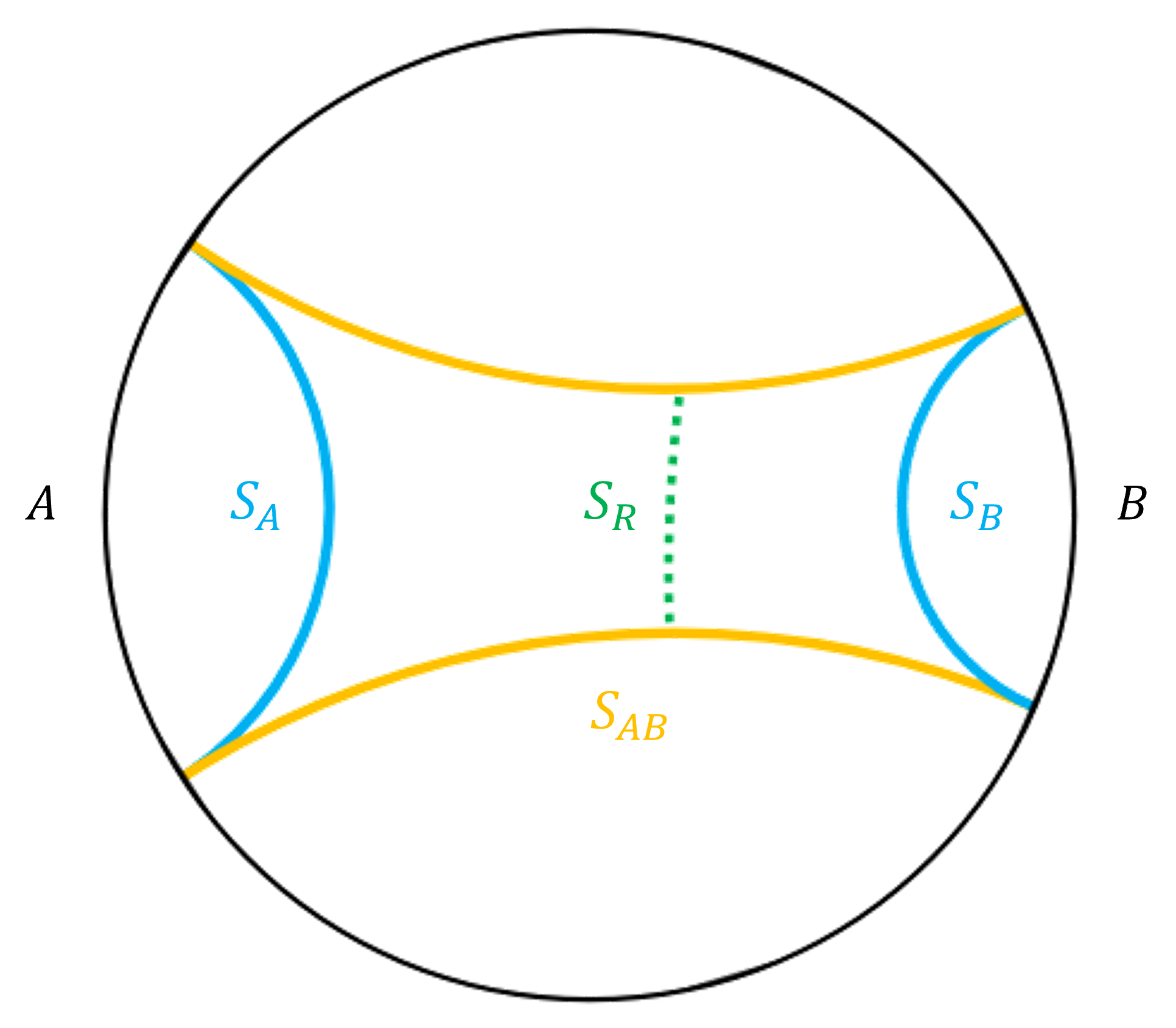}
    \caption{An example of a discontinuous phase transition in $R_3$: we consider vacuum AdS$_3$, and take $A$ and $B$ to be disjoint intervals. For large enough $A$ and $B$, $S_{AB}$ is given by the connected surface (boundary marked in yellow) and $R_3(A:B)$ is non-zero. As $A$ and $B$ shrink, there is a phase transition when the surface for $S_{AB}$ changes from the connected to the disconnected surface. $I(A:B)$ goes smoothly to zero, but $R_3(A:B)$ is bounded away from zero until we go through the transition by the argument of \cite{Hayden:2021gno}.}
    \label{fig:vac}
\end{figure}

A simple example of the continuous vs discontinuous transitions is the case where $A$ and $B$ are two disjoint intervals in the spatial boundary of vacuum AdS$_3$, and $C$ is the rest of the boundary. If the entanglement wedge is connected, $S_{AB}$ is given by the area of the surfaces joining one end of $A$ to one end of $B$, and $I(A:B)$ is non-zero, as shown in Figure \ref{fig:vac}. $S_R(A:B)$ is given by the entanglement wedge cross-section, and \cite{Hayden:2021gno} has shown that $R_3(A:B) = S_R(A:B) - I(A:B)$ is bounded away from zero. As we shrink $A$ and $B$ there is a phase transition at which the entanglement wedge of $AB$ changes to the disconnected one. $I(A:B)$ drops continuously to zero in this phase transition, but as $R_3(A:B)$ is bounded away from zero in the former phase, it must change discontinuously.\footnote{This example thus also illustrates that while $I=0$ implies $S_R=0$, it is not the case that $I \to 0$ implies $S_R \to 0$.}

\subsubsection{$I_3$ and $Q_4$}
\label{i3q3pt}

For $I_3$, it is easy to see that it vanishes if $EW(ABC)$ is disconnected. Suppose without loss of generality  $C$ is the disconnected party. Then $S_{ABC} = S_{AB} + S_C$, and entanglement wedge nesting implies $S_{AC} = S_A+ S_C$, $S_{BC} = S_B + S_C$. Hence
\begin{equation}
    I_3 = S_A + S_B + S_C - S_{AB} - S_{AC} - S_{BC} + S_{ABC} = 0. 
\end{equation}
By the permutation invariance, this implies that $I_3=0$ unless all three-party entanglement wedges are connected.

Similarly, $Q_4$ vanishes if $EW(ABC)$ is disconnected. Here we need to consider the cases where $C$ or $B$ are the disconnected party separately. If $EW(C)$ is a disconnected component in $EW(ABC)$, the doubled entanglement wedge in the purification has two components, and the trace over $CC^*$ is just throwing away a disconnected component. Hence the geometry dual to the canonical purification on $AA^*A_*A^*_*BB^*B_*B^*_*$ is simply two copies of the geometry dual to $AA^*BB^*$, and $S_{AA^*A_*A^*_*} = 2 S_{AA^*}$, so $Q_4=0$. If $EW(B)$ is a disconnected component in $EW(ABC)$, the part of the geometry containing $B$ is not connected to that containing $A$ in any of the purifications, so $S_{AA^*A_*A^*_*} = 2 S_{AA^*}=0$, and $Q_4=0$. 

For even $n$, the above argument straightforwardly generalizes to show that if some $n-1$ party entanglement wedge has a single disconnected component, $I_n=0$. If $A_n$ is the disconnected party, $S_{T A_n}= S_T + S_{A_n}$ for all $T \subset \{ A_1 \ldots A_{n-1} \}$, so \eqref{symdis} will hold up to additional factors of $S_{A_n}$, and these additional factors cancel out in $I_n$. 

\section{Multiparty quantities in vacuum AdS}
\label{vacads}

A simple setting to study these multiparty quantities holographically is the boundary subregions in the vacuum in global AdS. We consider the AdS$_3$ case, where these quantities can be calculated explicitly. 

The hyperbolic plane gives a constant time slice in vacuum AdS$_3$, which is described as the upper-half plane with metric 
\begin{equation}
    ds^2 = \frac{1}{z^2}(dz^2 + dx^2), 
\end{equation}
where we have set the AdS length scale $\ell_{\text{AdS}} = 1$. The geodesic between boundary points $x_1, x_2$ is a semi-circle $x^2 + z^2 = r^2$ of radius $r = \frac{1}{2} |x_2-x_1|$. This geodesic has length 
\begin{align}
    \ell = 2 \ln \frac{|x_2-x_1|}{\epsilon} ,
\end{align}
where $\epsilon$ is a radial cutoff.\footnote{The result for the multiparty quantities will be independent of the cutoff; in particular as the multiparty quantities all involve differences between geodesics with endpoints at the same point on the boundary, making the cutoff vary along the boundary does not change our results.}

\begin{figure}
    \centering
    \includegraphics[scale=0.6]{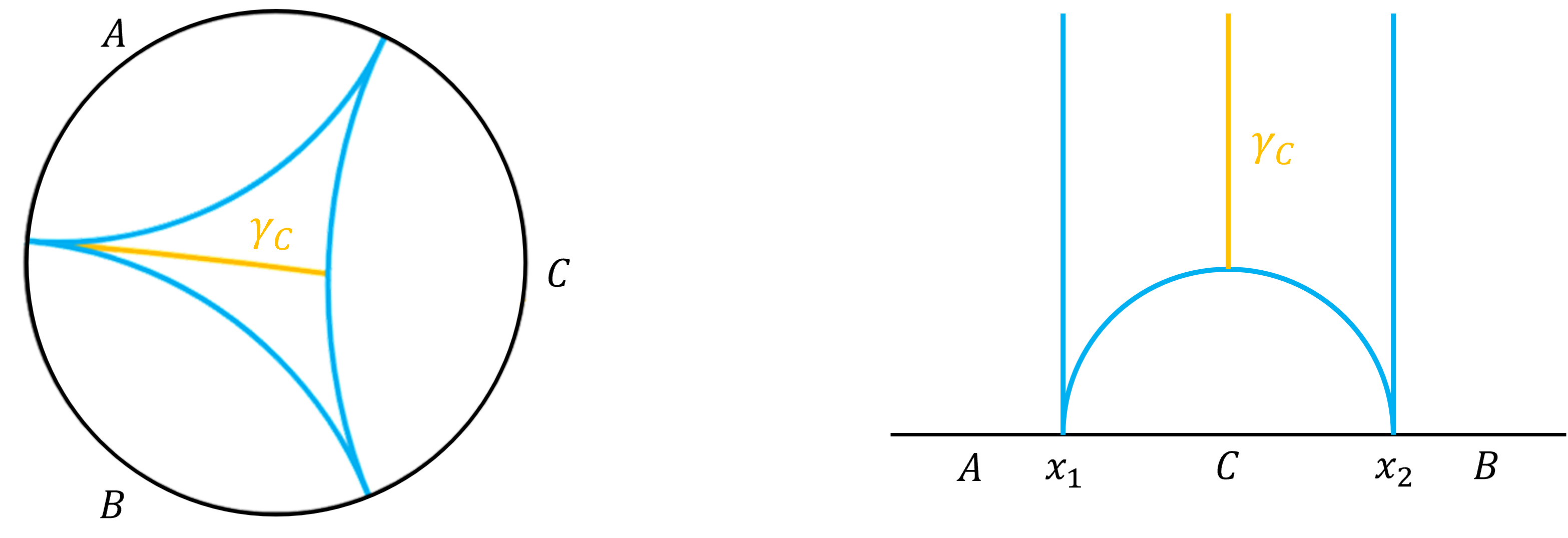}
    \caption{Left: Dividing the boundary of vacuum AdS into three regions. The entanglement wedge cross-section $\gamma_C$ is also shown. Right: The upper-half plane picture obtained by sending the point at $x_3$ between $A$ and $C$ to infinity.}
    \label{fig:3vac}
\end{figure}

Consider the division of the boundary of global AdS into three regions shown in Figure \ref{fig:3vac}. For any values of the moduli, $EW(AB)$ extends to the $C$ geodesic, and the entanglement wedge cross-section is given by $\gamma_C$. Thus, we have 
\begin{equation}
    R_3(A:B) = \frac{1}{4G}(2\lgc + \lc - \la - \lb). 
\end{equation}
To calculate the lengths, we consider the upper-half plane picture obtained by sending the boundary point between $A$ and $B$ to infinity; see the right panel in Figure \ref{fig:3vac}. It follows that 
\begin{equation}
    \la = \lb = \ln \frac{\zmax}{\epsilon}, 
    \qquad
    \lc = 2 \ln \frac{x_2-x_1}{\epsilon} ,
    \qquad
    \lgc = \ln \frac{\zmax}{r}, 
\end{equation} 
where $\zmax$ is an IR cutoff and $r = \frac{1}{2} (x_2-x_1)$ is the radius of the semicircular $C$ geodesic. Thus, 
\begin{equation}
    R_3(A:B) = \frac{1}{4G} \,  2 \ln \frac{x_2-x_1}{r} = \frac{1}{4G} \, \ln 4.
\end{equation}
This is constant, and saturates the bound in \cite{Hayden:2021gno}. The constancy is a consequence of conformal invariance. As the dependence on the UV cutoff cancels out, the resulting finite quantity is a conformally invariant function of the separations. However, there is no non-trivial conformal invariant for three points, so the result must be a constant.

\begin{figure}[H]
    \centering
    \includegraphics[scale=0.5]{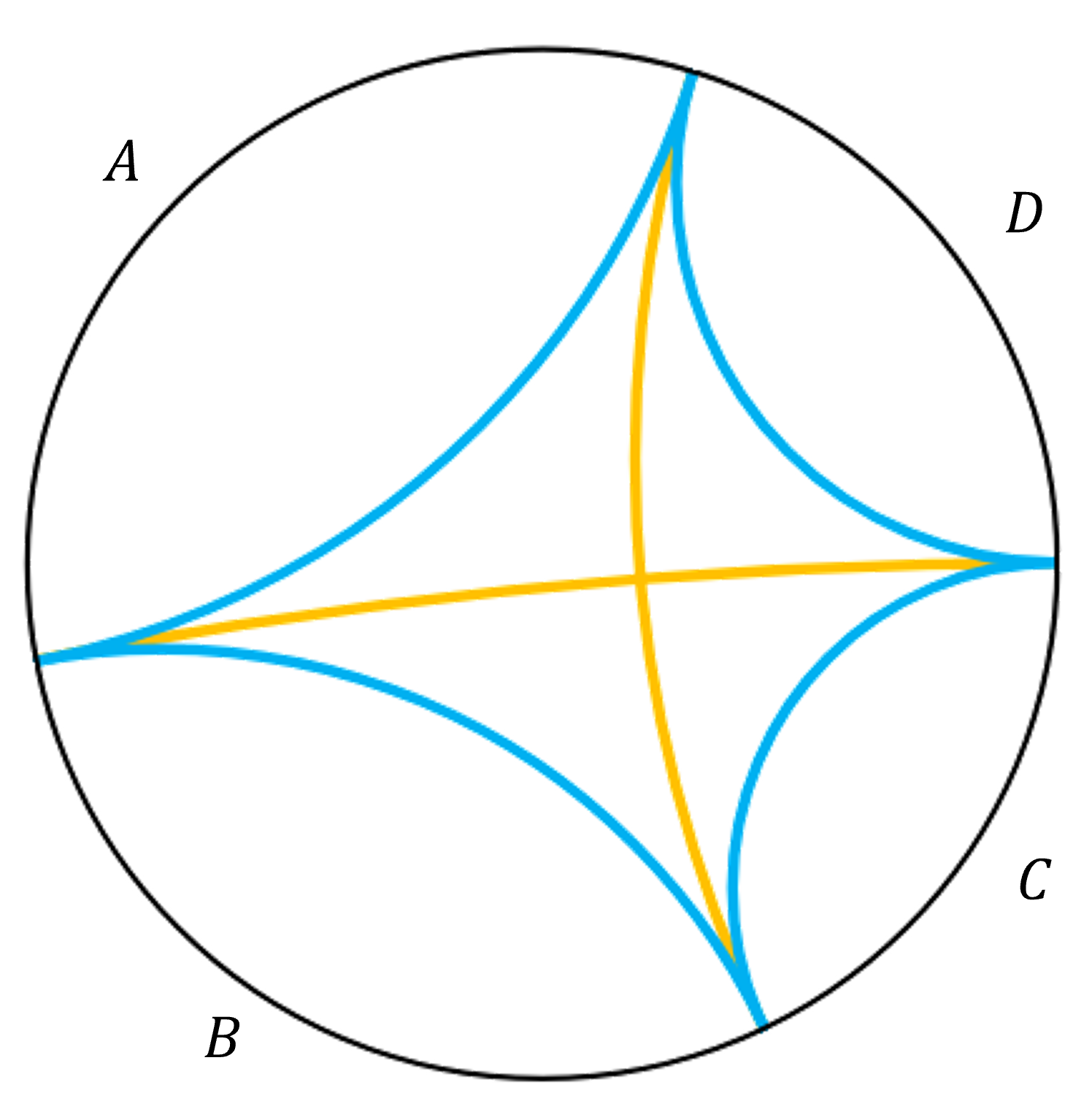}
    \caption{Dividing the boundary of vacuum AdS into four regions. The external geodesics spanning each of the boundary regions have lengths $\ell_A$, $\ell_B$, $\ell_C$, $\ell_D$. There are also two crossing geodesics whose lengths we call $\ell_{AB}$ and $\ell_{BC}$.}
    \label{fig:4vac}
\end{figure}

Consider the division of the boundary of global AdS into four regions as shown in Figure \ref{fig:4vac}. For any values of the moduli, we have  $S_i = \frac{1}{4G} \ell_i$ for $i = A,B,C,D$. We also have $S_{AB} = \frac{1}{4G} \ell_{AB}$, $S_{BC} = \frac{1}{4G} \ell_{BC}$, and $S_{AC} = \frac{1}{4G} \min(\la + \lc, \, \lb + \ld)$. Thus
\begin{equation} \label{i3vac}
    I_3 = \frac{1}{4G} [ \max \left( \la + \lc, \lb + \ld \right) - \ell_{AB} - \ell_{BC}].
\end{equation}
Thus $I_3$ is a difference between two external geodesics and two internal geodesics, where the internal geodesics intersect. 

\begin{figure}[H]
    \centering
    \includegraphics[scale=0.6]{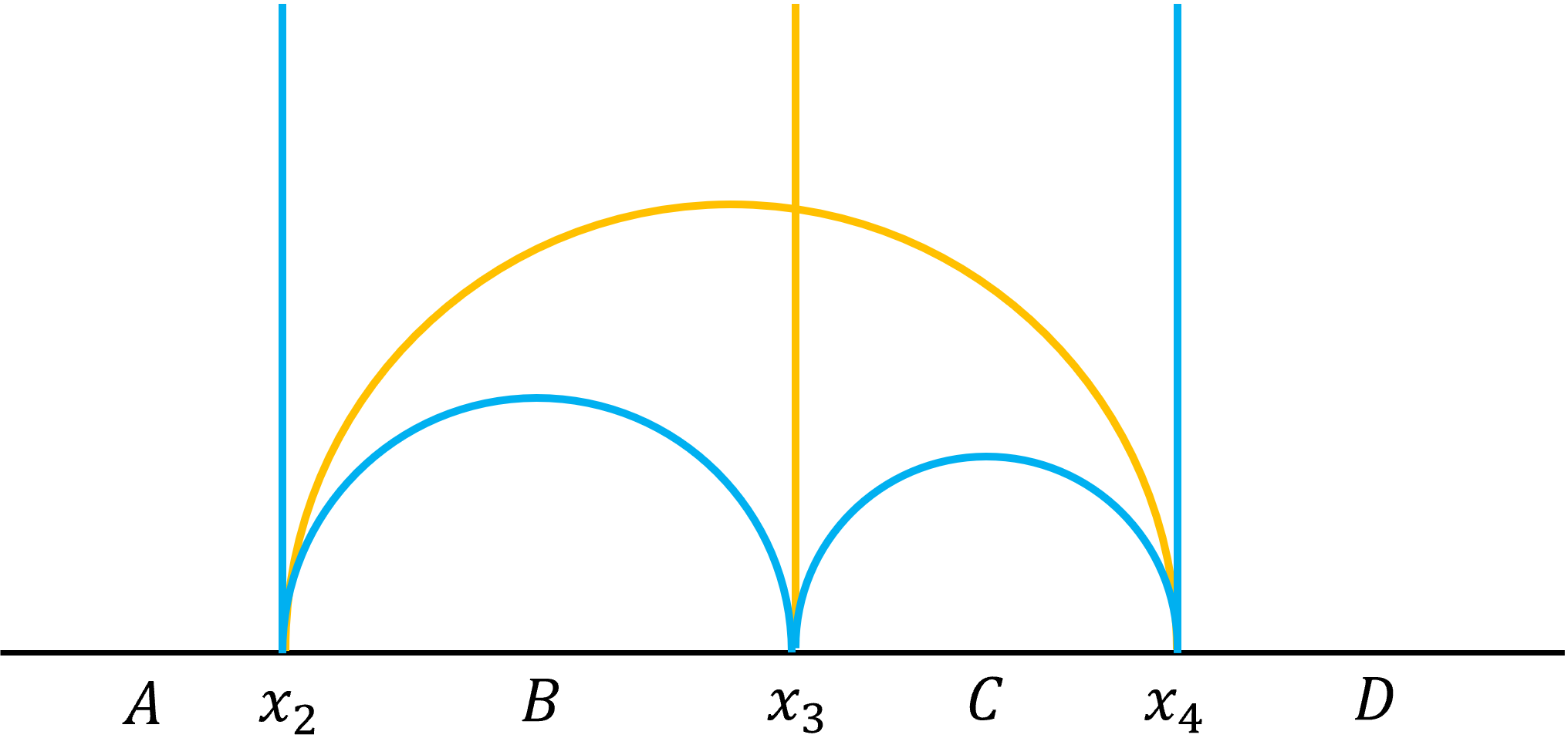}
    \caption{The four region case in the upper-half plane picture obtained by sending the point at $x_1$ between $A$ and $D$ to infinity.}
    \label{fig:4vac2}
\end{figure}

To calculate the lengths, we consider the upper-half plane picture obtained by sending the boundary point $x_1$ between $A$ and $D$ to infinity; see Figure \ref{fig:4vac2}. In these coordinates 
\begin{equation}
    \la = \ld = \ell_{AB} = \ln \frac{\zmax}{\epsilon}, \quad \lc = 2 \ln \frac{x_4-x_3}{\epsilon}, \quad \lb = 2 \ln \frac{x_3-x_2}{\epsilon},
    \quad \ell_{BC} = 2 \ln \frac{x_4-x_2}{\epsilon}.
\end{equation}
Thus,
\begin{equation}
    I_3 = \frac{1}{4G} [ \max( \lc, \lb )- \ell_{BC}] = \frac{1}{4G} \max \left( 2 \ln \frac{x_4-x_3}{x_4-x_2}, 2 \ln \frac{x_3-x_2}{x_4-x_2} \right). 
\end{equation}
\noindent
This is a function of the conformally invariant cross-ratio
\begin{equation}
    y = \frac{(x_2-x_1)(x_4-x_3)}{(x_3-x_1)(x_4-x_2)};
\end{equation}
as we have chosen to work in a frame with $x_1 = \infty$, $y = \frac{x_4-x_3}{x_4-x_2}$, so
\begin{equation}
    I_3 =  \frac{1}{4G}  \max \left( 2 \ln y, 2 \ln (1-y) \right). 
\end{equation}
Consider the symmetric case where the internal geodesics cross at right angles. In this case, $x_3= \frac{1}{2}(x_4+x_2)$, so $y = \frac{1}{2}$ and $I_3 = - \frac{1}{4G} \ln 4$. As earlier, this saturates the bound in \cite{Hayden:2021gno}. This gives the maximum value of $- I_3$. Moreover, in the limits where two of the $x_i$ coincide, $I_3 \to 0$. This matches the expectation that the four-party entanglement contribution vanishes when we shrink one of the regions to approach the three-party case.   

\begin{figure}
    \centering
    \includegraphics[scale=0.6]{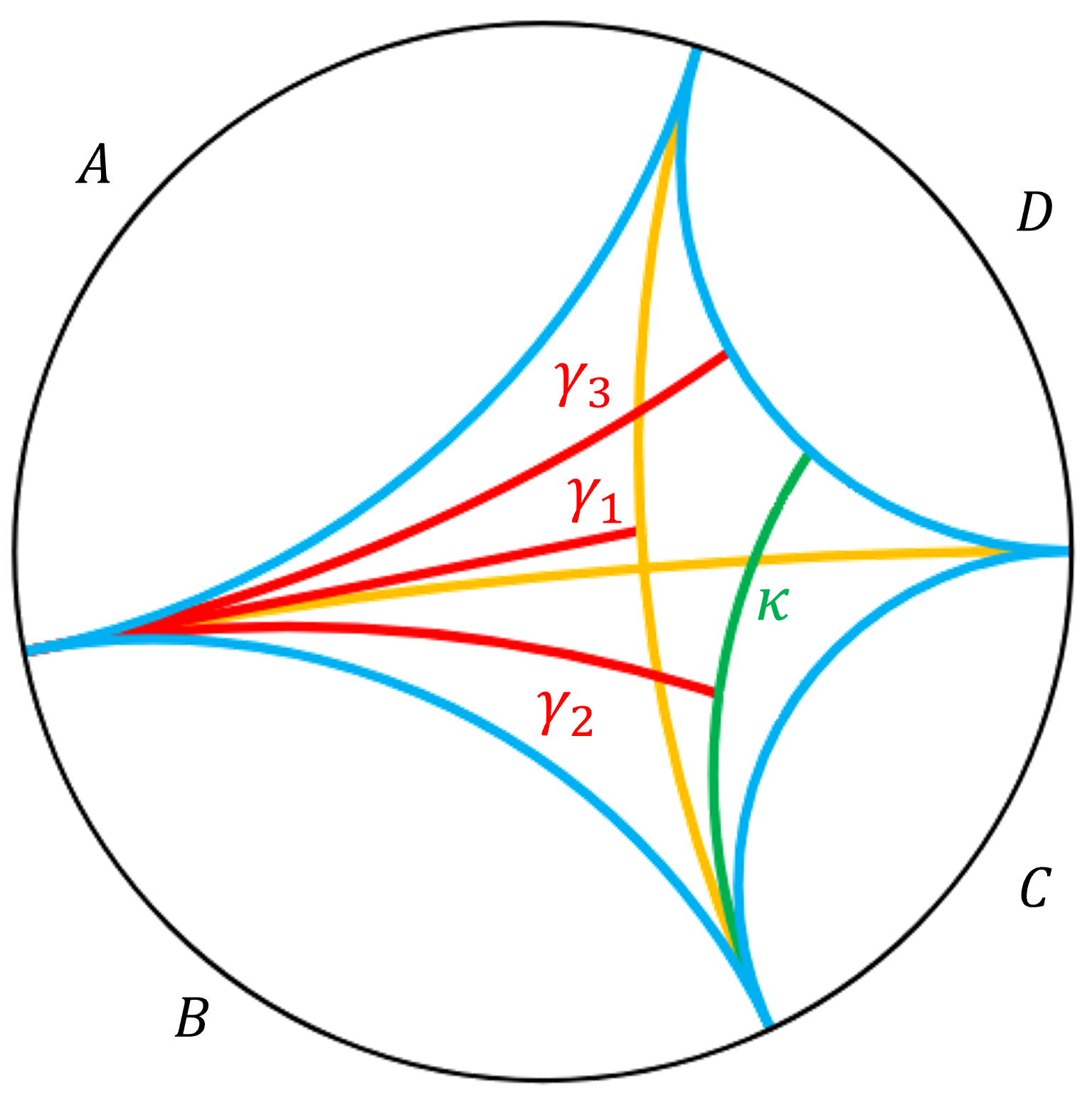}
    \caption{The minimal surfaces relevant for the calculation of $Q_4$. The boundary of $EW(ABA^*B^*)$ is the $\kappa$ geodesic. The reflected entropy is given by the length of the geodesic segment $\gamma_1$ ending on the $AB$ geodesic, and $S_{AA^*A_*A^*_*}$ is given by the length of the geodesic segment $\gamma_2$ ending on $\kappa$ or the length of a geodesic segment $\gamma_3$ ending on the $D$ geodesic.}
    \label{fig:4vacq4}
\end{figure}

Next, we consider $Q_4$. For any values of the moduli, we have $S_{AB} = \frac{1}{4G} \ell_{AB}$. The entanglement wedge cross-section in $EW(AB)$ which determines $S_{R}$ is the geodesic segment $\gamma_1$ meeting $\ell_{AB}$ orthogonally as shown in Figure \ref{fig:4vacq4}. The first purification in the doubled purification always joins two copies along the $D$ geodesic. Then $EW(ABA^*B^*)$ is bounded by an internal geodesic $\kappa$, which meets the $D$ geodesic orthogonally. 

\begin{figure}[H]
    \centering
    \includegraphics[scale=0.65]{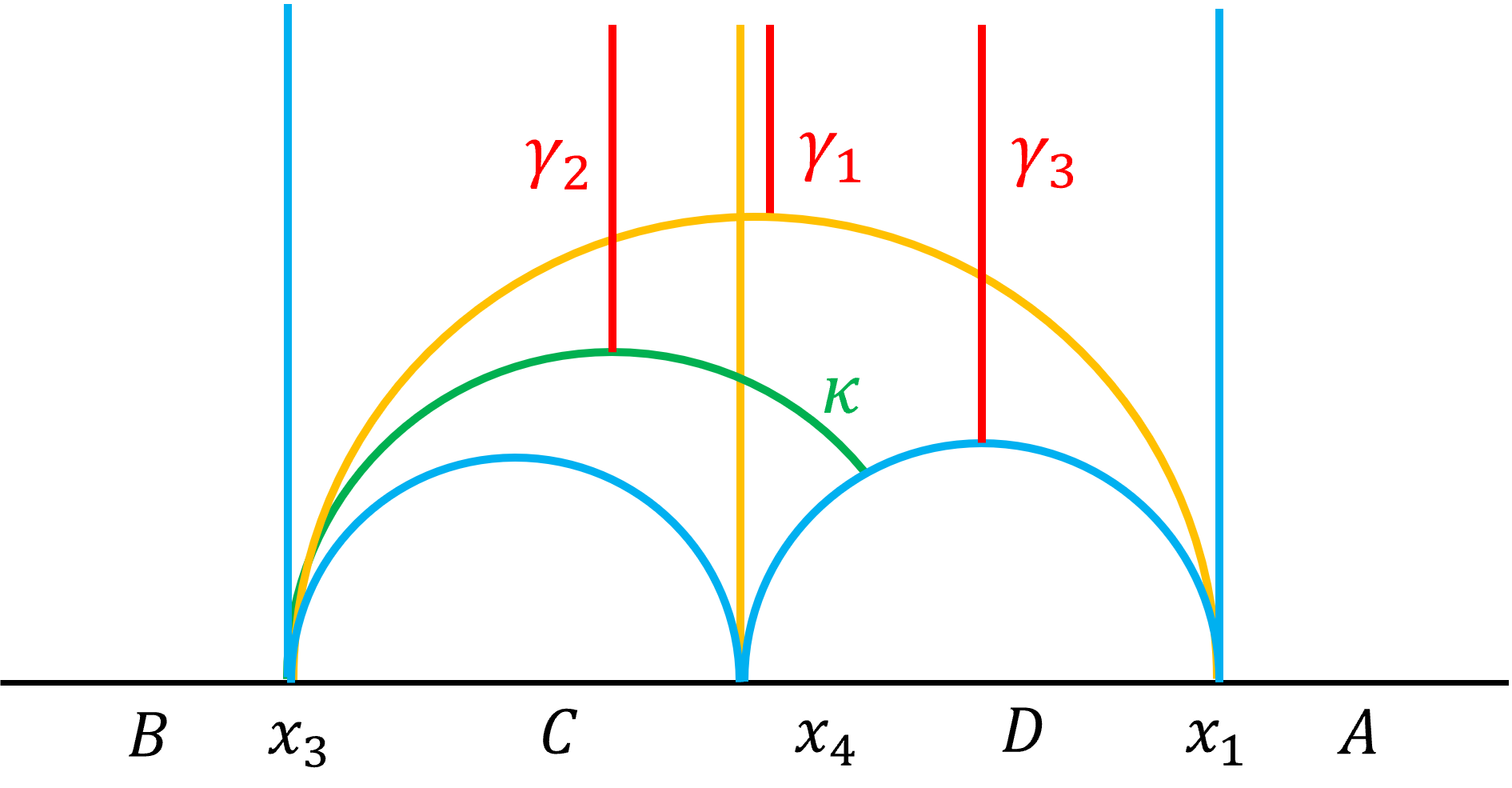}
    \caption{The upper-half plane picture in which we send $x_2 \to \infty$.}
    \label{fig:4vacq4a}
\end{figure}

\begin{figure}[H]
    \centering
    \includegraphics[scale=0.65]{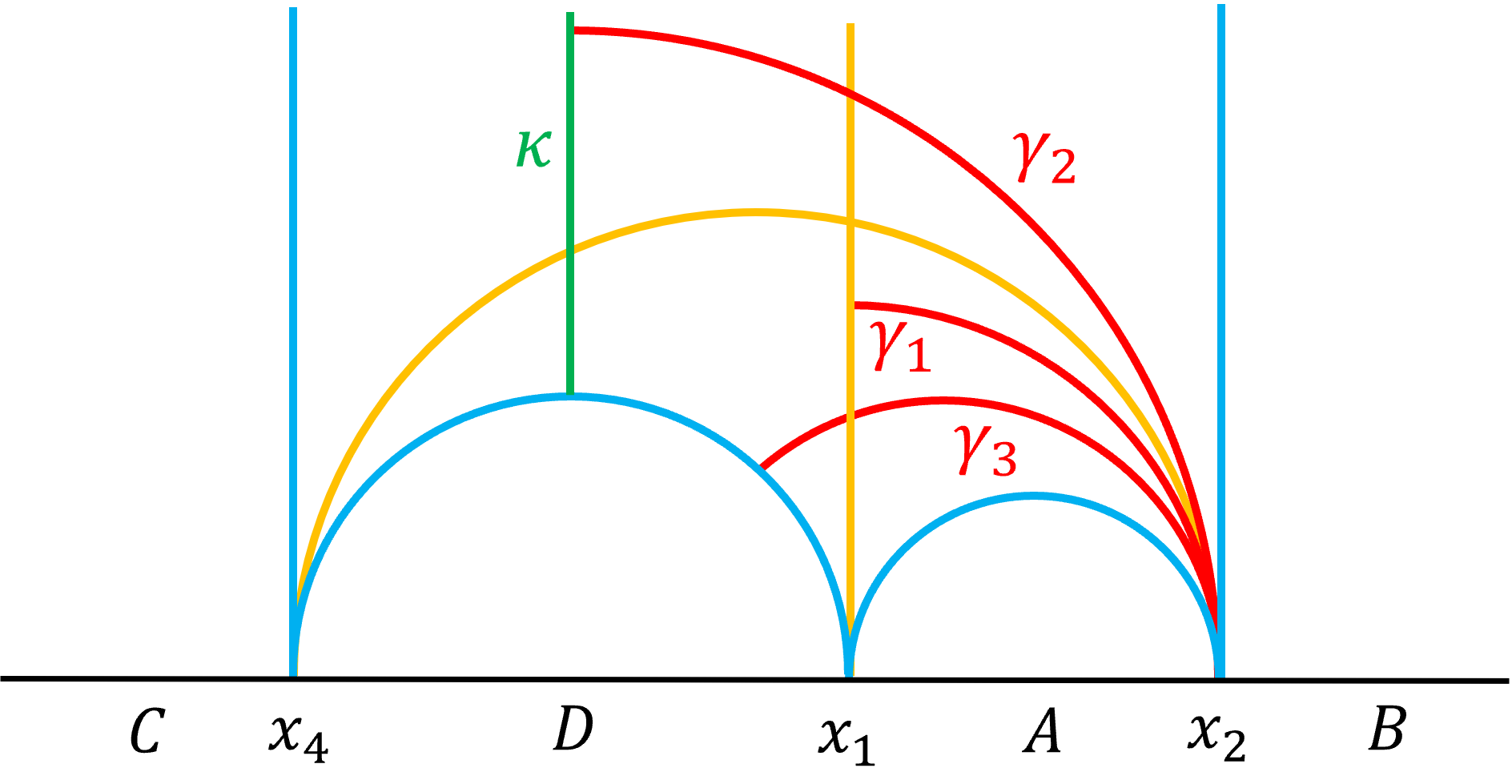}
    \caption{The upper-half plane picture in which we send $x_3 \to \infty$.}
    \label{fig:4vacq4b}
\end{figure}

The entanglement wedge cross-section in the double geometry is either a geodesic segment $\gamma_2$ ending on $\kappa$ orthogonally or a geodesic segment $\gamma_3$ ending on the $D$ geodesic orthogonally, and 
\begin{equation}
    Q_4 = \frac{1}{G} [\min \left( \ell_{\gamma_2}, \, \ell_{\gamma_3} \right) - \ell_{\gamma_1}]. 
\end{equation}
If $\gamma_3$ is the shorter geodesic, we consider the upper-half plane picture obtained by sending the point $x_2$ between $A$ and $B$ to infinity, see Figure \ref{fig:4vacq4a}. 
Then the lengths are
\begin{equation}
    \ell_{\gamma_1} = \ln \frac{2\zmax}{x_1-x_3}, \quad \ell_{\gamma_3} = \ln \frac{2\zmax}{x_1-x_4}, 
\end{equation}
so 
\begin{equation}
  Q_4 = \frac{1}{G} \ln \frac{x_1-x_3}{x_1-x_4} = \frac{1}{G} \ln \frac{1}{1-y}. 
\end{equation}
If $\gamma_2$ is the shorter geodesic, we consider the upper-half plane picture obtained by sending the point $x_3$ between $B$ and $C$ to infinity, see Figure \ref{fig:4vacq4b}. 
Then the lengths are
\begin{equation}
    \ell_{\gamma_1} = \ln \frac{2(x_2-x_1)}{\epsilon}, \quad \ell_{\gamma_2} = \ln \frac{2 (x_2 - \frac{1}{2}(x_1 + x_4))}{\epsilon}, 
\end{equation}
and hence, 
\begin{equation}
  Q_4 = \frac{1}{G} \ln \left( \frac{1}{2} + \frac{1}{2} \frac{x_2-x_4}{x_2-x_1} \right)  =  \frac{1}{G} \ln \frac{y+1}{2y}. 
\end{equation}
Thus we find that
\begin{equation}
  Q_4 =  \frac{1}{G} \max \left( \ln \frac{1}{1-y} , \ln \frac{y+1}{2y} \right) . 
\end{equation}
As for $I_3$, this goes to zero when $y \to 0$ or $y \to 1$, i.e., whenever one of the boundary regions shrinks. However, unlike $I_3$, the maximum here is not at the symmetric point, but at $y = \sqrt{2} -1$, where $Q_4 = \frac{1}{G}  \ln (1+\frac{1}{\sqrt{2}})$. 

In Appendix \ref{further}, we calculate the alternative four-party entanglement signal $D_4$ in this case. We find that $D_4$ is positive, going to zero as $y \to 0,1$, but it has a pair of degenerate maxima and $D_4=0$ at $y = \frac{1}{2}$. Thus, the qualitative behavior of the different four-party entanglement signals are similar, but they differ in detail. 

In Appendix \ref{vac56}, we extend this discussion to consider five and six boundary regions. This exemplifies the calculation of the higher multiparty entanglement signals introduced in Section \ref{higher}. The qualitative results are similar; the values of the signals are typically of order one in units of $1/G$. 

\section{Multiboundary wormholes}
\label{multbound}

Multiboundary wormhole geometries provide another holographic setting for studying the properties of these multiparty quantities. In this section, we review some aspects of these geometries. In the following two sections, we will then consider the multiparty quantities in the limits of small and large horizon size respectively. 

We begin by considering an Euclidean geometry with  metric
\begin{equation} \label{bulks}
    ds^2 = dt_E^2 + \cosh^2 t_E \, d\Sigma^2, 
\end{equation}
where $\Sigma$ is a genus--$g$ Riemann surface with $n$ boundaries, and $d\Sigma^2$ is a constant negative curvature metric (of unit radius) on $\Sigma$. 
This is a locally AdS$_3$ geometry, where the AdS length scale $\ell_{\text{AdS}} = 1$. Constant negative curvature metrics on $\Sigma$ are obtained by constructing $\Sigma$ as a quotient of the unit-radius hyperbolic plane $H^2$ by a discrete subgroup $\Gamma_\Sigma$ of $SL(2,\mathbb{R})$. These quotients of AdS$_3$ were first considered in \cite{Aminneborg:1997pz}, and their holographic study was initiated in \cite{Krasnov:2000zq,Krasnov:2003ye,Skenderis:2009ju}. A study of holographic entanglement properties was initiated in  \cite{Balasubramanian:2014hda,Marolf:2015vma}.

The solution has a conformal boundary which is isomorphic to two copies of $\Sigma$. The CFT path integral on one copy of $\Sigma$ defines a state in $n$ copies of the CFT on a circle, which we denote $|\Sigma\rangle$. If equation \eqref{bulks} is the dominant saddle-point for these boundary conditions, this state is dual to the slice at $t_E=0$, i.e., $|\Sigma\rangle$ is holographically dual to a connected spatial wormhole $\Sigma$ with $n$ asymptotic boundaries.  

In every asymptotic region of $\Sigma$, there is an outermost closed geodesic homologous to that boundary. We can thus divide $\Sigma$ into $n$ distinct cylindrical regions, connecting each asymptotic boundary to its corresponding outermost closed geodesic, and a central region with $n$ closed geodesic boundaries and genus $g$, which we denote $\overline{\Sigma}$. (There is one exceptional case, $n=2$ and $g=0$, where this central region is empty. Then, the two closed geodesics are identified, and we recover the Euclidean BTZ black hole \cite{Banados:1992wn}.)

Taking the $t_E=0$ slice as initial data for a Lorentzian solution, the Lorentzian spacetime looks like a black hole from the perspective of each asymptotic region, with the closed geodesic being the bifurcation surface of the event horizon. 
The geometry outside of the horizon in each asymptotic region is identical to an exterior region in the BTZ geometry.
The surface $\Sigma$ is then a generalization of the Einstein--Rosen bridge, connecting these $n$ asymptotic regions inside the black hole. 

In the state $|\Sigma \rangle$, we can construct a reduced density matrix over some subset $\mathcal{S}$ of the asymptotic boundaries by tracing over the remaining boundaries $\overline{\mathcal{S}}$.
The entropy of this reduced density matrix is determined holographically by a Ryu--Takayanagi surface in the bulk; this is a closed (but potentially not connected) geodesic homologous to $\mathcal{S}$, or equivalently, to $\overline{\mathcal{S}}$. 
These geodesics lie within (or on the boundary of) the central region $\overline{\Sigma}$. 
Thus, to understand the entanglement structure of $|\Sigma \rangle$, we wish to characterize the geometry of $\overline{\Sigma}$.  

This is a genus $g$ Riemann surface with $n$ geodesic boundaries. 
Any Riemann surface that admits a hyperbolic structure\footnote{The only ones that don't are the torus and the cylinder.} can be decomposed (in infinitely many ways) into $2g + n-2$ \emph{pairs-of-pants} surfaces with three geodesic boundaries, by cutting along $2g + n-3$ internal cycles. 
The Fenchel-Nielsen parameterization describes this surface's geometry by assigning a set of ``length'' and ``twist'' parameters to each geodesic boundary of these pairs-of-pants; the twist parameters describe how two pairs-of-pants are glued together along their shared internal geodesic boundary.
It follows that the moduli space is $4g+3n-6$ dimensional, given by the lengths of the $n$ external boundaries, and a length and twist at each of the $2g + n-3$ internal cycles which separate pairs-of-pants in our decomposition. 

The main difficulty comes from the fact that the relation between different Fenchel--Nielsen parametrizations of a given Riemann surface is generically not known. 
Hence, if we pick a particular decomposition of $\overline{\Sigma}$, the lengths of the geodesics we cut along are included in our parametrization, but we do not know the lengths of other closed geodesics in $\overline{\Sigma}$. 
This is been the main obstacle in determining the Ryu--Takayanagi surfaces in these geometries. In earlier works, this problem is addressed by focusing on two limits: 
\begin{itemize}
    \item the \emph{puncture limit}, where the external geodesics are much shorter than the AdS scale \cite{Balasubramanian:2014hda}, or
    \item the \emph{large $\ell$ limit}, where the external geodesics are much longer than the AdS scale \cite{Marolf:2015vma}.
\end{itemize}

In the puncture limit, the internal geodesics in $\overline{\Sigma}$ are longer than the external geodesics, and thus, become irrelevant for determining the Ryu--Takayanagi surfaces, at least away from pinching limits in the moduli space. 

In the large $\ell$ limit, $\overline{\Sigma}$ consists of thin strips. 
This is because the area of $\overline{\Sigma}$ is fixed by the Gauss-Bonnet theorem to be $2(n-2+2g) \pi$, so by taking the external geodesics to be long, their separations become small. Along most of their lengths, each portion of the external geodesics lies close to either a different portion of this geodesic or a portion of a different external geodesic. 
The entanglement structure of $|\Sigma \rangle$ in this limit is dominated by local bipartite entanglement between portions of the $n$ asymptotic boundaries. 

We will examine the behavior of the multiparty signals for three and four boundary wormholes in the puncture limit and the large $\ell$ limit in Sections \ref{punc} and \ref{largel}, respectively. For the latter, our central tool is the observation that the large $\ell$ simplifies the relations between the lengths of different geodesics in $\overline{\Sigma}$. 

\begin{figure}[H]
    \centering    \includegraphics[scale=0.6]{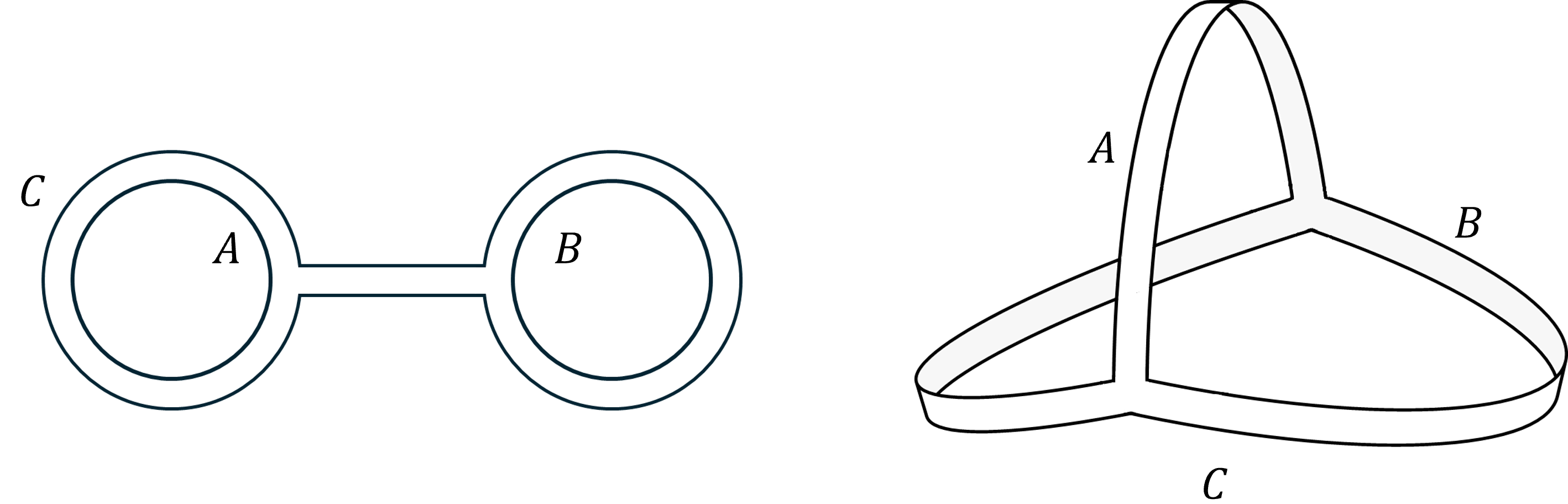}
    \caption{The $C$-eyeglass and pinwheel configurations for a pair-of-pants at large $\ell$.}
    \label{fig:pp}
\end{figure}

Consider a pair-of-pants whose geodesic boundaries are labeled as $A$, $B$, and $C$, with lengths $\la$, $\lb$, and $\lc$ in the large $\ell$ limit. 
If one of the geodesics is longer than the sum of the other two, say $\lc > \la + \lb$, the geometry of the pair-of-pants looks like the left picture in Figure \ref{fig:pp}, which is called an \emph{eyeglass configuration}.
We have three kinds of eyeglasses, which we label according to the longest geodesic as $A$-eyeglass, $B$-eyeglass, and $C$-eyeglass. 
Otherwise, the geometry looks like the right picture in Figure \ref{fig:pp}, which we call the \emph{pinwheel configuration}.

In the pinwheel configuration, the length of the geodesic segment $\gamma_C$ that runs from $C$ to itself between $A$ and $B$ is 
\begin{equation}
    \ell_{\gamma_C} \approx \frac{1}{2} (\la + \lb - \lc). 
\end{equation}
The approximation here means that this holds up to subleading corrections in the lengths $\la$, $\lb$, and $\lc$. 
Similarly, for an $A$-eyeglass we have 
\begin{equation}
    \ell_{\gamma_C} \approx \la - \lc, 
\end{equation}
and for a $B$-eyeglass 
\begin{equation}
    \ell_{\gamma_C} \approx \lb - \lc, 
\end{equation}
while for a $C$-eyeglass $\ell_{\gamma_C} \approx 0$. The lengths of these geodesic segments will be useful for our computations in later sections.

\subsection{Details of the four-boundary wormhole}
\label{4bd}

\begin{figure}
    \centering
    \includegraphics[scale=0.6]{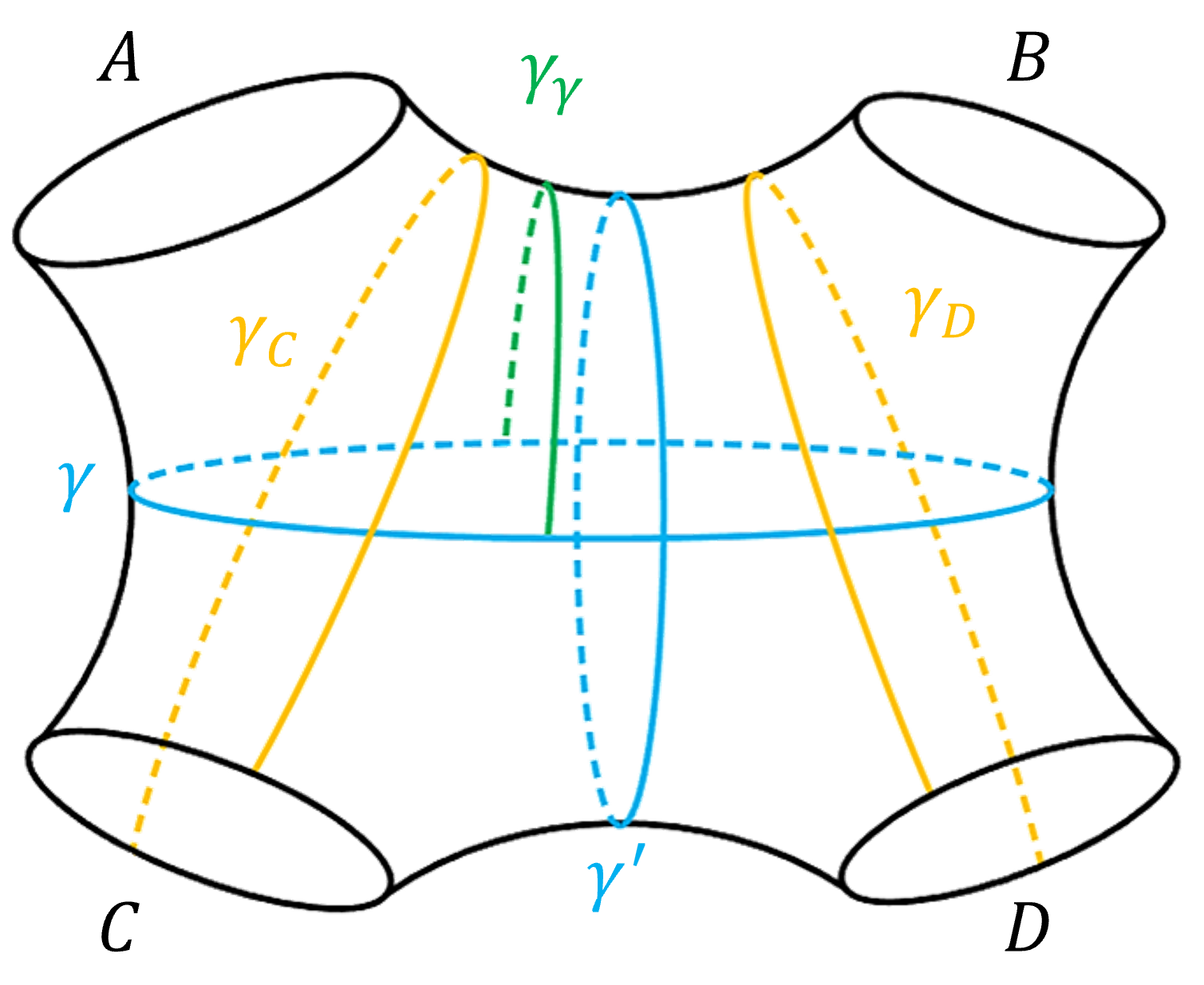}
    \caption{The decomposition of the region $\overline{\Sigma}$ in the four-boundary wormhole into two pairs-of-pants at zero twist, showing also the geodesic $\gamma'$, the geodesic segment $\gamma_\gamma$, and similar segments $\gamma_{C}$, $\gamma_{D}$ that will appear later.}
    \label{fig:4bdy}
\end{figure}

Consider a four-boundary wormhole with no handles. Its boundaries are labeled as $A, B, C, D$. This wormhole can be decomposed into two pairs-of-pants in infinitely many ways; among these, three decompositions are useful: 
\begin{enumerate}[\indent (i)]
\item split along the minimal geodesic $\gamma$ separating $A$ and $B$ from $C$ and $D$; we get $AB\gamma$ and $CD\gamma$ pairs-of-pants.
\item split along the minimal geodesic $\gamma'$ separating $A$ and $C$ from $B$ and $D$; we get$AC\gamma'$ and $BD\gamma'$ pairs-of-pants.
\item split along the minimal geodesic $\tilde \gamma$ separating $A$ and $D$ from $B$ and $C$; we get $AD\tilde \gamma$ and $BC\tilde \gamma$ pairs-of-pants.
\end{enumerate}
In any such decomposition, the geometry is specified by the lengths of the minimal geodesics in each asymptotic region, $\la, \lb, \lc$ and $\ld$, the length of the minimal geodesic where the two pairs-of-pants are joined together, and a twist $\tau \in [0, 2\pi)$ at the joining. In Figure \ref{fig:4bdy}, we show an example of a four-boundary wormhole where the twist in decompositions (i) and (ii) vanishes, showing the geodesics $\gamma$ and $\gamma'$ and some other segments that are relevant later.

We consider the large $\ell$ limit, i.e., 
\begin{align}
    \la, \, \lb, \, \lc , \, \ld >> 1 .
\end{align}
Since we have a topologically trivial interior, at least some portion of each geodesic must lie close to a distinct geodesic.
Then we can have two different scenarios. 

First, the set of four external geodesics can be split into two pairs so that for each pair some portion of the geodesics are close to each other. 
Without loss of generality, we assume that $A$ and $B$ are close, $C$ and $D$ are close.
The geometry splits into $AB\gamma$ and $CD\gamma$ pairs-of-pants along the minimal geodesic $\gamma$, separating $AB$ from $CD$. Since $A$ and $B$ are close, $C$ and $D$ are close, the $AB\gamma$ and $CD\gamma$ pairs-of-pants are not $\gamma$-eyeglasses, i.e.,
\begin{align}
    \lgm < \la + \lb , \quad \lgm < \lc + \ld .
\end{align}  

Second, all three of the geodesics only lie close to the fourth one. Without loss of generality, we pick $A$ to be the geodesic that is close to $B,\, C, \, D$. The geometry splits into $AB\gamma$ and $CD\gamma$ pair-of-pants along the minimal geodesic $\gamma$, separating $AB$ from $CD$. Since $A$ and $B$ are close, $AB\gamma$ is not a $\gamma$-eyeglass, i.e., $\lgm < \la + \lb$. In contrast, $C$ and $D$ are not close to each other; hence, $CD\gamma$ must be a $\gamma$-eyeglass, i.e., $\lgm > \lc + \ld$. Moreover, the portion of $\gamma$ along $B$ corresponds to a part of the bridge of this eyeglass. Since all of $B, C, D$ are along $A$,
\begin{align}
    \la > \lb + \lc + \ld .
\end{align} 
Using such choices of decompositions will considerably simplify our analysis in Section \ref{largel}.

Now suppose $AB\gamma$ and $CD\gamma$ are two pinwheels, and consider the other ways of splitting the four external boundaries into two pairs. Without loss of generality, consider the split of $AC$ from $BD$, where the minimal geodesic between them is $\gamma'$.
This $\gamma'$ can be constructed by connecting a geodesic segment $\gamma_\gamma$ across the $AB\gamma$ pinwheel to a similar segment across the $CD\gamma$ pinwheel. Hence, 
\begin{equation}
    \lgp \approx \frac{1}{2} (\la + \lb - \lgm) + \frac{1}{2} (\lc + \ld - \lgm) + \beta \lgm, \qquad \beta \in \left(0\,,\,1\right) ,
\end{equation}
where the last term comes from connecting the endpoints of the two geodesic segments along $\gamma$. It follows that 
\begin{equation}
    2\lgp \approx \la + \lb + \lc + \ld - 2(1 -  \beta) \lgm <  \la + \lb + \lc + \ld,
\end{equation}
which is inconsistent with $\lgp > \la + \lc$ and $\lgp > \lb + \ld$ simultaneously. 
Thus, when $AB\gamma$ and $CD\gamma$ are two pinwheels with large $\lgm$, $AC\gamma'$ and $BD\gamma'$ cannot both be $\gamma'$-eyeglasses. 

\subsection{Multiparty quantities in the puncture limit}
\label{punc}

The puncture limit is where the lengths of the external geodesics are much smaller than the AdS scale.  We will compute $R_3$ in three-boundary wormholes, and $I_4 = 2I_3$ and $Q_4$ in four-boundary wormholes.

\subsubsection{Three-boundary wormholes}
\label{3bpunc}

Consider a three-boundary wormhole, with boundaries $A, B, C$. The geometry is specified by fixing the lengths of the minimal geodesics in each asymptotic region, $\la, \lb, \lc$. Without loss of generality, we assume that $\la \geq \lb \geq \lc$. 
We consider the puncture limit, which for three boundaries amounts to assuming that each RT surface is either empty or one of the minimal geodesics.  

If $\la > \lb + \lc$, $S_A = \frac{1}{4G} (\lb + \lc)$, which implies that $EW(AB) = EW(A) \cup EW(B)$, $EW(AC) = EW(A) \cup EW(C)$, and $EW(BC)$ is disconnected.
We have 
\begin{align}
    I(A:B) = \frac{1}{4G} 2\lb ,
    \quad I(A:C) = \frac{1}{4G} 2\lc .
\end{align}
Hence, 
\begin{align}
    R_3(A:B) = R_3(A:C) = R_3(B:C) = 0 .
\end{align}
This is an example of the general result from Section \ref{sript} that the residual information vanishes when $EW(A) \cup EW(B) \cup EW(C)$ covers the whole slice.

If $\la < \lb + \lc$, we have 
\begin{equation}
    R_3(A:B) = R_3(A:C) = \frac{1}{4G} (\lb + \lc - \la), \quad
    R_3(B:C) = \frac{1}{4G} (\la + \lc - \lb). 
\end{equation}
In the limit $\la \to \lb + \lc$, we get $R_3(A:B) = R_3(A:C) \to 0$, but $R_3(B:C) \to \frac{1}{4G} 2\lc$. 
Let us model the holographic state by a state in a finite-dimensional Hilbert space with dimensions given by $\ln d_A = S_A$, $\ln d_B = S_B$, $\ln d_C = S_C$. In this case, the upper bound on $R_3(B:C)$ in terms of $d_C$ given by equation \eqref{r3bounds} is saturated.\footnote{Note that here we are considering $R_3(B:C)$, so the bound is $R_3(B:C) \leq 2 \ln d_C$.}
Thus, in this limit, the state dual to this geometry takes the form
\begin{equation}
    |\psi \rangle_{ABC} = |\psi_1 \rangle_{A_1 B_1} \otimes |\psi_2 \rangle_{A_2 B_2 C}, 
\end{equation}
where $|\psi_2 \rangle$ is a three-party entangled state which gives maximal $R_3(B:C)$. We need $\dim B_2 \geq d_C$ and $\dim A_2 \geq 2 d_C$ to be able to saturate the bound. Since $I(A:B) = \frac{1}{4G} (\la + \lb - \lc) \to \frac{1}{4G} 2 \lb$, the residual state $|\psi_1 \rangle$ must be maximally entangled between $A$ and $B$ and $|\psi_2 \rangle$ must also contribute maximally to this entanglement. 

There is a sharp transition at $\la = \lb + \lc$. 
While $R_3(A:B) = R_3(A:C)$ and the mutual informations vary smoothly, $R_3(B:C)$ changes discontinuously from its maximum value to zero at the transition. This signals some sudden change in the entanglement structure. However, as $R_3 = 0$ is a possible value for states with three-party entanglement, we do not know whether the three-party entanglement vanishes above the transition.

\subsubsection{Four-boundary wormholes}
\label{4bpunc}

Consider a four-boundary wormhole, with boundaries $A, \, B, \, C, \, D$. Without loss of generality, we assume that $\la \geq \lb \geq \lc \geq \ld$. 
If $\la > \lb + \lc + \ld$, $EW(BCD)$ is disconnected, so $I_3=0$. 
If $\la < \lb + \lc + \ld$, $S_i = \frac{1}{4G} \ell_i$ for $i = A,B,C,D$. 
Then $S_{AB} = \frac{1}{4G} (\lc + \ld)$, $S_{AC} = \frac{1}{4G} (\lb + \ld)$, and there are two cases for $I_3$ depending on the minimal surface for $S_{BC}$. 
If $\la + \ld > \lb + \lc$, $S_{BC} = \frac{1}{4G} (\lb + \lc)$ and 
\begin{equation}
    I_3 = \frac{1}{4G}(\la - \lb - \lc - \ld) .
\end{equation}
while if $\la + \ld < \lb + \lc$, $S_{BC} = \frac{1}{4G} (\la + \ld)$ and 
\begin{equation}
    I_3 = \frac{1}{4G}(-2\ld).
\end{equation}
If we model the holographic state in the latter case by a state in a finite-dimensional Hilbert space, this saturates the upper bound on $-I_3$ in equation \eqref{i3bounds}. Hence, the state takes the form
\begin{equation}
    |\psi \rangle_{ABCD} = |\psi_1 \rangle_{A_1 B_1 C_1 }\otimes |\psi_2 \rangle_{A_2 B_2 C_2 D}, 
\end{equation}
where $\dim A_2, \dim B_2, \dim C_2 \geq d_D$ and $|\psi_2\rangle$ is a four-party entangled state with maximal $I_3$. 

Now we proceed to the computation of $Q_4$. We can calculate variants of $Q_4$ by tracing over any pair of boundaries, and we will get in general different answers. We denote the two regions we keep as $X$ and $Y$, the one we trace over first as $Z$, and the one we trace over second as $W$. In the puncture limit, the entanglement wedge cross-sections are given by the external geodesics if the entanglement wedge is connected, and vanish if the entanglement wedge is disconnected. Thus, the only case where $Q_4$ is non-zero is when $EW(XX^*X_*X^*_*YY^*Y_*Y^*_*)$ in the double purification is connected and $EW(XX^*YY^*)$ in the single purification is disconnected. It follows that
\begin{equation}\label{eq:XYconds}
    \ell_X + \ell_Y < \ell_W + \ell_Z, \quad \ell_Z < \ell_X + \ell_Y + \ell_W, \quad \ell_W < \ell_X + \ell_Y. 
\end{equation}
If $\la > \lb + \lc + \ld$, all of the $Q_4$'s vanish.
Thus, in the puncture limit $Q_4$ is never non-zero when $I_3$ vanishes. 

Consider $\la < \lb + \lc + \ld$. If $\la + \ld > \lb + \lc$,  the first condition in equation \eqref{eq:XYconds} implies that $Q_4$ can be non-zero for the pair $(X,Y)$ being one of $(B,D)$, $(C,D)$, or $(B,C)$.
For the $(B,C)$ pair, if we trace over $A$ first and then $D$, 
\begin{equation}
    Q_4(B:C) = \frac{1}{4G} 4 \lc, 
\end{equation}
which saturates the upper bound on $Q_4$ in equation \eqref{q4bounds}. It follows that the state takes the form 
\begin{equation}
    |\psi\rangle_{ABCD} = |\psi_1 \rangle_{A_1B _1D_1} \otimes |\psi_2 \rangle_{A_2 B_2 C D_2}, 
\end{equation}
where $|\psi_2\rangle$ is a four-party entangled state with maximal $Q_4$ and $\dim B_2 \geq d_C$.
Note that $Q_4$ is not bounded by the dimension of the spaces that were traced out, so we have no bounds on $\dim A_i$ or $\dim D_i$. 

If $\la + \ld < \lb + \lc$, $Q_4$ can be non-zero for the pair $(X,Y)$ being one of $(B,D)$, $(C,D)$, or $(A,D)$. For the $(A,D)$ pair, 
\begin{equation}
    Q_4(A:D) = \frac{1}{4G} 4 \ld, 
\end{equation}
which saturates the upper bound on $Q_4$ in equation \eqref{q4bounds}. We note that this case also saturated the upper bound on $-I_3$, so the dual state must have a four-party entanglement structure that saturates both upper bounds. 

\subsubsection{Counterexample to holographic posivity}
\label{npos}

In the puncture limit, \cite{Balasubramanian:2014hda} gave an example demonstrating that $I_4$ is not sign-definite in holographic systems. We extend this to the general $I_n$. Consider $I_n$ evaluated for $n$ of the boundaries in an $m$-boundary wormhole with all horizons equal, and the horizons short enough that the minimal surface for any set of boundaries is given by a combination of horizons, never by an internal geodesic. Then each term in $S_i$ is the same. For $m \geq 2n$, we then have $S_i =  \begin{pmatrix} n \\ i \end{pmatrix} i S_{A_1}$ for each $i$, and $I_n=0$. If $m = 2n-1$, the only change is $S_n = (n-1) S_{A_1}$, so $I_n=(-1)^n S_{A_1}$. If $m=2n-3$, $S_n = (n-3)S_{A_1}$ and $S_{n-1} = n (n-2) S_{A_1}$, so $I_n = -(-1)^n (n-3) S_{A_1}$. Thus for all $n > 3$,  there are holographic cases where $I_n$ takes either sign. 

\subsection{Multiparty quantities in the large $\ell$ limit}
\label{largel}

The large $\ell$ limit is where the external geodesics are much longer than the AdS scale. 
We will compute $R_3(A:B)$ in three-boundary wormholes and $I_4 = 2I_3$ and $Q_4$ in four-boundary wormholes. 
We will see that in this limit the multiparty quantities do not grow with the horizon areas. This is somewhat similar to the case of regions in vacuum AdS, analyzed in Section \ref{vacads}, where the entropy of any single region is dominated by a divergent bipartite component; here too the large horizon areas seem to be due to bipartite entanglement as the multiparty quantities don't grow in this limit. Furthermore, we will argue that the multiparty quantities for more than three parties are small for generic moduli. This is consistent with the discussion in \cite{Marolf:2015vma}, which showed that the dual state in this large $\ell$ limit has mostly bipartite entanglement, with an order one amount of tripartite entanglement. 

\subsubsection{Three-boundary wormholes}
\label{sr3}

Consider a three-boundary wormhole, with boundaries $A, B, C$. The geometry is specified by fixing the lengths of the minimal geodesics in each asymptotic region $\la, \lb, \lc$. Without loss of generality, we assume that $\lb \leq \la$. There are two phases for the $AB$ entanglement wedge. First, if $\la + \lb < \lc$, $EW(AB)$ is disconnected, and $I(A:B) = S_R(A:B) = 0$. 
Second, if $\lc < \la + \lb$, $EW(AB)$ is connected. 
Assuming additionally that $\la > \lb + \lc$, we find $S_A = \frac{1}{4G} (\lb + \lc)$, $I(A:B) = 2 S_B = \frac{1}{4G} 2 \lb$, and $R_3(A:B)$ by the general argument of Section \ref{sript}. 

Thus, the non-trivial case is $ \lc <\la + \lb $ and $\la < \lb + \lc $. We also have $\lb < \la + \lc$ by our assumption that $\lb \leq \la$. Hence, $S_A = \frac{1}{4G} \la$ and $I(A:B) = \frac{1}{4G} (\la + \lb - \lc)$. There are two possible cases for the entanglement wedge cross-section which determines the reflected entropy: either $S_R(A:B) = 2\lb$ or $S_R(A:B) = 2\lgc$.  In the former case, 
\begin{equation} \label{2diff3a} 
R_3(A:B) = \frac{1}{4G} (\lb + \lc- \la) = I(B:C), 
\end{equation}
whereas in the latter case,
\begin{equation} \label{2diff3b} 
R_3(A:B) = \frac{1}{4G} (2\lgc + \lc - \la - \lb).  
\end{equation}

In the $A A^* B B^*$ geometry, we can cut and recombine the $C$ and $\gamma_C$ geodesics to obtain closed curves in the same homology class as the $A$ and $B$ geodesics. These must be longer than the $A$ and $B$ geodesics, and the difference has a lower bound \cite{Hayden:2021gno}:
\begin{align}
\label{eq:Haydenbound}
    2 \lgc + \lc - \la - \lb \geq 4 \ln 2 .
\end{align}

Now we take the large $\ell$ limit and consider the case where $R_3(A:B)$ is non-zero and the $ABC$ pair-of-pants is a pinwheel. In Section \ref{multbound}, we showed that
\begin{equation}
2 \ell_{\gamma_C} \approx \la + \lb - \lc . 
\end{equation}
As $\la < \lb + \lc$ and $\lgc < \lb$, the $\gamma_C$ geodesic determines the reflected entropy for generic parameters. We see from equation \eqref{2diff3b} that $R_3(A:B)$ is nonzero but small, i.e., it does not grow with $\ell$.
However, we see that it is $O(1/G)$ from equation \eqref{eq:Haydenbound}. 

When $\la = \lb + \lc$, we observe a phase transition with vanishing $R_3(A:B)$. At this transition point, by equation \eqref{eq:Haydenbound}, $2 \lgc > \la + \lb - \lc =  2\lb$. However, we know $\lgc < \lb$ if we are far from the phase transition; hence, there must be a switchover as we approach the phase transition. At this switchover, $R_3(A:B)$ changes from equation \eqref{2diff3b} to equation \eqref{2diff3a}, which can go smoothly to zero at the transition. Note that this transition shows that there are regions of the moduli space where $R_3$ is arbitrarily small but non-zero.

When $\lc = \la + \lb$, we also observe a phase transition with vanishing $R_3(A:B)$. By contrast, the $EW(AB)$ changes from being connected to disconnected, and $R_3(A:B)$ changes discontinuously. The corresponding transition in the puncture limit was discussed in Section \ref{3bpunc} (note that the external boundaries are labeled differently there).

At large $\ell$, this is a transition from a pinwheel to an eyeglass.
In both geometries, there are tri-junctions corresponding to local tripartite entanglement in the dual state.
For $\lc < \la + \lb$, these junctions entangle subregions in $A,B,C$, while for $\lc > \la + \lb$, they entangle a subregion in $A$ with two subregions in $C$ or a subregion in $B$ with two subregions in $C$. The latter corresponds to bipartite entanglement between $AC$ or $BC$. Thus, at large $\ell$, this transition does involve a change of the tripartite entanglement structure.   

We can extend this to consider wormholes with more boundaries and ask if $R_3(A:B)$ remains small at large $\ell$. For four-boundary wormholes we show it stays small in Appendix \ref{sr4}.

\subsubsection{$I_3$ for four-boundary wormholes}
\label{I34b}

Consider a four-boundary wormhole, with boundaries $A, \, B, \, C, \, D$.  In Section \ref{i3q3pt}, we argued in general that $I_3=0$ unless the three-party entanglement wedges are connected. This implies that $S_i = \ell_i / 4G$, for $i=A,B,C,D$, and that no single boundary is longer than the sum of the other three. 

Now consider the large $\ell$ limit. In Section \ref{4bd}, we showed that having no single boundary longer than the sum of the other three implies that there is a decomposition of the four-boundary wormhole into $AB\gamma$ and $CD\gamma$ pairs-of-pants which are not $\gamma$-eyeglasses; i.e., $\lgm < \la + \lb$, $\lgm < \lc + \ld$. Thus,
\begin{equation}
S_{AB} = \frac{1}{4G} \lgm.
\end{equation}
Connectedness of the three-party entanglement wedges further implies $\la < \lb + \lgm$, and similarly for the other boundaries. Hence, we deduce that $AB\gamma$ and $CD\gamma$ are pinwheels.

We now first consider zero twist $\tau$ between $AB\gamma$ and $CD\gamma$. The minimal geodesic $\gamma'$, which is a candidate minimal surface for $S_{AC}$, then has length
\begin{equation} 
\label{lgpll}
    \lgp \approx \frac{1}{2}(\la + \lb - \lgm) + \frac{1}{2} (\lc + \ld - \lgm) + \frac{1}{2} | \la + \ld - \lb - \lc |,
\end{equation}
where the first term is the length of the geodesic segment $\gamma_\gamma$ in the $AB\gamma$ pinwheel, the second term is the length of the similar segment in the $CD\gamma$ pinwheel, and the third term is the offset between the endpoints of these two segments along $\gamma$. It follows that
\begin{align}
    \lgp \approx \max (\la + \ld, \lb + \lc) - \lgm.
\end{align}
We have 
\begin{equation}
    S_{AC} = \frac{1}{4G} \min \left( \la + \lc,  \, \lgp, \, \lb + \ld \right).  
\end{equation}
Recall that $\la < \lb + \lgm$, and similarly for $B,C, D$. This implies that $\lgp < \min( \la + \lc,  \, \lb + \ld)$, so $S_{AC} = \frac{1}{4G} \lgp$. 

Consider the minimal geodesic $\tilde \gamma$ contributing to $S_{BC}$. For $\tau=0$, this has length 
\begin{equation}
\begin{aligned}
   \ltg 
   &\approx \frac{1}{2}(\la + \lb - \lgm) + \frac{1}{2} (\lc + \ld - \lgm) + \lgm \\
   &= \frac{1}{2} (\la + \lb + \lc + \ld), 
\end{aligned}  
\end{equation}
where the final term in the first line is present because $\tilde \gamma$ wraps around the $\gamma$ geodesic to achieve the correct homology. This is always intermediate between the other options for $S_{BC}$, so $S_{BC} = \frac{1}{4G} \min (\la+\ld,  \, \lb + \lc)$. 
Thus we see that, with zero twist, 
\begin{equation} \label{i34bdy}
    I_3 = \frac{1}{4G} \left[ \max (\la + \ld, \lb + \lc)  - \lgm -\lgp \right].
\end{equation}
Since $\lgm + \lgp \approx \max (\la+\ld, \lb + \lc) $, we have $I_3 \approx 0$. This formula is similar to the expression for four boundary regions in the vacuum AdS in equation \eqref{i3vac}: in both cases, we get a difference between two external geodesics and two intersecting internal geodesics.  

Now we consider a non-zero twist. Initially, adding a twist does not change $\lgp$, but it decreases $\ltg$.
The $\tilde \gamma$ geodesic remains non-minimal until an endpoint of the $\gamma_{\gamma}$ geodesic in the $AB\gamma$ pinwheel coincides with an endpoint of the analogous geodesic in the $CD\gamma$ pinwheel. At this point 
\begin{equation}
\begin{aligned}
    \ltg &\approx \frac{1}{2}\left( \la + \lb - \lgm \right) + \frac{1}{2} \left( \lc + \ld - \lgm \right) + \lgm - \frac{1}{2} \left| \la - \lc + \ld -\lb \right| \\
    &= \min \left( \la + \ld,  \, \lb + \lc \right),
\end{aligned}
\end{equation}
and $S_{BC}$ is now determined by $\tilde \gamma$. As we increase the twist, $\ltg$ is decreasing, but $\lgp$ is increasing by the same amount, so we continue to have $I_3 \approx 0$. When the other pair of endpoints cross, $\ltg$ becomes constant, and $\gamma'$ no longer determines $S_{AC}$. Thus, $I_3$ can never grow larger than order one at large $\ell$. 

Now we determine how large $I_3$ can be as a function of the moduli. In the case where $\la = \lc$ and $\lb = \ld$, the two pairs-of-pants are symmetric. In this case, $\gamma'$ consists precisely of two copies of $\gamma_\gamma$, which meet $\gamma$ orthogonally. Hence, the lower bound of \cite{Hayden:2021gno} applies, and $-I_3 \geq \ln 2 / G$. However, $\gamma'$ will not meet $\gamma$ orthogonally in general, and hence, we do not necessarily get the lower bound in \cite{Hayden:2021gno}. The difference $\lgg - \frac{1}{2} (\la+\lb-\lgm)$ can still be bounded, but the approximation for $\lgp$ in equation \eqref{lgpll} is an over-estimate. 
In the offset region, $\gamma'$ will run approximately along $\gamma$, so when the offset is significant, the angle at which $\gamma'$ and $\gamma$ meet will be small. In this small angle regime, it is reasonable to assume that the difference $\lgm + \lgp - \max (\la +\ld, \lb +\lc)$ is also small, as seen in the vacuum region case in Section \ref{vacads}.

The conclusion is that in the large $\ell$ limit, $I_3$ for the four-boundary wormhole is at most of order one in units of $1/G$. There are choices of the moduli for which it is of order one, but for generic moduli, we conjecture that it is small. This would be consistent with the structure of entanglement in large $\ell$, observed in \cite{Marolf:2015vma}. The pair-of-pants at large $\ell$ consists of a set of thin strips joined together at two tri-junctions. At generic moduli, these tri-junctions do not line up, so they give a local tripartite contribution to the entanglement structure of the dual state. There is no local tetrapartite contribution, as the four boundary regions are never simultaneously close. 
 
\subsubsection{$Q_4$ for four-boundary wormholes}
\label{Q34b}

\vspace{3\baselineskip}

\begin{figure}[H]
    \centering
    \includegraphics[scale=0.6]{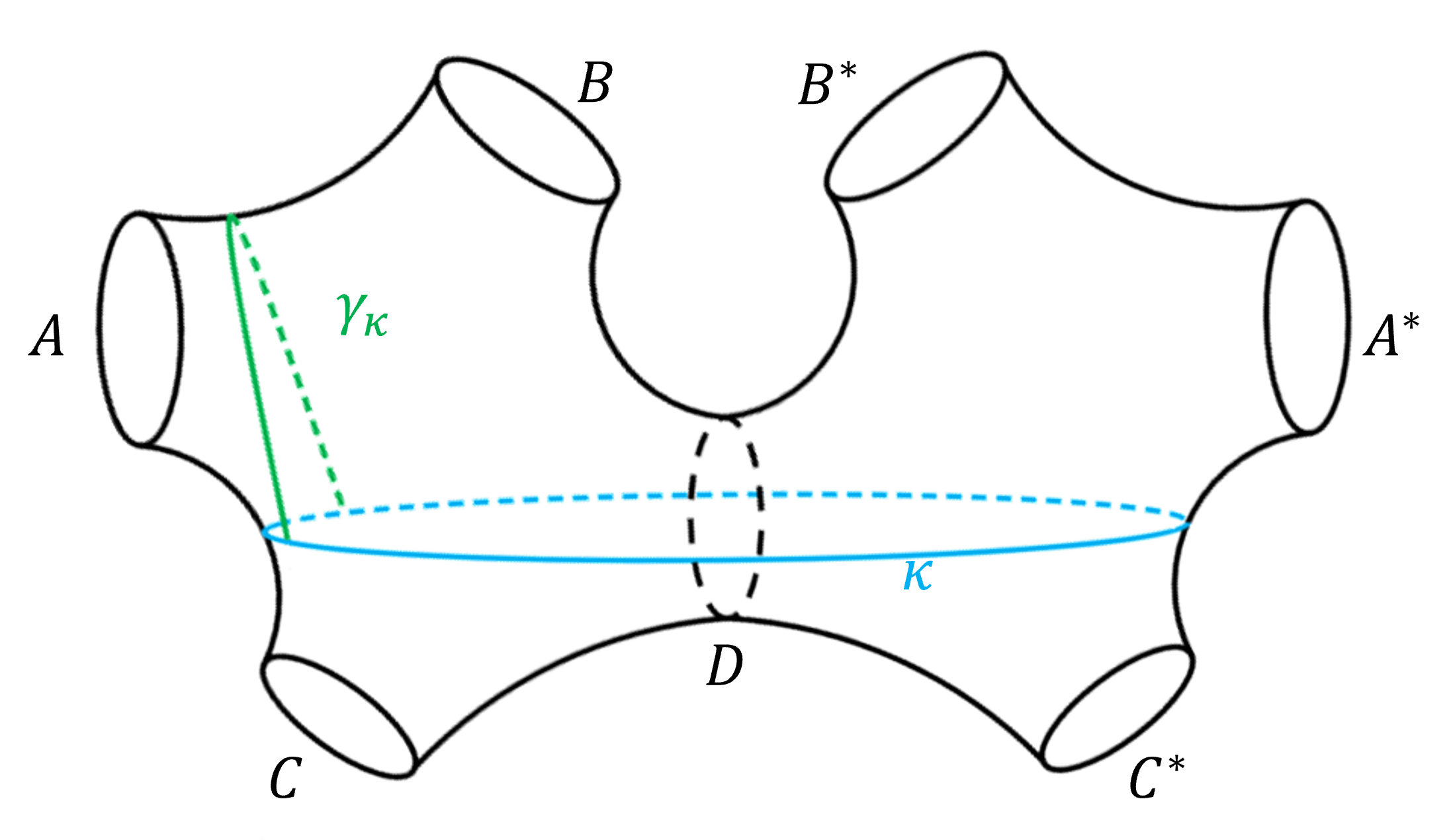}
    \caption{The geometry dual to the first purification, where we have traced over $D$ but not yet $C$. The geodesic $\kappa$ is a possible boundary of $EW(ABA^*B^*)$ in this state.}
    \label{fig:4bdpur}
\end{figure}

\newpage

Now we consider $Q_4$, defined in equation \eqref{eq:q3}, in four-boundary wormholes. We focus on the cases with non-zero $Q_4$, which as argued in Section \ref{i3q3pt}, requires $EW(ABC)$ to be connected. We then have $S_{ABC} = S_D = \frac{1}{4G} \ld$. Hence, the first purification in the calculation of $S_{AA^*A_*A^*_*}$ will be two copies of $EW(ABC)$ joined along the $D$ geodesic. This is depicted in Figure \ref{fig:4bdpur}. The reflected entropy $S_R(A:B)$ depends on $EW(AB)$. There are five cases:

\begin{enumerate}
    \item 
    The $EW(AB)$ is bounded by the $C$ and $D$ geodesics, and thus $S_{AB} = \frac{1}{4G} (\lc + \ld)$. In this case, the geometry involved in computing $S_R(A:B)$ is two copies of $EW(AB)$ joined at the $C$ and $D$ geodesics. Due to entanglement wedge nesting, $EW(ABA^*B^*)$ contains $EW(AB)$. Thus, $EW(ABA^*B^*)$ is bounded by the two $C$ geodesics; the doubled geometry is four copies of the $AB$ entanglement wedge joined at the $C$ and $D$ geodesics. We have $S_{A A^* A_* A^*_*} = 2 S_R(A:B)$, and hence $Q_4=0$. 

    \item The $EW(AB)$ is bounded by $\gamma$, so $S_{AB} = \frac{1}{4G} \lgm$. Then the geometry involved in computing $S_R(A:B)$ is two copies of $EW(AB)$ joined at $\gamma$; hence, we find that $S_R(A:B) = \frac{1}{2G} \min ( \lb, \, \lgg )$, where $\gamma_\gamma$ is shown in Figure \ref{fig:4bdy}. 
    If $\lb < \lgg$, the minimal surface for $S_{AA^*A_*A^*_*}$ is given by four copies of $\lb$, and thus $S_{AA^*A_*A^*_*} = 2S_R(A:B)$ and $Q_4$ vanishes.
    Hence, for non-zero $Q_4$, we must have $S_R(A:B) = \frac{1}{2G} \lgg$.

    Now we consider the second purification, which yields three possibilities:
    \begin{enumerate}[(a)]
        \item The $EW(ABA^*B^*)$ is bounded by two copies of $\gamma$; thus, $S_{A A^* B B^*} = \frac{1}{2G} \lgm$. Then the double geometry is given as two copies of the $S_R(A:B)$ geometry, so $S_{A A^* A_* A^*_*} = 4 \lgg = 2 S_R(A:B)$ and $Q_4$ vanishes.
        
        \item The $EW(ABA^*B^*)$ is bounded by two copies of $C$, so $S_{A B A^* B^*} = \frac{1}{2G} \lc$. Then, the doubled geometry is similar to case 1, but $S_R(A:B)$ is different. This case has positive $Q_4$.
        
        \item The $EW(ABA^*B^*)$ is bounded by the internal geodesic $\kappa$, as pictured in Figure \ref{fig:4bdpur}. Then, $S_{A B A^* B^*} = \frac{1}{4G} \lk$. This case has positive $Q_4$.
    \end{enumerate}

  \item The $EW(AB)$ is bounded by the $A$ and $B$ geodesics, $S_{AB} = \frac{1}{4G} (\la + \lb)$. In this case, the $AB$ entanglement wedge is disconnected, so $S_R(A:B) = 0$. In the double purification, there are five possibilities:

  \begin{enumerate}[(a)]
      \item The $EW(ABA^*B^*)$ is bounded by two copies of the $A$ and $B$ geodesics; hence, $S_{A B A^*  B^*}  = \frac{1}{2G} ( \la +  \lb)$. The entanglement wedge is not connected, so $Q_4$ vanishes.
      \item The $EW(ABA^*B^*)$ is bounded by two copies of the $C$ geodesic. Hence, $S_{A B A^*  B^*} = \frac{1}{2G} \lc$ and the doubled geometry is similar to case 1, while $S_R(A:B)$ differs. This case has positive $Q_4$.
      \item The $EW(ABA^*B^*)$ is bounded by the internal geodesic $\kappa$, hence, $S_{A B A^*  B^*} = \frac{1}{4G} \lk$. The doubled geometry is identical to the case 2(c), and $Q_4$ is positive. 
      \item The $EW(ABA^*B^*)$ is bounded by two copies of the $A$ and $\gamma'$ geodesics; hence, $S_{ABA^*B^*} = \frac{1}{2G} (\la + \lgp)$. The entanglement wedge is not connected, so $Q_4$ vanishes.
      \item The $EW(ABA^*B^*)$ is bounded by two copies of the $B$ and $\tilde \gamma$ geodesics; hence, $S_{ABA^*B^*} = \frac{1}{2G} (\lb + \ltg)$. Again, the entanglement wedge is not connected and $Q_4$ vanishes.
  \end{enumerate}

    \item The $EW(AB)$ is disconnected, but has  $S_{AB} = \frac{1}{4G}(\la + \lgp + \ld)$. As $EW(AB)$ is disconnected, $S_R(A:B)=0$. Because $EW(ABA^*B^*)$ cannot overlap with $EW(C)$, the only possibility is $S_{A B A^*  B^*} = \frac{1}{2G}(\la + \lgp)$. Hence, this entanglement wedge is also disconnected and $Q_4$ vanishes.  

    \item The $EW(AB)$ is disconnected, but has $S_{AB} = \frac{1}{4G}(\lb + \ltg + \ld)$. Since $EW(ABA^*B^*)$ cannot overlap with $EW(C)$, the only possibility is $S_{A B A^*  B^*} = \frac{1}{2G}(\lb + \ltg)$. Hence, this entanglement wedge is also disconnected and $Q_4$ vanishes. 
\end{enumerate}
Note that $S_{AB} = \frac{1}{4G} (\lb + \lgp + \lc)$ or $S_{AB} = \frac{1}{4G} (\la + \ltg + \lc)$ are forbidden by the connectedness of $EW(ABC)$. 
From this analysis, we observe that $Q_4$ is non-zero only in cases 2(b), 2(c), 3(b), and 3(c). 

Now we want to show that if $I_3$ vanishes, $Q_4$ also vanishes.
For $I_3$ to vanish, one of the three-party entanglement wedges must be disconnected as shown earlier.
There are three cases:
\begin{enumerate}[\indent (i)]
    \item When $EW(ABD)$ is disconnected, $EW(C)$ extends beyond the $C$ geodesic. Hence, we cannot be in any of the four cases with positive $Q_4$, so $Q_4$ vanishes.
    \item When $EW(BCD)$ is disconnected, $EW(A)$ extends beyond the $A$ geodesic. Then we cannot be in case 3. Since $\lb <\lgg$, in the case 2 scenario, we would necessarily have $S_R(A:B) =\frac{2}{4G} \lb$, so  $Q_4$ vanishes.
    \item When $EW(ACD)$ is disconnected, $EW(B)$ extends beyond the $B$ geodesic. Similarly to (ii), $Q_4$ vanishes.
\end{enumerate}

We now consider the large $\ell$ limit. We first note that 2(b) never arises at large $\ell$. To see this, note that to be in case 2(b), we must have both
\begin{equation} \label{2bcond}
S_{AB} = \frac{1}{4G} \lgm \quad \Rightarrow\quad \lgm < \lc + \ld,     
\end{equation}
and 
\begin{equation}
    S_{ABA^*B^*} = \frac{1}{2G} \lc \quad \Rightarrow\quad \lc < \lgm, \quad 2 \lc < \lk, 
\end{equation}
but these lead to an inconsistency, as follows. 
Similar to $\lgp$ in Section \ref{I34b}, we find the approximate length of the $\kappa$ geodesic at large $\ell$  to be
\begin{equation}
\label{lkapprox}
    \lk \approx \lc + \lgm - \ld.
\end{equation}
Thus, $ 2 \lc < \lk \, \Rightarrow \, \lc + \ld < \lgm$, which contradicts equation \eqref{2bcond}. Hence, 2(b) cannot arise at large $\ell$. 

Now consider case 2(c). For non-zero $Q_4$, we must have $S_R(A:B) = \frac{1}{2G} \lgg$. In the double geometry, $S_{A A^* A_* A^*_*}$ is given by the minimal length geodesic which separates $A A^* A_* A^*_*$ from $B B^* B_* B^*_*$. Because of the symmetry in the double geometry, this geodesic consists of four copies of a geodesic segment in the four-boundary wormhole. 
There are four cases for this geodesic segment:
\begin{enumerate}[\indent (i)]
    \item the $B$ geodesic,
    \item $\gamma_{D}$ with both ends on the $D$ geodesic as in Figure \ref{fig:4bdy},
    \item $\gamma_{\kappa}$ with both ends on the $\kappa$ geodesic as in Figure \ref{fig:4bdpur},
    \item $\gamma_{\kappa D}$ with one end on $\kappa$ and the other end on the $D$ geodesic.
\end{enumerate}

We can bound the length of the segments $\gamma_{\kappa}$ and $\gamma_{\kappa D}$ by the length of segments ending on the $C$ geodesic. As $\gamma_{\kappa}$ is the shortest segment with both ends on $\kappa$, the portion of a segment $\gamma_{C}$ above $\kappa$ cannot be shorter than $\gamma_{\kappa}$; i.e. $\lgk < \lgc$ and similarly, $\lgkd < \lgcd$. Putting them into $Q_4$, we find
\begin{equation}
    Q_4 = \frac{1}{G} \left[ \min \left( \lb, \, \lgk, \, \lgkd, \, \lgd \right) - \lgg \right] \leq \frac{1}{G} \left[ \min \left( \lgc, \,  \lgcd, \, \lgd \right) - \lgg \right]. 
\end{equation}
We have $\lgm < \lc + \ld$, so the $CD\gamma$ pair-of-pants is not a $\gamma$-eyeglass. Each endpoint of $\gamma_\gamma$ on $\gamma$ is close to either $C$ or $D$, so one of $\gamma_{C}$, $\gamma_{CD}$ or $\gamma_{D}$ is no longer than the segment formed by extending $\gamma_\gamma$ across the strip. This adds at most an order one length, so at large $\ell$, $\min (\lgc, \, \lgcd, \, \lgd) \approx \lgg$, and $Q_4 \approx 0$. In fact, for most points on $\gamma$, its distance from the nearest point on either $C$ or $D$ is exponentially small. Hence, $Q_4$ will be exponentially small for generic values of the moduli.

In case 3, $\la+\lb < \lgm$, so $AB \gamma$ is a $\gamma$-eyeglass. In this case, $S_R(A:B) = 0$, so $Q_4 = S_{AA^*A_*A^*_*}$, and our aim is to show that there is a minimal geodesic of small length in the doubled geometry that separates the copies of $A$ from the copies of $B$. 

For case 3(b), we require 
\begin{equation}
    S_{ABA^*B^*} = \frac{1}{2G} \lc \quad \Rightarrow 2 \lc < \lgk \Rightarrow \lc + \ld < \lgm,
\end{equation}
so $CD\gamma$ is also a $\gamma$-eyeglass.
It follows that either $A$ is close to $C$ and $B$ is close to $D$ or vice-versa. The connectedness of the three-party entanglement wedge implies that we have no $A,B,C,D$-eyeglasses. If $A$ is close to $C$, both $AC\gamma'$ and $BD\gamma'$ are pinwheels. Similarly, if $A$ is close to $D$, both $AD\tilde \gamma$ and $BC \tilde \gamma$ are pinwheels. 

Consider the former case with $\lgp$ being large. The argument in Section \ref{4bd} tells us that we cannot have a decomposition into two $\gamma$-eyeglasses. However, case 3(b) requires that $AB\gamma$ and $CD\gamma$ to be $\gamma$-eyeglasses.
Hence, we have a contradiction and $\lgp$ is $O(1)$. Note that this implies that case 3(b) only arises in a small region of the moduli space of four-boundary wormholes, where an internal geodesic is pinching. Similarly, $\ltg$ is small in the other case. In the doubled geometry, four copies of $\lgp$ or $\ltg$ separate the copies of $A$ from the copies of $B$. Thus we find that
\begin{equation}
    Q_4 = S_{AA^*A_*A^*_*}\leq  \frac{1}{G} \min \left( \lgp, \, \ltg \right) \approx 0. 
\end{equation}

For case 3(c), we have at large $\ell$
\begin{equation}
    S_{ABA^*B^*} = \frac{1}{2G} \lc \quad \Rightarrow\quad \lgk < 2 \lc \quad \Rightarrow \quad \lgm < \lc + \ld \, ,
\end{equation}
where in the last step we have used equation \eqref{lkapprox}. Hence, $CD\gamma$ is a pinwheel. Similarly to the case 2(c), we find
\begin{equation}
    Q_4 = \frac{1}{G} \min \left( \lb,  \, \lgk,  \, \lgkd,  \, \lgd \right) \leq \frac{1}{G} \min \left( \lb,  \, \lgc,  \, \lgcd,  \, \lgd \right) \, ,
\end{equation}
and we extend $\gamma_\gamma$ to construct a candidate for $\gamma_{C}$, $\gamma_{CD}$ or $\gamma_{D}$. Thus,
\begin{equation}
    Q_4 \leq \frac{1}{G} \min \left( \lgc,  \, \lgcd,  \, \lgd \right) \approx  \frac{1}{G} \lgg  \approx 0 .
\end{equation}
At generic moduli, $Q_4$ is exponentially small. 

Thus, in all possible cases, we have shown that $Q_4$ does not grow at large $\ell$, and is exponentially small at generic moduli. 

\section{Discussion}
\label{disc}

We have studied signals of multiparty entanglement. The residual information $R_3 = S_R-I$ and the triple information $I_3$ were previously identified as signals of three- and four-party entanglement \cite{Cui:2018dyq,Akers:2019gcv}. We have proposed a new signal of four-party entanglement, the residual entropy $Q_4$, using constructions based on canonical purification that extend the reflected entropy.   Appendix~\ref{further} presents another such quantity, $D_4$, and explains how to calculate it. We showed that the $n$-informations $I_n$ provide a signal of $n$-party entanglement for even $n$, and generalized the construction of $R_3$ to obtain the $n$-residual informations $R_n$, which provide a signal of $n$-party entanglement for odd $n$. In Appendix~\ref{further}, we also generalized $D_4$ to obtain a signal of $n$-party entanglement for all $n$. One reason it is useful to have several different quantities is that no single quantity is a universal test for multipartiteness: the four-party and higher quantities are not sign-definite, so they vanish on a codimension one subspace of the space of states, whereas the subspace with no $n$-party entanglement should have higher codimension. We would expect that the zero locus of different quantities would be different, so taking their intersection would increase the codimension. These quantities are not multipartite entanglement {\it measures}, because they can vanish for states that have significant multipartite entanglement, and they fail to satisfy some of the axioms normally required for an entanglement measure. But they are {\it signals} of multipartite entanglement: they are non-zero only if the state we are considering has multipartite entanglement. 

We have derived explicit bounds for the three- and four-party quantities in terms of the dimensions of the Hilbert spaces of the subsystems. The bounds on $R_3$ and $-I_3$ are tight, but the lower bound on $Q_4$ probably is not. Similar bounds will apply for the higher-party quantities, although we have not discussed them explicitly. The significance of these bounds is that if we model an $n$-party state as a tensor product of states with no $n$-party entanglement and a component with $n$-party entanglement, the value of a multipartite entanglement signal on the overall state provides a lower bound on the size of the subspace that the component with $n$-party entanglement lives in. 

We saw that $R_3=0$ for generalized GHZ states. The GHZ state also saturates the lower bound on $-I_3$. This provides interesting further evidence that GHZ-like states are not useful in modelling the multiparty entanglement in holographic states. We showed that for three qubits, the states for which $R_3=0$ are precisely the bipartite entangled states and the generalized GHZ state. For larger systems we were not able to identify the full set of states such that $R_3=0$, and even for four qubits we could not identify the states with vanishing four-party signals. This is an interesting direction for future work, including understanding the intersections between the sets of states with $I_3=0$, $Q_4=0$ and $D_4=0$.\footnote{This is difficult to investigate in the holographic context, where we only probe the value of these quantities to leading order in $1/G$, and there are open regions in the moduli space where the different quantities vanish. We found that holographically $Q_4=0$ whenever $I_3=0$, so considering $Q_4$ did not refine our understanding of which states might not have four-party entanglement.} These are several different constraints, as there are different versions of $Q_4$ and $D_4$ depending on which systems we trace out and in what order. It is also interesting to identify the states saturating the bounds on these quantities. 

One central aspect of our work was investigating the behavior of these multiparty entanglement signals in AdS$_3/$CFT$_2$ holography. We showed that $Q_4$ is positive holographically, and noted that the higher-party quantities are not, as observed in \cite{Balasubramanian:2014hda}. We showed that $R_3=0$ to leading order in $1/G$ if $EW(AB) = EW(A) \cup EW(B)$, suggesting a simple geometric picture of the significance of three-party entanglement; we need three-party entanglement to have a part of $EW(ABC)$ which is not in the entanglement wedge of one of the components. We similarly saw that $I_3$ and $Q_4$ vanish if the entanglement wedge for three out of the four parties is disconnected. 

Considering the vacuum of the CFT and dividing the boundary into $n$ regions, we saw that the multiparty quantities were of order one in units of $1/G$. In the dual CFT, this corresponds to values of order the central charge. For four or more parties, the entanglement signals are functions of the size of the regions; they go to zero as one or more regions shrink, and are maximized at some point in the interior of the parameter space. We see that the CFT vacuum state involves multiparty entanglement: for a generic division of the space into subregions, it has $n$-party entanglement of order the central charge for all values of $n$. The results are consistent with the Markov gap bound of \cite{Hayden:2021gno}, but for higher $n$ we find that the $n$-party entanglement signals can be arbitrarily small for special choices of the parameters, so there is no analogous gap. 

\begin{figure}
    \centering
    \includegraphics[scale=0.8]{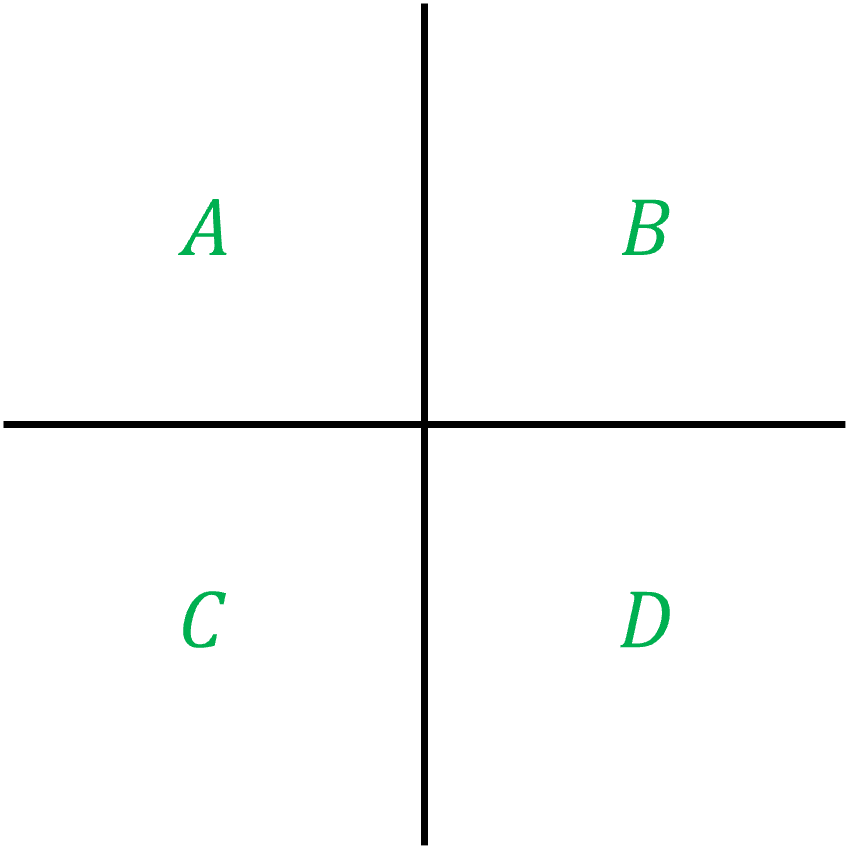}
    \caption{In a CFT$_3$, we can consider dividing the plane into four quadrants.}
    \label{fig:quad}
\end{figure}

The multiparty signals were finite in all of our AdS$_3/$CFT$_2$ examples, but this is not universal. Consider for instance AdS$_4$/CFT$_3$ in the vacuum state for the CFT on the plane, and take the regions $A, B, C, D$ to be quadrants in the plane, as shown in Figure \ref{fig:quad}. Then the entanglement entropies have a leading divergence associated with the length of the boundary, but also a subleading divergence coming from the cusp where the quadrants meet \cite{Hirata:2006jx}, so e.g. 
\begin{equation}
    S_A = \frac{1}{4G} \left( \frac{L_A}{\epsilon} + f \left( \frac{\pi}{2} \right) \ln \epsilon + \mbox{finite} \right),
\end{equation}
where $L_A$ is the length of the boundary of region $A$, and $f(\Omega)$ is a function of the opening angle at the cusp, where we are considering the special case where $\Omega = \frac{\pi}{2}$, taking the regions to be equal quadrants. The boundary of $AB$ is a smooth line, so $S_{AB}$ has the leading linear divergence but no $\ln \epsilon$ term, and similarly for $S_{AC}$, while $S_{BC} = S_B + S_C$, so 
\begin{equation}
I_3 = S_A + S_B + S_C + S_D - S_{AB} - S_{AC} - S_{BC} = S_A + S_D - S_{AB} - S_{AC}. 
\end{equation}
The leading linear divergence will cancel, but we will be left with a log divergence, so 
\begin{equation}
I_3 = \frac{1}{2G} f \left( \frac{\pi}{2} \right) \ln \epsilon + \mbox{finite},
\end{equation}
indicating a divergent four-party entanglement contribution associated with the corner.\footnote{A similar divergence was discussed in a Fermi liquid system in \cite{Tam:2022tpy}.}

We considered the multiparty entanglement signals in the states dual to multiboundary wormhole geometries, focusing on the puncture limit considered in \cite{Balasubramanian:2014hda} and the large $\ell$ limit considered in \cite{Marolf:2015vma}. Here we only explicitly discussed three- and four-boundary wormholes. It would be conceptually straightforward to extend the discussion to include higher numbers of boundaries, but the set of candidate minimal surfaces in the bulk will grow rapidly, making explicit calculations challenging. In the puncture limit, we saw that these states involved significant multiparty entanglement, including cases that saturated the upper bounds on these quantities. This provides a strong motivation for understanding the structure of the states that saturate these bounds, which we leave as an open problem for the future. In particular, there was an open region in the moduli space of four-boundary wormholes where the upper bound on both $-I_3$ and $Q_4$ were saturated. It would be very interesting to find states that saturate both bounds. 

The structure we found in our studies of the three- and four-boundary wormholes at large $\ell$ is consistent with the analysis of \cite{Marolf:2015vma}. We saw that at large $\ell$ $R_3$ does not grow with $\ell$, but it is lower bounded by the argument of \cite{Hayden:2021gno}, so the amount of tripartite entanglement must be at least of order one in units of $1/G$ to account for the value of $R_3$. The result for $R_3$ does not exclude having a larger amount of tripartite entanglement, as there are tripartite entangled states for which $R_3=0$, but the picture of \cite{Marolf:2015vma} suggests that the amount of tripartite entanglement is indeed order one. The four-party measures $-I_3$ and $Q_4$ were also found not to grow with $\ell$. There are special choices of moduli where they are of order one, but we argued that for generic moduli they would be small, consistent with the picture of \cite{Marolf:2015vma}. This analysis could be extended to consider higher-party entanglement measures like $I_n$, and we conjecture the results would be similar. 

We have obtained a range of explicit results on the multiparty entanglement signals for holographic states, notably those dual to the multiboundary wormholes. In the future, we hope to understand what these results imply for the structure of the states. This will involve further understanding the behavior of these signals in simple model states. Perfect tensor states saturate the upper bound for $-I_3$, and this has been used in the past to argue that perfect tensors are useful models in discussing the entanglement structure of holographic states. However, perfect tensor states do not saturate the upper bound on $Q_4$, while some of the holographic states we considered do. Networks of stabilizer states have been used as models to study multiparty entanglement in \cite{Nezami:2016zni}, but tripartite stabilizer states are up to local unitaries composed of bipartite and GHZ states, so they will always have $R_3=0$. A tensor network model for the holographic states we have considered here would therefore need to involve some more general type of tensor. 

\section*{Acknowledgements}

V.B., M.J.K., and C.M. are supported in part by the DOE through grant DE-SC0013528 and QuantISED grant DE-SC0020360. V.B. was also supported by the  Eastman Professorship at Balliol College, Oxford University. S.F.R. is supported in part by STFC under grant ST/T000708/1.
V.B., M.J.K., and C.M. thank the Aspen Center for Physics (ACP) for their hospitality during this work.  The ACP is supported by NSF under grant PHY-2210452.  M.J.K.'s work at the ACP was additionally supported by a grant from the Simons Foundation (1161654, Troyer).
\appendix

\section{Further purification based quantities}
\label{further}

In this appendix, we construct another quantity that provides a useful signal of four-party entanglement using repeated canonical purifications. Unlike $Q_4$, this construction generalizes to higher numbers of parties. 

Consider a four-party state $|\psi\rangle_{ABCD}$, and focus on $AB$. We can consider as before the reduced density matrix $\rho_{AB}$ and its mutual information  $I(A:B)$, and the canonical purification $|\sqrt{\rho_{AB}} \rangle$ and the reflected entropy $S_R(A:B)$. However, we also have the reduced density matrices obtained by tracing out just one of the other parties, $\rho_{ABC}, \rho_{ABD}$, and their corresponding canonical purifications, $|\sqrt{\rho_{ABC}} \rangle_{ABCA^*B^*C^*} $, $|\sqrt{\rho_{ABD}} \rangle_{ABDA^*B^*D^*}$. In the context of these canonical purifications, the mutual informations $I(AA^*:BB^*)$ are sensitive to correlations between $A$ and $B$. Let's define $I_D(A:B) = I_{ABCA^*B^*C^*}(AA^*:BB^*)$ and $I_C(A:B) = I_{ABDA^*B^*D^*}(AA^*:BB^*)$. We thus have four different quantities that reflect correlations between $A$ and $B$. 

We define
\begin{equation}
D_4(A:B) = I_C (A:B) + I_D(A:B) - 2 S_R(A:B) - 2 I(A:B) . 
\end{equation}
This quantity is insensitive to all bipartite and tripartite entanglement between $A, \, B, \, C, \, D$. To see this, note that each mutual information is insensitive to entanglement that does not involve both $A$, $B$, so we need only consider bipartite entanglement between $A, B$, tripartite entanglement between $A, B, C$, and tripartite entanglement between $A, B, D$. The bipartite entanglement can be seen as a subcase of the tripartite, so we only need to show that $D_4$ vanishes for the tripartite entangled states.

For purely tripartite entanglement between $A, B, C$, we have 
\begin{equation}
|\psi_2\rangle_{ABCD} = |\psi \rangle_{ABC} \times |\chi\rangle_D . 
\end{equation} 
The reduced density matrix is $\rho_{AB} = Tr_C (  |\psi \rangle_{ABC} \langle \psi|_{ABC})$, which has some mutual information $I(A:B)$. The purification of this density matrix gives some state $|\sqrt{\rho_{AB}} \rangle_{ABA^*B^*}$, with reflected entropy $S_R(A:B)$. In $D_4$, the first partial trace is 
\begin{equation}
\rho_{ABC} = |\psi \rangle \langle \psi |, \quad  |\sqrt{\rho_{ABC}} \rangle = |\psi \rangle |\psi \rangle, \end{equation} 
so the reduced density matrix will be $\rho_{ABA^*B^*} = \rho_{AB} \rho_{A^* B^*}$, and $I_C(A:B) = 2 I(A:B)$. The second partial trace is 
\begin{equation}
\rho_{ABD} = \rho_{AB} \times |\chi \rangle \langle \chi |, \quad  |\sqrt{\rho_{ABD}} \rangle = |\sqrt{\rho_{AB}} \rangle \times |\chi \rangle \times |\chi \rangle, \end{equation} 
so the reduced density matrix will be $\rho_{ABA^*B^*} =  |\sqrt{\rho_{AB}}\rangle \langle \sqrt{\rho_{AB}}|$, so $I_D(A:B) = 2 S_R(A:B)$. Hence  $D_4 = 0$ as desired; this combination is insensitive to this tripartite entanglement. The argument for tripartite entanglement between $A, B, D$ goes through similarly interchanging $C, D$. 

A general state with only bipartite and tripartite entanglement is some tensor product of states with purely bipartite or tripartite entanglement on some subset of the parties. Since the quantities above are all additive under tensor product, it follows that $D_4=0$ for the general state with only bipartite and tripartite entanglement as claimed. 

Thus, this is a signal of four-party entanglement. We have not been able to find an argument that it is sign-definite, even holographically. On the four-party GHZ state, $D_4 = -2\ln 2$. 

To illustrate this construction, consider calculating $D_4$ in the division of the boundary of global AdS into four regions considered in Section \ref{vacads}. We have $I(A:B) = S_A + S_B - S_{AB} = \frac{1}{4G} (\la + \lb - \ell_{AB})$, and $S_R(A:B) = \frac{1}{4G} 2 \ell_{\gamma_1}$, where $\gamma_1$ is pictured in Figure \ref{fig:4vacq4}. For $I_D(A:B)$, we consider the doubled geometry dual to the canonical purification of $\rho_{ABC}$. In this doubled geometry, $AA^*$ and $BB^*$ form single boundary regions, so $S_{AA^*} = \frac{1}{4G} 2\ell_{\gamma_3}$ and $S_{AA^*BB^*} = \frac{1}{4G} 2 \ell_\kappa$, while $B$ and $B^*$ are disjoint regions, so there is a connected and a disconnected candidate minimal surface,
\begin{equation}
    S_{BB^*} = \frac{1}{4G} \min \left( 2 \lb,  \, 2 \ell_{\gamma_3} + 2\ell_{\kappa} \right), 
\end{equation}
where $\gamma_3$ and $\kappa$ are the geodesic segments pictured in Figure \ref{fig:4vacq4}. Thus
\begin{equation}
    I_D(A:B) = S_{AA^*} + S_{BB^*} - S_{AA^* BB^*} = \frac{1}{4G} \left[ 2\ell_{\gamma_3} + \min \left( 2 \lb, \, 2 \ell_{\gamma_3} + 2\ell_{\kappa} \right) - 2 \ell_\kappa \right]. 
\end{equation}
For $I_C$, there is a similar calculation involving geodesic segments ending on the $C$ geodesic. Let's call the geodesic from the point between $A$ and $D$ to the $C$ geodesic $\tilde \kappa$ and the one from the point between $A$ and $B$ to the $C$ geodesic $\tilde \gamma_3$. Then 
\begin{equation}
    I_C(A:B) = \frac{1}{4G} \left[ 2\ell_{\tilde \gamma_3} + \min \left( 2 \la,  \, 2 \ell_{\tilde \gamma_3} + 2\ell_{\tilde \kappa} \right) - 2 \ell_{\tilde \kappa} \right]. 
\end{equation}

\begin{figure}[H]
    \centering
    \includegraphics[scale=0.6]{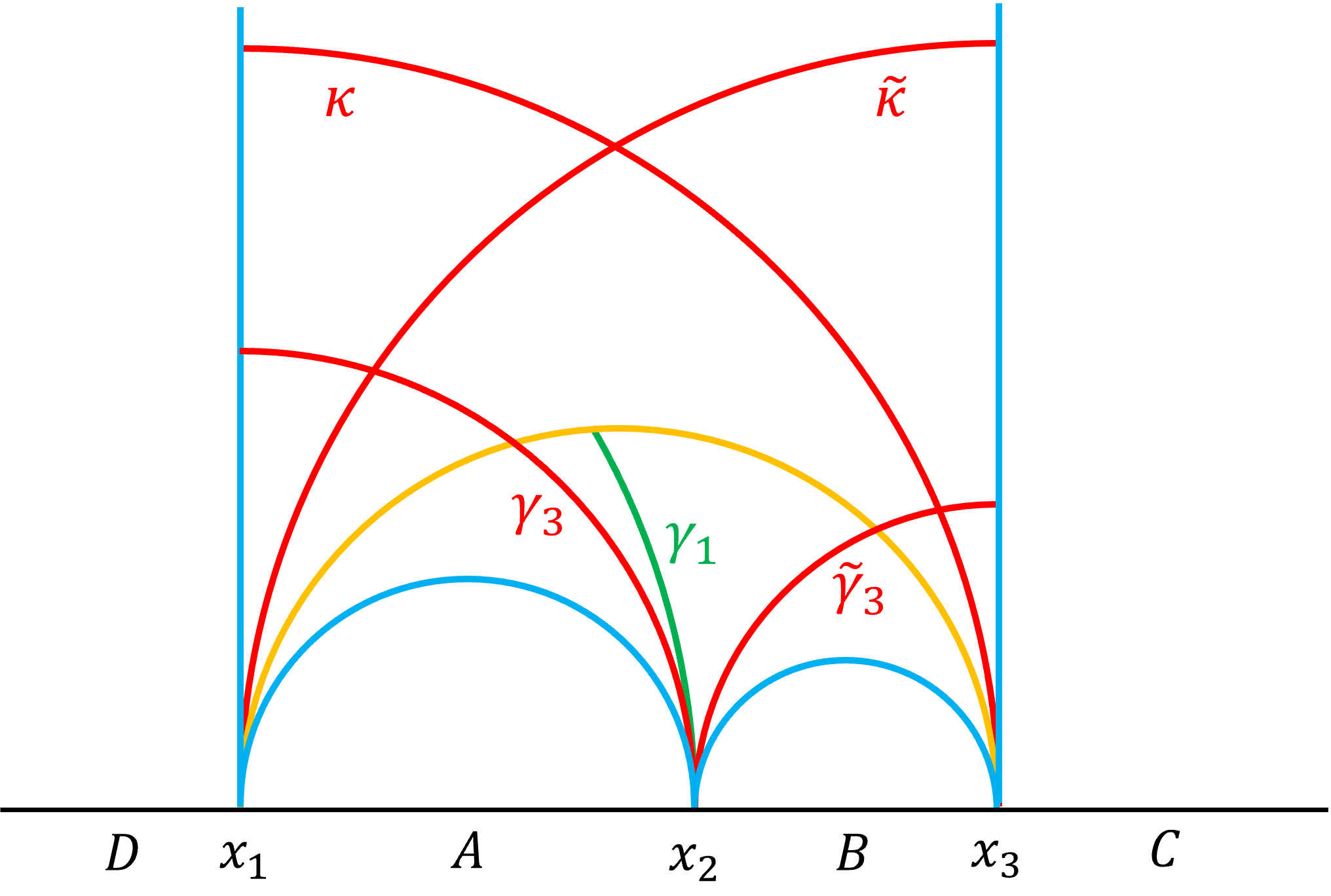}
    \caption{The upper-half plane picture in which we take $x_4 \to \infty$, showing the geodesics involved in the calculation of $D_4$.}
    \label{fig:4vacd4}
\end{figure}

\begin{figure}[H]
    \centering
    \includegraphics[scale=0.77]{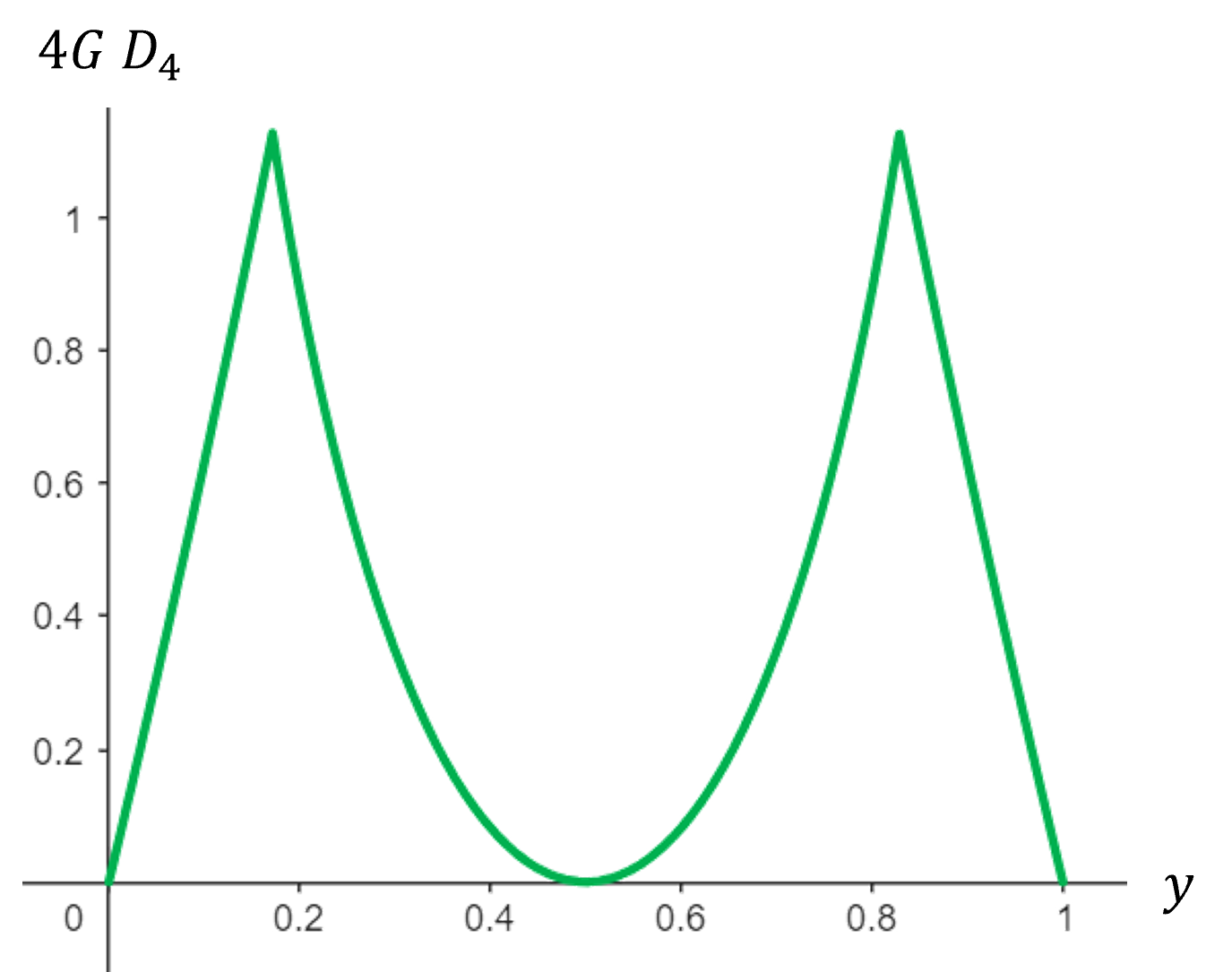}
    \caption{A plot of $D_4$ for four subregions in global AdS.}
    \label{fig:d4plot}
\end{figure}

To calculate the geodesic lengths, we consider the upper-half plane picture obtained by sending the point $x_4$ between $C$ and $D$ to infinity as shown in Figure \ref{fig:4vacd4}. We also set $x_1= 0$ and $x_3=1$, so $x_2=y \in (0,1)$. In this frame, the lengths are
\begin{equation}
\begin{aligned}
    \la = 2 \ln \frac{y}{\epsilon},
    \qquad
    \lb &= 2 \ln \frac{1-y}{\epsilon},
    \qquad
    \ell_{AB} = 2 \ln \frac{1}{\epsilon} \\
    \ell_{\gamma_3} = \ln \frac{2y}{\epsilon} , 
    \qquad
    \ell_{ \tilde{\gamma}_3} =& \ln \frac{2(1-y)}{\epsilon}, 
    \qquad
    \ell_{\kappa} = \ell_{\tilde{\kappa}} = \ln \frac{2y}{\epsilon}
    \\
    \ell_{\gamma_1} &= \ln  \frac{2y(1-y)}{\epsilon} .
\end{aligned}
\end{equation}
Thus, we get
\begin{equation}
    D_4 = - \frac{1}{4G} \left[ 3 \ln y(1-y) + \left| \ln \frac{4 y}{(1-y)^2} \right| + \left| \ln \frac{4 (1-y)^2}{y} \right| \right]. 
\end{equation}
This is plotted in Figure \ref{fig:d4plot}. The novelty here relative to $I_3$ and $Q_4$ is that $D_4$ involves two minimizations. As a result, there are two maxima, at
\begin{align}
    y = 3 - 2\sqrt{2}\quad\text{and}\quad y = 2 \sqrt{2} -2 ,
\end{align}
and we find $D_4=0$ at the symmetric point where $y = \frac{1}{2}$.

\subsection{Generalization to more parties}

If we write $S_R(A:B) = I_{CD}(A:B)$, the mutual information in the canonical purification where we have traced over $C$ and $D$, then we can rewrite 
\begin{equation}
    D_4(A:B) = I_C(A:B) + I_D(A:B) - I_{CD}(A:B) - 2I(A:B). 
\end{equation}
If $D$ is disentangled, the first and third and second and fourth terms cancel, while if $C$ is disentangled it's the first and fourth and second and third terms that cancel. This suggests a direct generalization: on a pure state on $ABC_1 \ldots C_n$,  
\begin{equation}
    D_{n+2}(A:B) = \sum_{s \subset \{ C_1, \dots C_n \} } (-1)^{|s|+1} \, I_{s}(A:B)  - 2I(A:B)\, ,
\end{equation}
where the sum is over all non-empty subsets of $C_1 \ldots C_n$. If $C_i$ is disentangled, the terms involving $C_i$ cancel with the corresponding term omitting $C_i$, with the term with just $C_i$ canceling against the final term. Thus, $D_{n+2}$ is non-zero only for states with genuine $(n+2)$-party entanglement. It is worth noting that $D_3$ matches with the residual information as
\begin{equation}
    D_3(A:B) = 2 S_R(A:B) - 2 I(A:B) = 2 R_3(A:B) .
\end{equation}

\section{Zero set of the residual information}
\label{zeros} 

In Section \ref{simstates}, we saw that there are states with three-party entanglement which nonetheless have $R_3=S_R-I=0$. It would be useful to understand how widespread a phenomenon this is; that is, to characterize the zero set of $R_3$ in $\mathcal H_A \times \mathcal H_B \times \mathcal H_C$, and determine how far it extends beyond bipartite entangled states. We will not be able to do this in general, but we can do it in the simplest case where $A, B, C$ are single qubits. 

First, we observe that the positivity proof in \cite{Dutta:2019gen} proceeded by writing
\begin{equation}
\begin{aligned}
    R_3(A:B) 
    &= S_R(A:B) - I(A:B) \\
    &= S_{AA^*} - S_{A} - S_{B} + S_{AB} \\
    &= S_{AA^*} + S_{AB} - S_{A} - S_{AA^*B},
\end{aligned}  
\end{equation}
where we used that in the canonical purification on $\mathcal H_A \times \mathcal H_B  \times \mathcal H_A^* \times \mathcal H_B^*$, $S_B = S_{B^*} = S_{AA^*B}$. Then $R_3(A:B) \geq 0$ follows from strong subadditivity on $AA^*B$. In \cite{Hayden:2003ss}, the structure of states saturating strong subadditivity was determined. Their results show that $R_3=0$ if and only if there is a decomposition $\mathcal H_A = \oplus_i \mathcal H_A^{L_i} \otimes \mathcal H_A^{R_i}$ such that 
\begin{equation}
    \rho_{AA^*B} = \oplus_i q_i\, \rho_{A^{L_i}A^*} \otimes \rho_{A^{R_i} B}. 
\end{equation}
Thus, the question of which states $|\psi\rangle_{ABC}$ have $R_3=0$ amounts to asking which leads to a reduced density matrix $\rho_{AA^*B}$ of this form. 
This simplifies markedly if we consider the case of three qubits, as the Hilbert spaces are all two-dimensional, so the options for decomposing $\mathcal H_A$ are limited. The density matrices $\rho_{AA^*B}$ of this form are then, up to a choice of basis in $A$:
\begin{equation}
    \rho_{AA^*B} =  p |0\rangle \langle 0|_A \rho_{0B} \rho_{0A^*} \oplus (1-p) |1\rangle \langle 1|_B \rho_{1B} \rho_{1A^*}
    \label{eq:firstdecomp}
\end{equation}
for some density matrices $\rho_{iA^*,B}$, or
\begin{equation}
    \rho_{AA^*B} = \rho_{AA^*} \rho_B,
     \label{eq:seconddecomp}
\end{equation}
or
\begin{equation}
    \rho_{AA^*B} = \rho_{AB} \rho_{A^*}.
         \label{eq:thirddecomp}
\end{equation}

Now the reduced density matrix $\rho_{AA^*B}$ in each case is supposed to come from a pure state $|\psi\rangle_{AA^*BB^*}$ which arose from a canonical purification. Since $B^*$ is a single qubit, this implies that $\rho_{AA^*B}$ has rank at most 2. The state is also invariant under swapping $AB$ with $A^*B^*$ because it arose from a canonical purification. In the first case, this implies, up to a choice of basis in $B$, 
\begin{equation}
    \rho_{AA^*B} =  p |000\rangle \langle 000|_{AA^*B}  \oplus  (1-p) |111\rangle \langle 111|_{AA^*B},
\end{equation}
which arises from the generalized GHZ state $|\psi\rangle_{ABC} = \sqrt{p} |000\rangle + \sqrt{1-p} |111\rangle$. 

In the second case, the maximum rank condition implies either $\rho_{AA^*}$ or $\rho_B$ is pure. The former comes from a canonical purification state of the form $|\psi\rangle_{AA^*BB^*} = |\chi\rangle_{AA^*} \otimes |\phi \rangle_{BB^*}$, which comes from a reduced density matrix $\rho_{AB} = \rho_A \otimes \rho_B$. Again, since $C$ is a single qubit, $\rho_{AB}$ has rank at most two, so one of $\rho_{A,B}$ is pure, and this form of $\rho_{AA^*B}$ comes from states with bipartite entanglement between $AC$ or $BC$. 

If instead $\rho_B$ is pure, $\rho_{AA^*B}$ must come from a canonical purification state of the form $|\psi\rangle_{AA^*BB^*} =  |\chi\rangle_{AA^*B^*} \otimes |\phi \rangle_{B}$, and invariance under interchanging $AB$ and $A^*B^*$ implies this must actually be of the form $|\psi\rangle_{AA^*BB^*} = |\chi\rangle_{AA^*} \otimes |\phi \rangle_{B} \otimes |\phi \rangle_{B^*}$, reducing to a special case of the previous case. 

In the third case, the maximum rank condition similarly implies either $\rho_{AB}$ or $\rho_{A^*}$ is pure. If $\rho_{AB}$ is pure, the canonical purification state is of the form $|\psi\rangle_{AA^*BB^*} = |\psi\rangle_{AB} \otimes |\psi\rangle_{A^*B^*}$, coming from a three-qubit state $|\psi\rangle_{ABC} = |\psi \rangle_{AB} \otimes |\chi \rangle_C$, with just bipartite entanglement between $AB$. The case with $\rho_{A^*}$ is again a special case, similar to the previous case with $\rho_B$ pure. 

Thus, for three qubits, the states with $R_3=0$ consist precisely of the bipartite entangled states and the generalized GHZ states, which can be written up to a choice of basis in $A, B, C$ as 
\begin{equation}
    |\psi\rangle_{ABC} = \sqrt{p} |000\rangle + \sqrt{1-p} |111\rangle. 
\end{equation}

This line of reasoning clearly does not extend easily to higher-dimension Hilbert spaces; we made essential use of the fact that the one-qubit Hilbert space is two-dimensional, so its decompositions are limited. It seems reasonable to conjecture that in higher-dimensional cases there are additional tripartite entangled states which have $R_3=0$ which we have not yet identified. 

\section{Five and six boundary regions in vacuum AdS}
\label{vac56}

Here we continue the discussion of vacuum AdS in Section \ref{vacads}, considering cases with five and six regions. This illustrates the use of the higher multiparty entanglement signals. As in the main text, we find that the multiparty entanglement signals are finite and order one in units of $1/G$. The lack of sign definiteness of the higher multiparty entanglement signals is also seen here; there are regions of the parameters where they are both positive and negative. 

\subsection{Five regions}

\begin{figure}[t]
    \centering
    \includegraphics[scale=0.6]{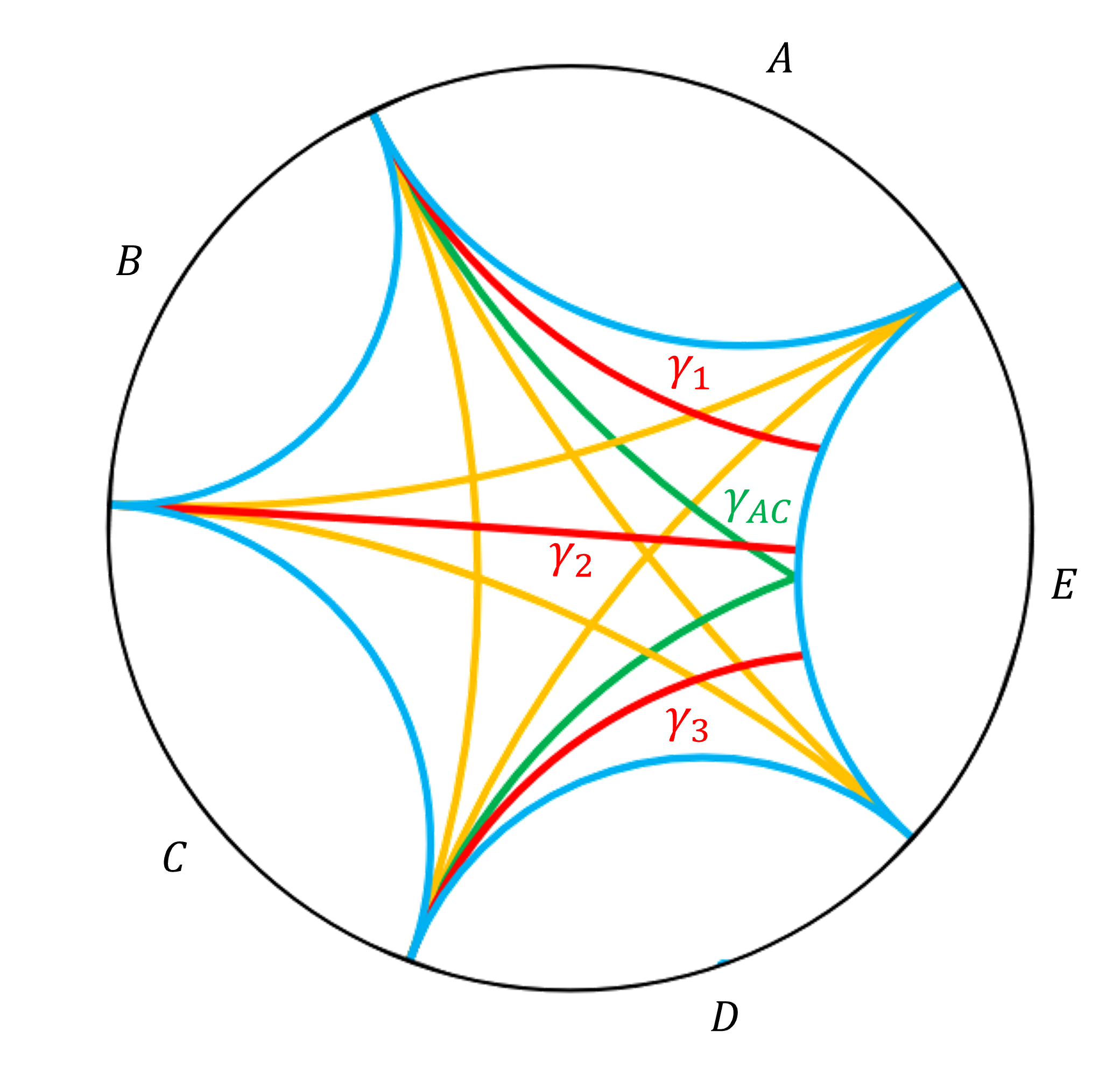}
    \caption{A division into five regions, showing the minimal surfaces relevant to the calculation of $R_5$. }
    \label{fig:5vac}
\end{figure}

Consider the calculation of $R_5$ for the division of the boundary into five regions pictured in Figure \ref{fig:5vac}. Suppose we define $R_5$ by tracing out region $E$. Then we have 
\begin{equation}
    R_5(A:B:C:D) = \frac{1}{2} I_4(AA^*:BB^*:CC^*:DD^*) - I_4(A:B:C:D), 
\end{equation}
where
\begin{equation}
\begin{split}
     I_4(A:B:C:D) = &S_A + S_B + S_C + S_D - (S_{AB} + S_{AC} + S_{AD} + S_{BC} + S_{BD} + S_{CD})\\ & + S_{ABC} + S_{ABD} + S_{ACD} + S_{BCD} - S_{ABCD}, 
\end{split}
\end{equation}
and
\begin{equation}
\begin{split}
    \frac{1}{2} I_4(AA^*:BB^*:CC^*:DD^*) 
    &=  I_3(AA^*:BB^*:CC^*) \\ 
    &= S_{AA^*} + S_{BB^*} + S_{CC^*} + S_{DD^*}  \\ 
    &\qquad - (S_{AA^*BB^*} + S_{AA^*CC^*} + S_{BB^*CC^*} ).
\end{split}
\end{equation}

All of the single-interval entropies are given by the length of the corresponding geodesic, so $S_A = \frac{\ell_A}{4G}$ etc. We also have $S_{ABCD} = S_E = \frac{\ell_E}{4G}$. Two adjacent intervals form a single boundary region, so the entropy for the combined region is given by the corresponding spanning geodesic, e.g. $S_{AB} = \frac{\ell_{AB}}{4G}$, and similarly for three adjacent intervals, e.g. $S_{ABC} = \frac{\ell_{ABC}}{4G}$. When we consider two non-adjacent intervals, there is a disconnected and a connected candidate minimal surface, so e.g. 
\begin{equation}
 S_{AC} = \frac{1}{4G} \min \left( \la + \lc ,  \, \lb + \ell_{ABC} \right).
\end{equation}
Similarly for the sets of three intervals when one is non-adjacent, we have e.g. 
\begin{equation}
 S_{ABD} = \frac{1}{4G} \min \left( \ell_{AB} + \ld ,  \, \lc + \ell_E \right).
\end{equation}
Putting this all together we have
\begin{multline}
    I_4 = \frac{1}{8G} \bigg[ \la+\lb+\lc+\ld -\ell_E - \ell_{AB} - 3 \ell_{BC} - \ell_{CD} + \ell_{ABC} + \ell_{BCD} \\ 
    + \left| \la + \lc - \lb - \ell_{ABC} | + |\la + \ld - \ell_{BC} - \ell_E \right| + \left|\lb + \ld - \ell_{BCD}  - \lc \right| \\ 
    - \left| \la + \ell_{CD} - \lb -\ell_E \right| - \left|\ell_{AB} + \ld - \lc - \ell_E \right| \bigg]. 
\end{multline}

In the calculation of $I_3(AA^*:BB^*:CC^*)$, $AA^*$, $AA^*BB^*$ and $DD^*$ each form a single boundary region in the doubled geometry when we trace over $E$, so the entanglement entropy is given by the length of the spanning geodesic, 
\begin{equation}
 S_{AA^*} = \frac{2 \ell_{\gamma_1}}{4G}, \quad S_{AA^*BB^*} =  \frac{2 \ell_{\gamma_2}}{4G}, \quad 
 S_{AA^*BB^*CC^*} =S_{DD^*} = \frac{2 \ell_{\gamma_3}}{4G}. 
\end{equation}
Other cases form a pair of boundary regions, so there is a connected and a disconnected candidate for the minimal surface, and 
\begin{align}
\begin{aligned}
    S_{BB^*} = & \frac{1}{4G} \min \left( 2 \lb, \, 2\ell_{\gamma_1} + 2 \ell_{\gamma_2} \right), \quad S_{CC^*} = \frac{1}{4G} \min \left( 2 \lc, \, 2\ell_{\gamma_2} + 2 \ell_{\gamma_3} \right),\\
    &\qquad\qquad S_{BB^*CC^*} = \frac{1}{4G} \min \left( 2 \ell_{BC}, \, 2 \ell_{\gamma_1} + 2 \ell_{\gamma_3} \right). 
\end{aligned}
\end{align}
Finally, $AA^*CC^*$ forms three disjoint regions in the doubled geometry ($AA^*$ is a single region and $C$ and $C^*$ are disjoint regions) so there are a connected, a disconnected, and three partially connected possibilities for the minimal surface. Two of the partially connected possibilities, where $AA^*$ is connected to $C$ or to $C^*$, are related by symmetry, so there are four distinct possibilities for the entanglement entropy, 
\begin{equation}
 S_{AA^*CC^*} = \frac{1}{4G} \min \left( 2\lc+2 \ell_{\gamma_1}, \, 2 \lb + 2 \ell_{\gamma_3}, \, 2 \ell_{\gamma_1} + 2 \ell_{\gamma_2} + 2 \ell_{\gamma_3}, \, \lc + \lb + \ell_{\gamma_{AC}} \right). 
\end{equation}
Putting this all together we have 
\begin{multline}
     I_3 = \frac{1}{4G} \bigg[ \lb + \lc + 2 \ell_{\gamma_1} + 2 \ell_{\gamma_3} - \ell_{BC} \\ 
     - 2 \min  \left( 2\lc+2 \ell_{\gamma_1} , \, 2 \lb + 2 \ell_{\gamma_3}, \, 2 \ell_{\gamma_1} + 2 \ell_{\gamma_2} + 2 \ell_{\gamma_3}, \, \lc + \lb + \ell_{\gamma_{AC}} \right) \\ 
     + \left|\ell_{BC} - \ell_{\gamma_1} - \ell_{\gamma_3} \right| - \left| \lb - \ell_{\gamma_1} - \ell_{\gamma_2} \right| - \left| \lc - \ell_{\gamma_2} - \ell_{\gamma_3} \right| \bigg]. 
\end{multline}

\begin{figure}[H]
    \centering
    \includegraphics[scale=0.6]{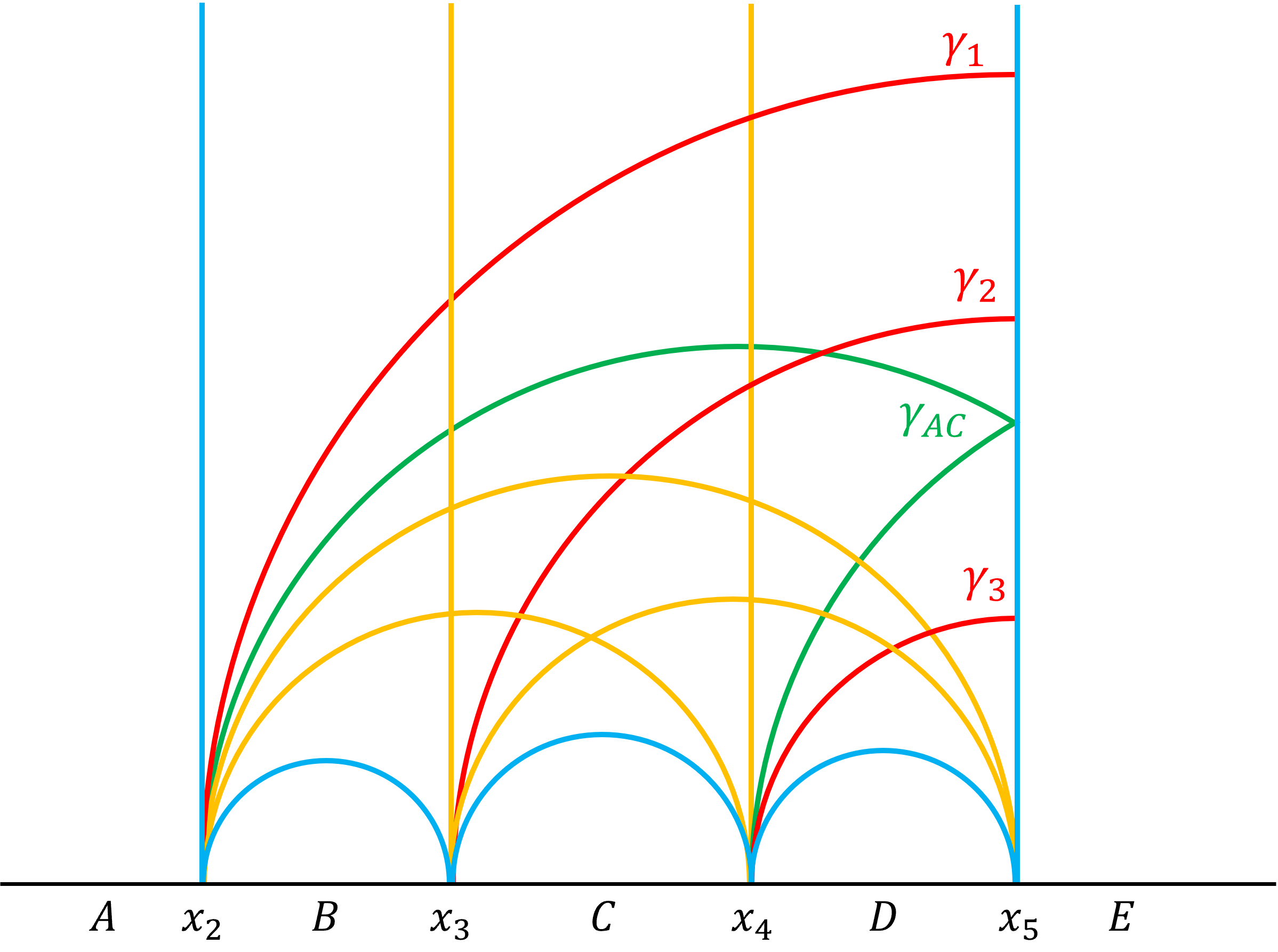}
    \caption{The division into five regions in an upper-half plane picture, where the point $x_1$ between $A$ and $E$ has been sent to infinity.}
    \label{fig:5vacuhp}
\end{figure}
\begin{figure}[H]
    \centering
    \includegraphics[scale=0.72]{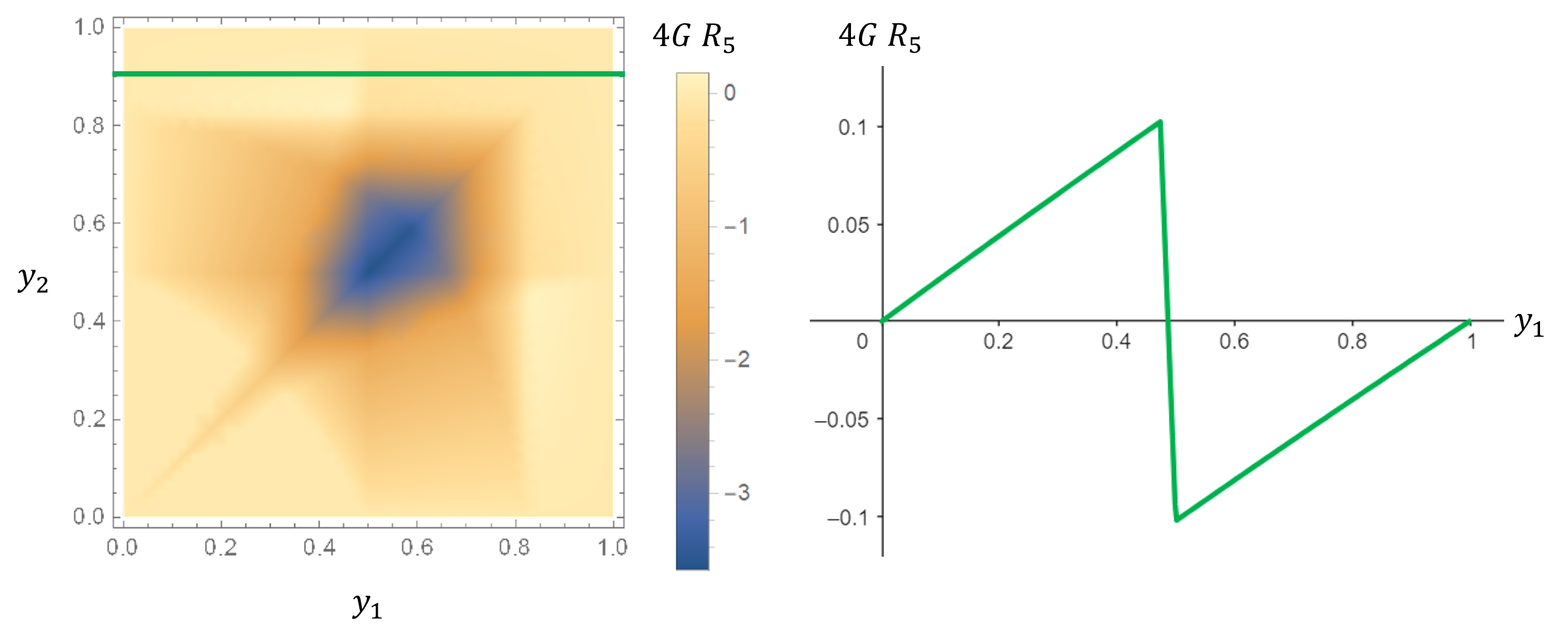}
    \caption{Left: $R_5$ as a function of $y_1$ and $y_2$. Right: $R_5$ as a function of $y_1$ for $y_2=0.9$ shown as a green line on the left. Note that $R_5$ takes both positive and negative values.}
\label{fig:R5plot}
\end{figure}

To calculate the lengths, it's useful to work in the frame where we send the point between $A$ and $E$ to infinity, as shown in Figure \ref{fig:5vacuhp}. This simplifies the form of the geodesics $\gamma_i$ which meet $E$ orthogonally. The only complicated case is $\ell_{\gamma_{AC}}$: passing to the doubled space we find 
\begin{equation}
    \ell_{\gamma_{AC}}= 2 \ln \left( \frac{2 x_5- x_4 -x_2}{\epsilon} \right). 
\end{equation}
As with the previous cases, $R_5$ will only depend on the boundary points modulo conformal transformations, which we can use to set $x_3=0$ and $x_5=1$. The result should be symmetric under the reflection about $x_3$, so it's convenient to parametrise the remaining points as $x_2 = \frac{y_1}{y_1-1}$, $x_4 = y_2$, with $y_1, y_2 \in (0,1)$. We then expect $R_5$ to be a symmetric function of $y_1, y_2$. We finally have 
\begin{multline}
    R_5 = \frac{1}{4G} \Bigg[ \ln(16 y_1 y_2 (1- y_1)(1-y_2)(y_1 + y_2 - y_1 y_2) ) - 2 \min \bigg( \ln(2 y_1^2 (1-y_2)), \\
    \ln(2 y_2^2 (1-y_1)),\, \ln (8(1-y_1)(1-y_2)),\, \ln (y_1 y_2 (2-y_1 -y_2 + y_1 y_2)) \bigg) \\ 
    + \left| \ln \frac{y_1}{1-y_1} \right|+ \left| \ln \frac{y_2}{1-y_2} \right| + \left| \ln \frac{(y_1+y_2-y_1 y_2)^2}{4(1-y_1)(1-y_2)} \right|   - \left| \ln \frac{y_1^2}{4(1-y_1)} \right| \\
    - \left| \ln \frac{y_2^2}{4(1-y_2)} \right| - \left| \ln \frac{y_1}{y_2(1-y_1)} \right| - \left| \ln \frac{y_2}{y_1(1-y_2)} \right| - \left| \ln \frac{(y_1+y_2-y_1 y_2)}{(1-y_1)(1-y_2)}\right|  \Bigg].
\end{multline}
As expected, this is a symmetric function of $y_1, y_2$. This is plotted in Figure \ref{fig:R5plot}. It is mostly negative, with a minimum at $y_1 = y_2 = 1/2$, where $R_5 = - \frac{1}{4G}  \ln 36$. However, there are points where $R_5$ is positive, as we can see in the plot of the cross-section at $y_2 = 0.9$ in Figure \ref{fig:R5plot}.  

\subsection{Six regions}

\begin{figure}
    \centering
    \includegraphics[scale=0.6]{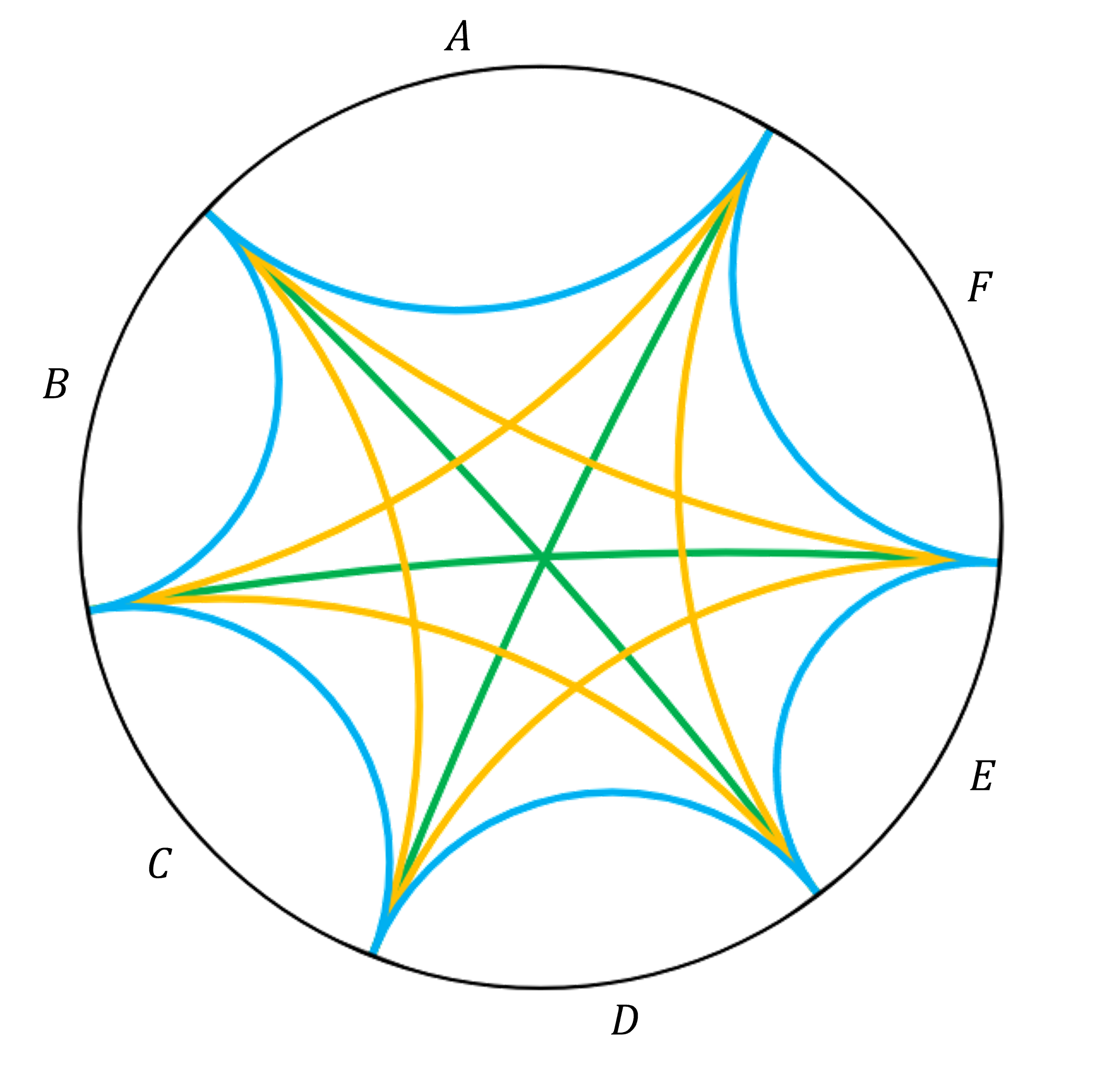}
    \caption{A division into six regions, showing the minimal surfaces relevant to the calculation of $I_5$. The blue geodesics have lengths $\ell_A, \ell_B, \ldots $, the yellow ones have lengths $\ell_{AB}, \ell_{BC}, \ldots$, and the green ones have lengths $\ell_{ABC}, 
    \ell_{BCD}, \ell_{CDE}$.}
    \label{fig:6vac}
\end{figure}

We now consider the calculation of $I_5$ for the division of the boundary into six regions pictured in Figure \ref{fig:6vac}. We have 
\begin{equation}
    I_5 = \frac{1}{2}I_6 = S_1 - S_2 + S_3, 
\end{equation}
where $S_1$ is the sum of all single-interval entropies, $S_2$ is the sum of the entropies of all pairs of regions, and $S_3$ is the sum of the entropies of all sets of three regions. 
All of the single-interval entropies are given by the length of the corresponding geodesic, so e.g. $S_{A} = \frac{\ell_A}{4G}$. Two adjacent intervals form a single boundary region, so the entropy for the combined region is given by the corresponding spanning geodesic, so e.g. $S_{AB} = \frac{\ell_{AB}}{4G}$, and similarly for three adjacent intervals, the entropy is given by the corresponding spanning geodesic, so e.g.  $S_{ABC} = \frac{\ell_{ABC}}{4G}$. For two non-adjacent intervals, there is a disconnected and a connected candidate minimal surface, so e.g. 
\begin{equation}
 S_{AC} = \frac{1}{4G} \min \left( \la + \lc ,  \, \lb + \ell_{ABC} \right), \quad S_{AD} = \frac{1}{4G} \min \left( \la + \ld ,  \, \ell_{BC} + \ell_{EF} \right).
\end{equation}
For three intervals, if two are adjacent we have two boundary regions and there is again a disconnected and a connected surface, so e.g. 
\begin{equation}
 S_{ABD} = \frac{1}{4G} \min \left( \ell_{AB} + \ld ,  \, \lc + \ell_{EF} \right).
\end{equation}
For $S_{ACE}$, we have three disjoint boundary regions, so in addition to the disconnected and connected candidate minimal surfaces, there are three partially connected possibilities, so 
\begin{multline}
    S_{ACE} = \frac{1}{4G} \min \big( \la + \lc + \ell_E ,  \, \lb + \ld + \ell_F,  \, \ell_B + \ell_{ABC} + \ell_E , \\
    \ell_F + \ell_{BCD} + \ell_C , \, \ell_D + \ell_{CDE} + \ell_A \big).
\end{multline} 
Putting this all together we have
\begin{multline}
     I_5 = \, \frac{1}{8G} \Big[ 2 \min \Big( \la + \lc + \ell_E, \, \lb + \ld + \ell_F, \, \ell_B + \ell_{ABC} + \ell_E, \, \ell_F + \ell_{BCD} + \ell_C, \\
     \ell_D + \ell_{CDE} + \ell_A \Big) 
     - \ell_{AB} - \ell_{BC} - \ell_{CD} - \ell_{DE} - \ell_{EF} - \ell_{AF}  \\ 
     + \left| \la + \lc - \lb - \ell_{ABC} \right| + \left| \la + \ld - \ell_{BC} - \ell_{EF} \right| + \left| \la + \ell_E - \ell_{BCD} - \ell_F \right| \\ + \left| \lb + \ld - \lc - \ell_{BCD} \right| + \left| \lb + \ell_E - \ell_{CD} - \ell_{AF} \right| + \left| \lb + \ell_F - \la - \ell_{CDE} \right| \\ 
     + \left| \lc + \ell_E - \ld - \ell_{CDE} \right| + \left| \lc + \ell_F - \ell_{DE} - \ell_{AB} \right| + \left| \ld + \ell_F - \ell_E - \ell_{ABC} \right| \\  
     - \left| \ell_{AB} + \ld - \ell_{EF} - \lc \right| - \left| \ell_{AB} + \ell_E - \ell_{CD} - \ell_F \right| - \left| \ell_{CD} + \la - \ell_{EF} - \lb \right| \\ 
     - \left| \ell_{AF} + \lc - \ell_{DE} - \lb \right|  - \left| \ell_{DE} + \la - \ell_{BC} - \ell_F \right| - \left| \ell_{AF} + \ld - \ell_{BC} - \ell_E \right|   \Big].
\end{multline}

\begin{figure}[H]
    \centering
    \includegraphics[scale=0.6]{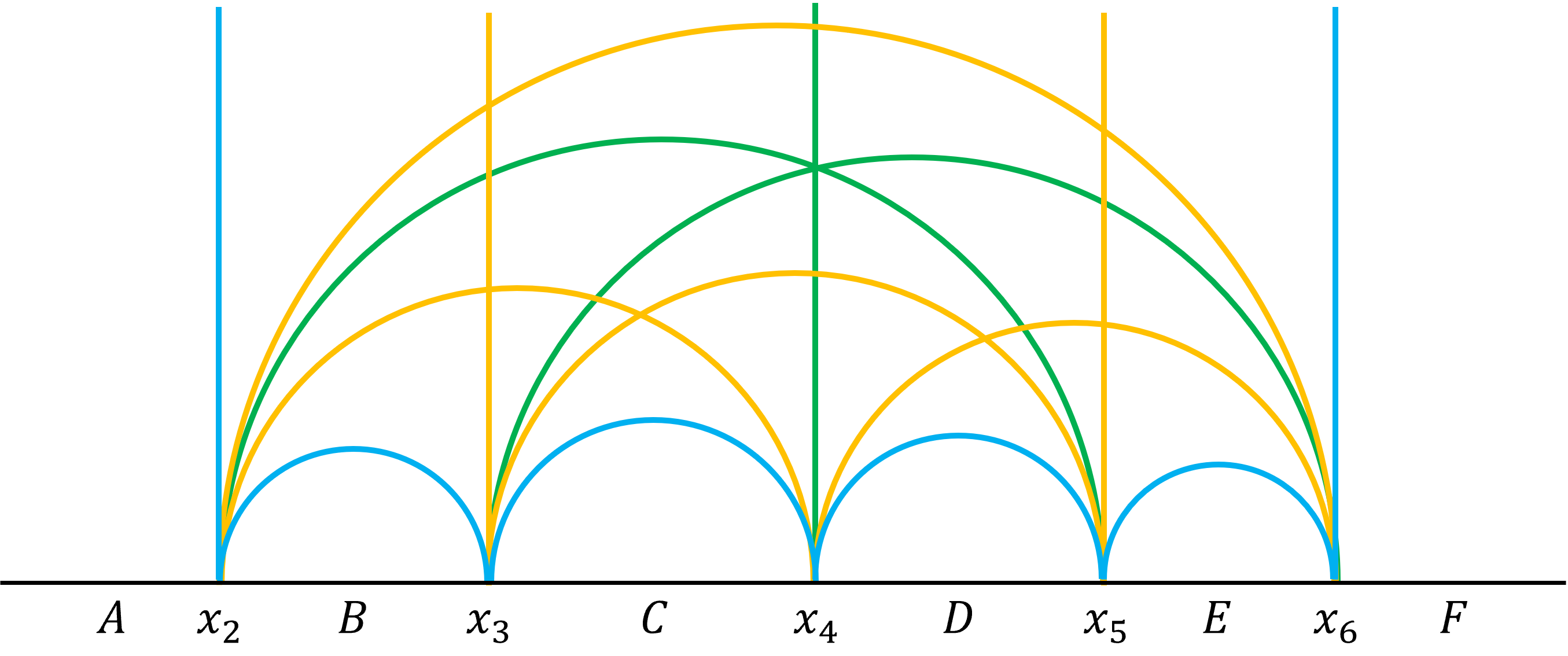}
    \caption{The division into six regions in an upper-half plane picture, where the point $x_1$ between $A$ and $F$ has been sent to infinity.}
    \label{fig:6vacuhp}
\end{figure}

We can calculate the lengths in the frame where we send the point $x_1$ between $A$ and $F$ to infinity, as shown in Figure \ref{fig:6vacuhp}. 
We can simplify further by choosing $x_3=0$, $x_5=1$, and writing $x_2 = 1 - \frac{1}{y_1}$, $x_4=y_2$, and $x_6 = \frac{1}{1-y_3}$ to rewrite in terms of the cross ratios $y_1, y_2, y_3 \in (0,1)$. Then 
\begin{multline}
     I_5 = \,  \frac{1}{4G} \Bigg[  \ln \frac{ (1-y_1)^2  (1-y_2)^2  (1-y_3)^2}{(1-y_1 + y_1 y_2) (1- y_2 + y_2 y_3) (1 - y_3 + y_1 y_3)}    \\
     + 2 \min \bigg( 0, \, \ln  \frac{y_1 y_2 y_3}{(1-y_1)(1-y_2)(1-y_3)}, \, \ln  \frac{y_1}{(1-y_1)(1-y_3)} , \\
     \ln  \frac{y_2}{(1-y_2)(1-y_1)} , \, \ln  \frac{y_3}{(1-y_3)(1-y_2)} \bigg)
     + \left| \ln \frac{y_1 y_2}{1-y_1} \right| + \left| \ln \frac{y_1 y_3}{1-y_3} \right| \\ 
     + \left| \ln \frac{y_2 y_3}{1-y_2} \right| + \left| \ln \frac{y_1}{(1-y_1)(1-y_3)} \right| + \left| \ln \frac{y_2}{(1-y_1)(1-y_2)} \right| \\
      + \left| \ln \frac{y_3}{(1-y_2)(1-y_3)} \right| 
     + \left| \ln \frac{y_1(1-y_2)}{(1-y_1+y_1y_2)} \right| + \left| \ln \frac{y_2(1-y_3)}{(1-y_2+y_2y_3)} \right| \\ 
     + \left| \ln \frac{y_3(1-y_1)}{(1-y_3+y_1y_3)} \right| 
     - \left |\ln \frac{y_1}{1-y_1} \right| - \left |\ln \frac{y_2}{1-y_2} \right| - \left |\ln \frac{y_3}{1-y_3} \right| \\  
     -\left| \ln \frac{y_1(1-y_2+y_2 y_3)}{(1-y_3)(1-y_1+y_1 y_2)} \right|
     - \left| \ln \frac{y_2 (1-y_3 + y_1 y_3)}{(1-y_1) (1- y_2 + y_2 y_3)} \right| \\
     -\left| \ln \frac{y_3(1-y_1+y_1y_2)}{(1-y_2)(1-y_3+y_1y_3)} \right| 
     \Bigg].
\end{multline}

\begin{figure}[H]
    \centering
    \includegraphics[scale=0.72]{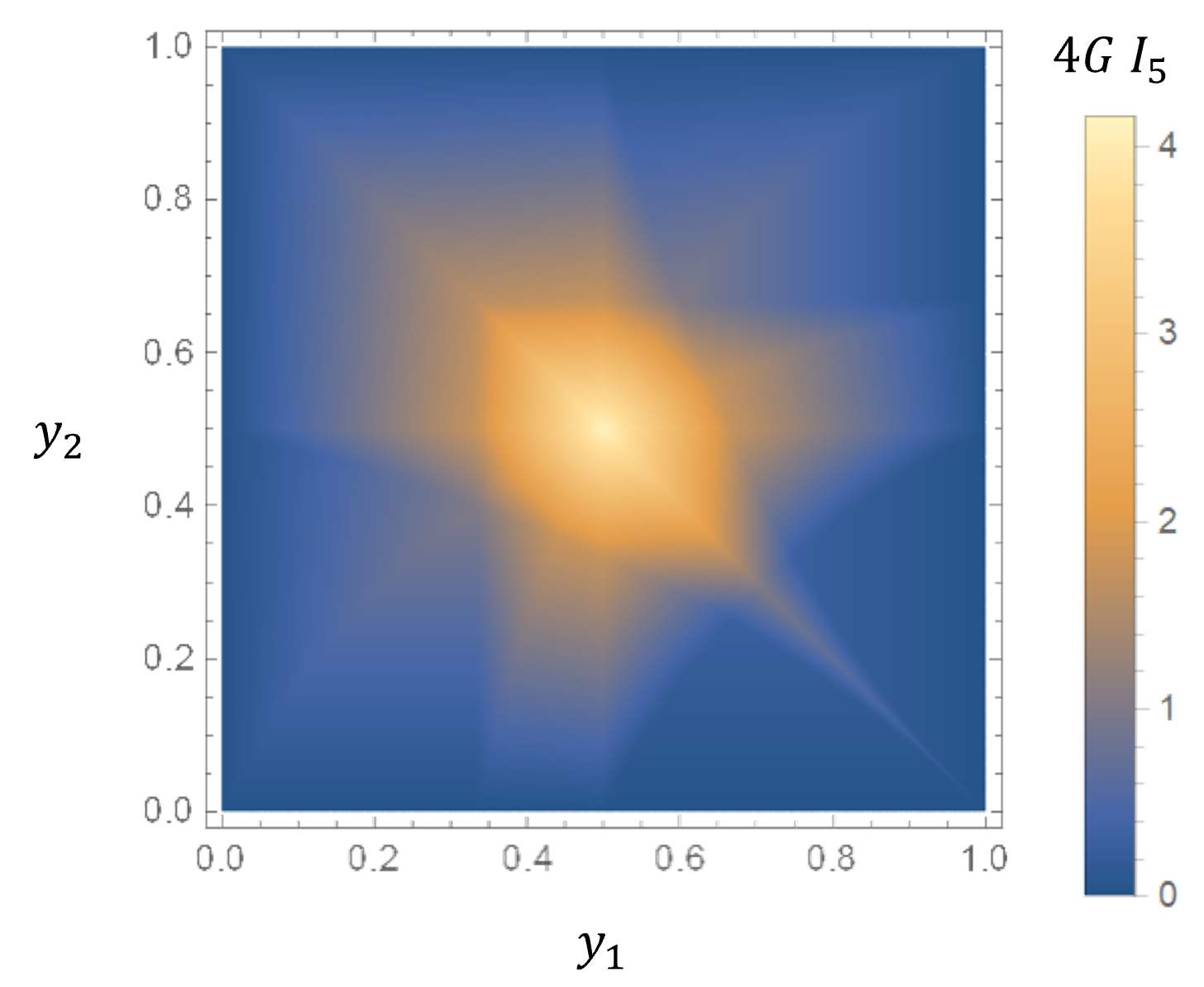}
    \caption{$I_5$ as a function of $y_1$ and $y_2$ for $y_3 = \frac{1}{2}$.}
\label{fig:I5plot}
\end{figure}
This is invariant under cyclic permutation of the $y_i$, as expected. It has a maximum at  $y_1 = y_2 = y_3 = \frac{1}{2}$, where $I_5 = \frac{1}{4G} \ln 64$.  A cross-section at $y_3 = \frac{1}{2}$ is plotted in Figure \ref{fig:I5plot}. We see that it's positive for most values of the $y_i$, but there are values where it's negative, e.g. $y_1 = 0.54, y_2 = 0.91, y_3 = 0.43$ gives $4G I_5 = -0.09$. Moreover, there are open regions where $I_5=0$.

It is interesting to note that while the higher entanglement quantities are not sign-definite, some of them are mostly positive and some are mostly negative. While $R_3$ was always positive, $R_5$ is mostly negative, and while $I_3$ was holographically negative, $I_5$ is mostly positive. It's not clear if this pattern is special to the case we considered here, of the CFT vacuum state, or applies more generally.

\section{Residual information for a four-boundary wormhole at large $\ell$}
\label{sr4}

In Section \ref{sr3}, we showed that $R_3=S_R-I$ does not grow with $\ell$ at large $\ell$ for three-boundary wormholes. We can extend this discussion to the four-boundary wormhole. We will consider $R_3(A:B)$, so we split the wormhole into an $AB\gamma$ and a $CD\gamma$ pair-of-pants. 

There are now several different possibilities for $R_3$. The first question is what is the entanglement wedge of $AB$, then within that the entanglement wedge of $A$, then the entanglement wedge of $B$, and finally the choice of entanglement wedge cross-section if any. 
\begin{enumerate}
\item  If $S_{AB}  = S_A + S_B$, the entanglement wedge is disconnected, and $S_R(A:B) = I(A:B) = 0$. 
\item If $S_{AB} = \frac{1}{4G} \lgm$, $\rho_{AB}$ looks like the three-party case, with $\lgm$ playing the role of $\lc$. 
\begin{enumerate}[\indent (a)]
\item If $S_A = \frac{1}{4G} (\lb + \lgm)$ , $S_R(A:B) = I(A:B) =  \frac{1}{4G} (2 \lb)$, so $R_3(A:B)=0$.
\item If $S_A = \frac{1}{4G} \la$, we have in general two cases for the cross-section: 
\begin{enumerate}[\indent (i)]
\item If $S_R(A:B) = \frac{1}{4G} (2 \lb)$, $R_3(A:B) = \frac{1}{4G} (\lgm + \lb - \la) = I(B:CD)$. 
\item If $S_R(A:B) = \frac{1}{4G} (2\lgg)$, $R_3(A:B) = \frac{1}{4G} (2\lgg + \lgm - \la - \lb)$. 
\end{enumerate}
Of these two cases, only the latter appears generically for large $\ell$. 
\end{enumerate}
Thus as in the three-party case, in case 2, $R_3(A:B)$ does not grow with $\ell$ at large $\ell$.
\item If $S_{AB} = \frac{1}{4G} (\lc + \ld)$, the doubled entanglement wedge is glued together along the $C$, $D$ minimal geodesics, and has non-trivial topology. The entanglement wedge cross-section could be the $B$ geodesic, $\gamma'$, $\tilde \gamma$, $\gamma_{C}$, $\gamma_{CD}$, or $\gamma_{D}$. There are three possibilities for the $A$ entanglement wedge: 
\begin{enumerate}[\indent (a)]
\item If $S_A = \frac{1}{4G} (\lb + \lc + \ld)$, then $E(AB) = E(A) \cup E(B)$, and $R_3(A:B) = 0$ by the general argument. 
\item If $S_A = \frac{1}{4G}(\lgp + \lc)$, there are two possibilities for the $B$ entanglement wedge:
\begin{enumerate}[\indent (i)]
\item If $S_B = \frac{1}{4G}(\lgp + \ld)$, then $E(AB) = E(A) \cup E(B)$, and $R_3(A:B) = 0$ by the general argument. 
\item If $S_B = \frac{1}{4G} \lb$, then $I(A:B) = \frac{1}{4G}(\lb + \lgp - \ld)$.
\end{enumerate}
\item If $S_A =  \frac{1}{4G} \la$, we have the same two possibilities for  the $B$ entanglement wedge:
\begin{enumerate}[\indent (i)]
\item If $S_B = \frac{1}{4G}(\lgp + \ld)$ then $I(A:B) =\frac{1}{4G}  (\la + \lgp - \lc)$. 
\item If $S_B = \frac{1}{4G} \lb$, then $I(A:B) = \frac{1}{4G}(\la + \lb - \lc - \ld)$.
\end{enumerate}
\end{enumerate}
\end{enumerate}

We now want to consider large $\ell$ and ask if there are any cases in 3(b) or 3(c) that could lead to large $R_3(A:B)$. At large $\ell$, we have  
\begin{equation}
2\lgc \approx \la + \lgp - \lc, \quad 2\lgd \approx \lb + \lgp - \ld.
\end{equation}
Thus, in case 3(b)(ii), $S_R(A:B) \leq \frac{1}{4G} (2\lgd) \approx I(A:B)$ so $R_3(A:B) \approx 0$, and in 3(c)(i), $S_R(A:B) \leq \frac{1}{4G} (2\lgc) \approx I(A:B)$ so $R_3(A:B) \approx 0$. 

In 3(c)(ii), we observe that $\lc + \ld < \lgm$, so $CD\gamma$ is in a $\gamma$-eyeglass.  Consider first the case where $\la + \lb < \lgm$, so $AB\gamma$ is also a $\gamma$-eyeglass. Then $A$ is not close to $B$, so each of $A$ and $B$ must be close to one of $C$ or $D$, and each of $C$ or $D$ must be close to one of $A$ and $B$. Suppose without loss of generality $A$ is close to $C$ and $B$ is close to $D$, so $AC\gamma'$ and $BD\gamma'$ are not $\gamma'$-eyeglasses. $S_A = \frac{1}{4G} \la$ implies $\la < \lc + \lgp$, so $AC\gamma'$ is not an $A$-eyeglass, and $S_B = \frac{1}{4G} \lb$ implies $\lb < \ld + \lgp$, so $BD\gamma'$ is not an $B$-eyeglass. Then $EW(CD)$ is bounded by the $C$ and $D$ geodesics, so by entanglement wedge nesting $EW(C)$ is bounded by the $C$ geodesic, so $\lc < \la + \lgp$ and $AC\gamma'$ is not an $C$-eyeglass, and similarly $BD\gamma'$ is not an $D$-eyeglass. Thus $AC\gamma'$ and $BD\gamma'$ are pinwheels.

We saw in Section \ref{4bd} that if the four-boundary wormhole has a decomposition into two pinwheels where the separating geodesic is long, it can't also split along a minimal geodesic into two $\gamma$-eyeglasses. Thus as in case 3(b) in Section \ref{Q34b}, to have $AB\gamma$ and $CD\gamma$ be $\gamma$-eyeglasses, $\lgp$ must be small, and $S_R(A:B) < \frac{1}{2G} \lgp \approx 0$. This might seem surprising as $I(A:B)$ is not obviously small; but to be in this case we must have $\la < \lc + \lgp$ and $\lb < \ld + \lgp$, so $I(A:B) \leq \frac{1}{2G} \lgp$ as well. Thus, if both are eyeglasses $S_R(A:B) \approx 0$, $I(A:B) \approx 0$ and hence, $R_3(A:B) \approx 0$. 

Consider now $\la + \lb > \lgm$, so $AB\gamma$ is not a $\gamma$-eyeglass. As $S_A = \frac{1}{4G} \la$ and $S_B = \frac{1}{4G} \lb$, it is also not an $A$-eyeglass or a $B$-eyeglass, so it must be a pinwheel. We have
\begin{equation}
    S_R(A:B) = \frac{1}{2G} \,  \min \left( \lgp, \, \ltg, \, \lgc, \, \lgcd, \, \lgd \right) \approx \frac{1}{2G} \left( \lgg + X \right) \approx \frac{1}{4G} \left[ \left( \la + \lb - \lgm \right) + 2X \right],  
\end{equation}
where $X$ is a potential contribution from the offset of the endpoints in the $CD\gamma$ $\gamma$-eyeglass. If the endpoints of $\lgg$ are in $C$ or $D$, $\min (\lgc,  \, \lgcd,  \, \lgd) \approx \lgg$ so $X=0$. Thus the largest contribution comes from cases where both endpoints are in the bridge. The maximal value is the length of the bridge, $X = \frac{1}{2}(\lgm - \lc -\ld)$, which gives  
\begin{equation}
    S_R(A:B) \approx  \frac{1}{4G} (\la + \lb - \lc -\ld ) \approx I(A:B),  
\end{equation}
so $R_3(A:B) \approx 0$. 

Thus, at large $\ell$, the four-boundary wormhole always has $R_3$ between any two of the boundaries at most of order one, never growing with $\ell$.

\bibliographystyle{utphys}
\bibliography{bib}

\end{document}